\documentclass[preprint,3p,times]{elsarticle}
\RequirePackage{multirow,booktabs,subfigure,color,array,hhline,makecell}
\usepackage{amssymb}
\usepackage{amsmath}
\usepackage{graphicx}
\usepackage{amsthm}
\usepackage{mathrsfs}
\usepackage{indentfirst}
\usepackage[colorlinks,citecolor=blue,urlcolor=blue]{hyperref}
\allowdisplaybreaks
\usepackage[table]{xcolor}
\usepackage{tikz-network}
\usepackage{pgf}
\usepackage{tikz}
\usepackage{subfigure}

\usepackage{bbm}

\usetikzlibrary{arrows, decorations.pathmorphing, backgrounds, positioning, fit, petri, automata}
\usetikzlibrary{shadows,arrows,positioning}
\RequirePackage{algorithm,algpseudocode}
\allowdisplaybreaks
\usepackage{changes}

\newtheorem{thm}{Theorem}
\newtheorem{defin}{Definition}
\newtheorem{lem}{Lemma}
\newtheorem{assum}{Assumption}
\newtheorem{rem}{Remark}
\newtheorem{cor}{Corollary}

\allowdisplaybreaks[4]
\AtBeginDocument{%
	\providecommand\BibTeX{{%
			\normalfont B\kern-0.5em{\scshape i\kern-0.25em b}\kern-0.8em\TeX}}}
\journal{~}
\begin{document}
\begin{frontmatter}
\title{Community detection in multi-layer bipartite networks}
\author[label1]{Huan Qing\corref{cor1}}
\ead{qinghuan@u.nus.edu~\&~qinghuan07131995@163.com~\&~qinghuan@cqut.edu.cn}
\cortext[cor1]{Corresponding author.}
\address[label1]{School of Economics and Finance, Lab of Financial Risk Intelligent Early Warning and Modern Governance,  Chongqing University of Technology, Chongqing, 400054, China}
\begin{abstract}
The problem of community detection in multi-layer undirected networks has received considerable attention in recent years. However, practical scenarios often involve multi-layer bipartite networks, where each layer consists of two distinct types of nodes. Existing community detection algorithms tailored for multi-layer undirected networks are not directly applicable to multi-layer bipartite networks. To address this challenge, this paper introduces a novel multi-layer degree-corrected stochastic co-block model specifically designed to capture the underlying community structure within multi-layer bipartite networks. Within this framework, we propose an efficient debiased spectral co-clustering algorithm for detecting nodes' communities. We establish the consistent estimation property of our proposed algorithm and demonstrate that an increased number of layers in bipartite networks improves the accuracy of community detection. Through extensive numerical experiments, we showcase the superior performance of our algorithm compared to existing methods. Additionally, we validate our algorithm by applying it to real-world multi-layer network datasets, yielding meaningful and insightful results.
\end{abstract}
\end{frontmatter}
\section{Introduction}\label{sec1}
In network analysis, multi-layer networks have emerged as ubiquitous structures in the real world, encompassing diverse systems from social networks to biological interactions \citep{de2013mathematical,kivela2014multilayer,boccaletti2014structure,de2015structural,de2016physics,pilosof2017multilayer}. These networks, characterized by multiple layers, offer a rich framework for understanding the complex relationships and interactions within and across various systems. For instance, users connect with each other on different platforms such as email, messaging, Facebook, Twitter, Instagram, and WeChat \citep{papalexakis2013more}, where each platform denotes a layer. Similarly, in biological systems, different layers record genes' co-expression relationships of the animal at different developmental stages \citep{narayanan2010simultaneous,bakken2016comprehensive,zhang2017finding}.

In recent years, the problem of community detection in multi-layer networks has garnered significant attention. Community detection aims to identify groups of nodes that are tightly interconnected within each group but relatively sparsely connected to nodes in other groups \citep{girvan2002community,newman2004finding,fortunato2010community,papadopoulos2012community,fortunato2016community,javed2018community}. The significance of community detection in multi-layer networks lies in its ability to reveal hidden structures and patterns that may not be evident from a single-layer analysis \citep{lei2023bias}. By accounting for the richness of interactions in multiple layers, researchers can gain a more comprehensive understanding of the system's overall behavior \citep{kim2015community,huang2021survey}. This, in turn, has led to a surge in research efforts aimed at developing effective and efficient algorithms for community detection in multi-layer networks.

An extensively studied version of this problem assumes that each layer of the multi-layer network is derived from the classical stochastic block model (SBM) \citep{holland1983stochastic}. This model assumes that the probability of an edge forming between two nodes is solely determined by their respective communities, with nodes belonging to the same community exhibiting similar connectivity patterns. \citep{han2015consistent} conducted an asymptotic analysis of a spectral clustering algorithm and a maximum likelihood method by increasing the number of layers while fixing the number of nodes within the context of the multi-layer SBM. \citep{paul2020spectral} focused on the consistent community detection of spectral and matrix factorization methods under the multi-layer SBM.  \citep{lei2020consistent} proposed a least-squares estimation technique and provided its theoretical guarantees of consistency within the multi-layer SBM framework. \citep{jing2021community} introduced a mixture multi-layer SBM and developed a tensor-based approach for identifying the community memberships of both nodes and layers. \citep{fan2022alma} proposed an alternating minimization algorithm that outperforms the tensor-based method in \citep{jing2021community} both theoretically and numerically under the mixture multi-layer SBM introduced in \citep{jing2021community}. \citep{chen2022communityINS} introduced a multi-layer weighted SBM to model multi-layer weighted networks and fitted this model by a variational expectation-maximization algorithm. \citep{lei2023bias} developed an efficient debiased spectral clustering method that significantly outperforms methods considered in \citep{paul2020spectral} and established its theoretical guarantees within the multi-layer SBM framework.

\begin{figure}
\centering
\subfigure[]{\includegraphics[width=0.32\textwidth]{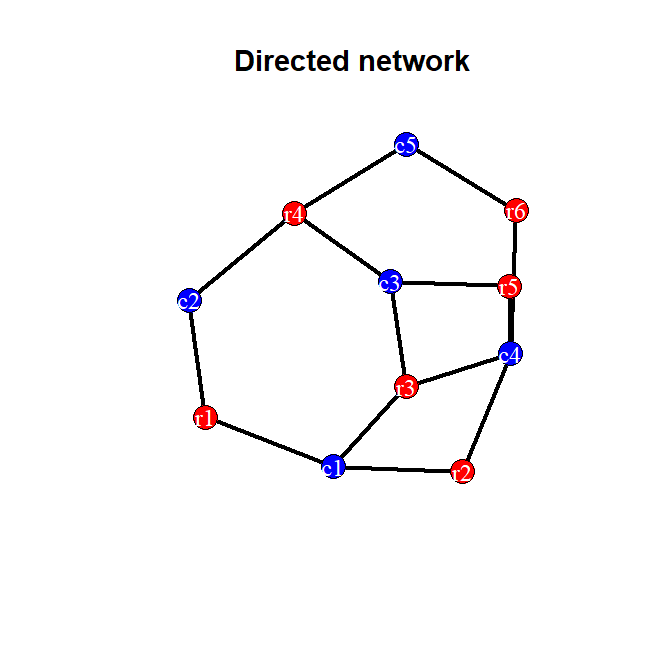}}
\subfigure[]{\includegraphics[width=0.32\textwidth]{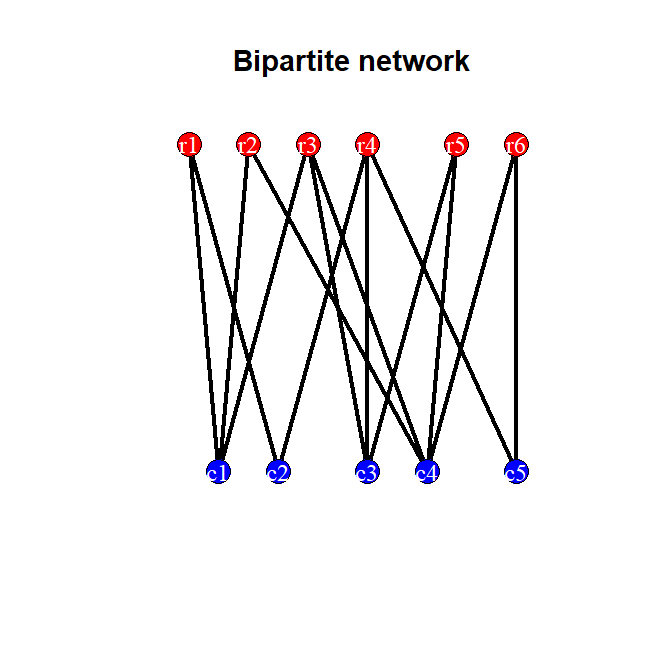}}
\caption{A toy example of transforming a directed network in panel (a) to a bipartite network in panel (b). In both panels, red dots and blue dots represent row nodes and column nodes, respectively.}
\label{DBA} 
\end{figure}

\begin{figure}
\centering
\resizebox{\columnwidth}{!}{
\subfigure[]{\includegraphics[width=0.2\textwidth]{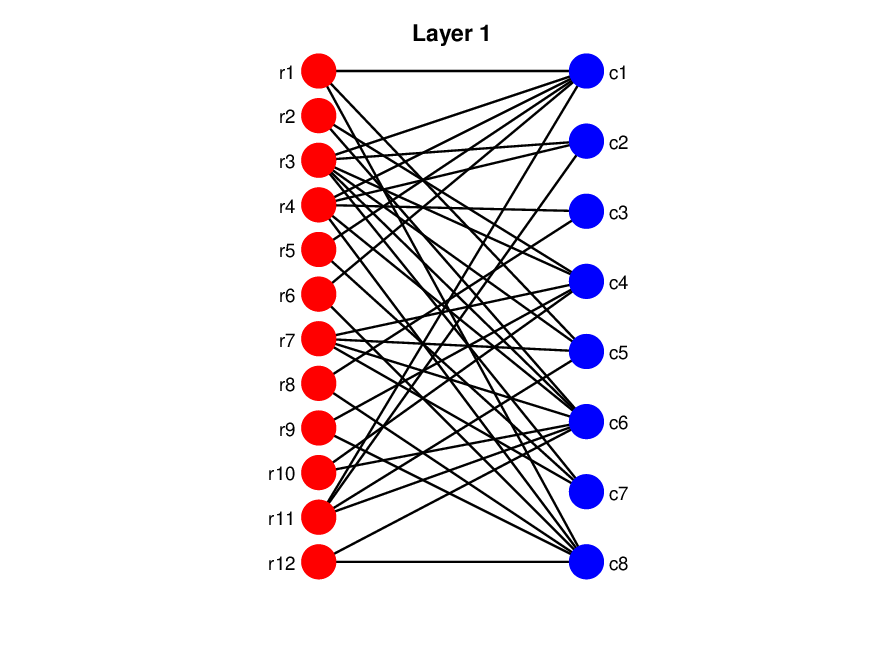}}
\subfigure[]{\includegraphics[width=0.2\textwidth]{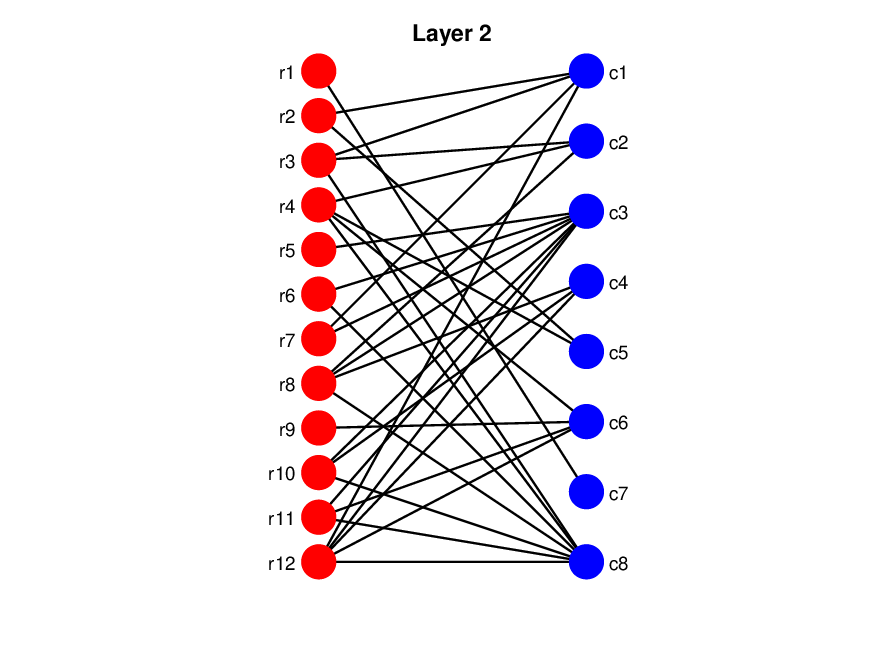}}
\subfigure[]{\includegraphics[width=0.2\textwidth]{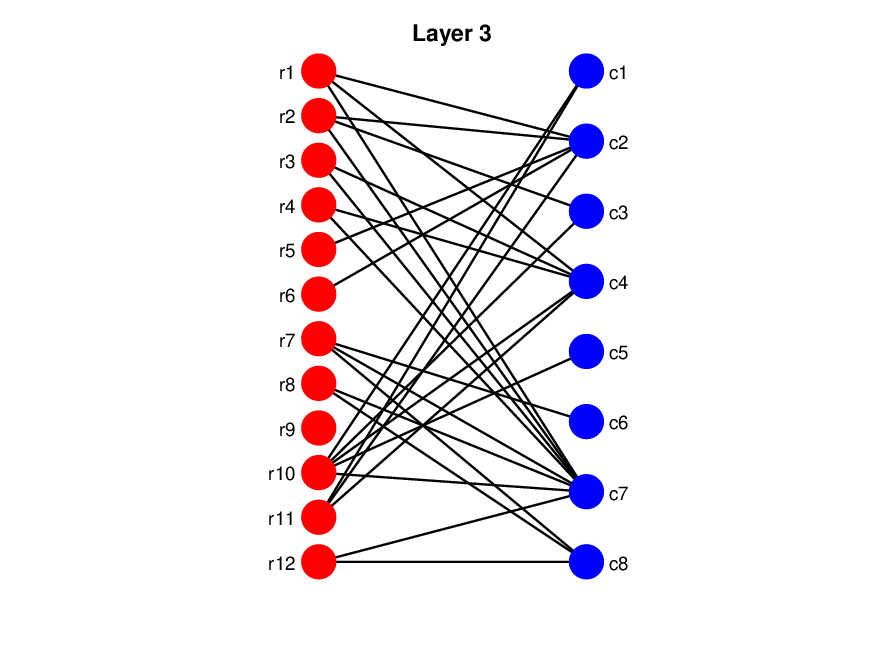}}
}
\caption{A multi-layer bipartite network with 12 row nodes, 8 column nodes, and 3 layers.}
\label{BipartiteNLayer3} 
\end{figure}

While the multi-layer SBM is powerful in modeling multi-layer networks with latent community structures, it assumes that the interactions between nodes are symmetric or undirected, limiting its ability to capture community patterns in multi-layer directed networks. To take into the asymmetry in multi-layer directed networks, \citep{su2023spectral} proposed a multi-layer stochastic co-block model (multi-layer ScBM) and developed an efficient, theoretically consistent debiased spectral co-clustering approach to detect communities. This model assumes that each layer of the multi-layer directed network is generated from the stochastic co-block model (ScBM) introduced in \citep{rohe2016co}. On the one hand, since directed networks naturally generalize to bipartite networks (also known as two-mode networks), where a bipartite network contains two types of nodes, and edges exist solely between nodes of different types \citep{latapy2008basic,rohe2016co, vasques2018degree}, theoretical results in \citep{su2023spectral} are only suited for multi-layer directed networks and inapplicable for multi-layer bipartite networks. Figure \ref{DBA} provides an illustrative example of transforming a directed network into a bipartite network. Figure \ref{BipartiteNLayer3} illustrates an example of a multi-layer bipartite network, consisting of three layers. On the other hand, the multi-layer ScBM assumes that nodes within the same communities share similar connectivity patterns and does not consider the degree heterogeneity of nodes while nodes have various degrees in real-world networks \citep{karrer2011stochastic}. To address this, the degree-corrected stochastic block model (DCSBM) proposed by \citep{karrer2011stochastic} introduced degree heterogeneity parameters, extending SBM to networks with heterogeneous degrees. Similarly, the degree-corrected stochastic co-block model (DC-ScBM) in \citep{rohe2016co} extended DCSBM to bipartite networks. Given these limitations of the multi-layer ScBM considered in \citep{su2023spectral}, this paper focuses on the problem of community detection in multi-layer bipartite networks by assuming that each layer of a multi-layer bipartite network is generated from DC-ScBM and developing an efficient method to detect nodes' communities. Our main contributions are as follows:
\begin{itemize}
  \item We introduce the multi-layer degree-corrected stochastic co-block model (multi-layer DC-ScBM), a highly flexible network model designed to capture the latent community structures within multi-layer bipartite networks. This model encompasses numerous previous models, including the multi-layer ScBM, multi-layer SBM, DC-ScBM, ScBM, DCSBM, and SBM, as specific cases.
  \item To detect nodes' communities within the framework of the multi-layer DC-ScBM for multi-layer bipartite networks, we propose an efficient debiased spectral co-clustering method. Specifically, we apply the K-means algorithm to the row-normalized versions of a select few top eigenvectors of two debiased matrices, enabling us to accurately identify the community memberships of nodes. To the best of our knowledge, our study is the first to investigate the community detection problem within the context of the multi-layer DC-ScBM for multi-layer bipartite networks.
  \item We establish the consistency of our method by providing theoretical upper bounds for its clustering errors as the number of nodes and/or layers increases under the multi-layer DC-ScBM. Our theoretical findings underscore the benefits of incorporating multiple layers in community detection tasks.
  \item Through simulated data, we demonstrate that our method significantly surpasses state-of-the-art community detection methods for multi-layer bipartite networks. Furthermore, real-world data examples illustrate the practical utility of our proposed method, yielding meaningful and insightful results.
\end{itemize}

The rest of this paper is structured as follows. Section \ref{sec2} describes the model. Section \ref{sec3} presents the proposed method. Section \ref{sec4} establishes its consistency. Section \ref{sec5} offers simulations to demonstrate its effectiveness. Section \ref{sec6realdata} includes applications using real-world data. Section \ref{sec7} concludes. Technical proofs are given in \ref{SecProofs}.

\emph{Notation.} For brevity, we denote the set $\{1,2,\ldots,m\}$ as $[m]$ and the $m\times m$ identity matrix as $I_{m\times m}$ for any positive integer $m$. We employ $\|x\|_{q}$ to represent the $l_{q}$ norm of any vector $x$. For any matrix $X$, $X'$ denotes its transpose, $\|X\|$ and $\|X\|_{F}$ represent the spectral and Frobenius norms, respectively, $\mathrm{rank}(X)$ denotes its rank, $X(i,:)$ indicates its $i$-th row, and $\lambda_{k}(X)$ stands for its $k$-th largest eigenvalue in magnitude. Additionally, $\mathbb{E}[\cdot]$ and $\mathbb{P}(\cdot)$ are used to represent expectation and probability, respectively.
\section{The multi-layer degree-corrected stochastic co-block model}\label{sec2}
\begin{figure}
\centering
\resizebox{\columnwidth}{!}{
\subfigure[]{\includegraphics[width=0.2\textwidth]{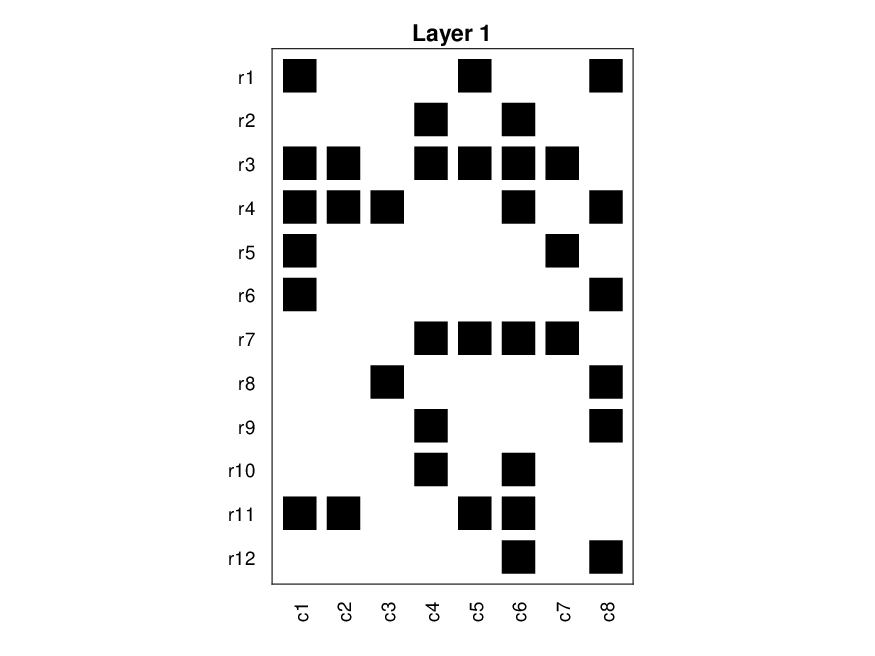}}
\subfigure[]{\includegraphics[width=0.2\textwidth]{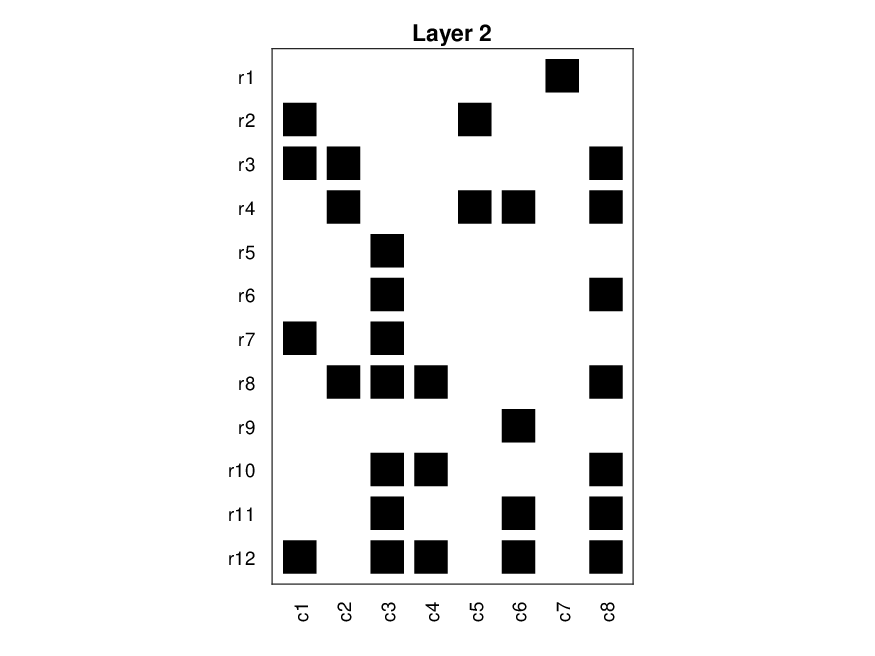}}
\subfigure[]{\includegraphics[width=0.2\textwidth]{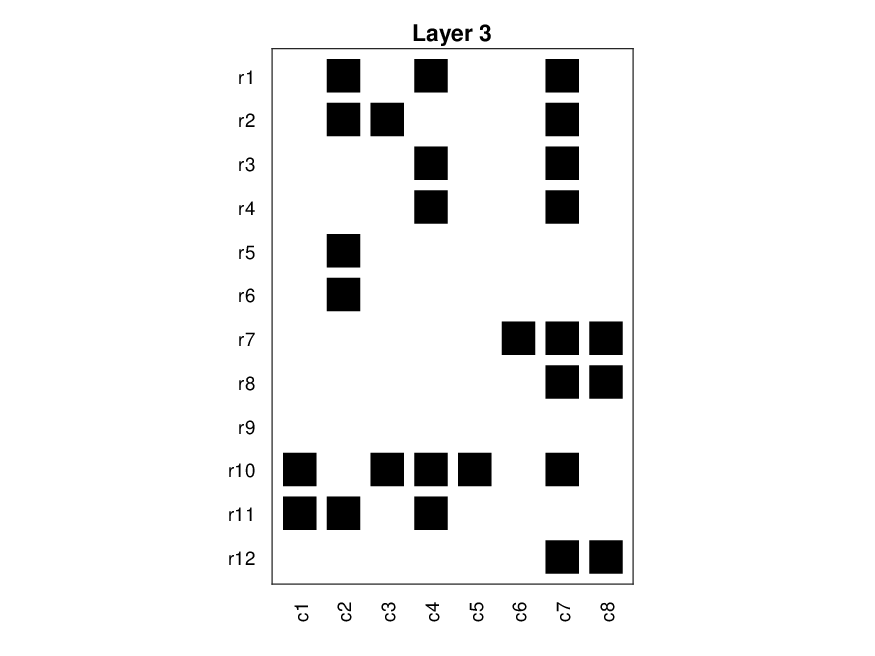}}
}
\caption{The adjacency matrices corresponding to the 3 distinct layers of the multi-layer bipartite network shown in Figure \ref{BipartiteNLayer3}, where black square represents 1.}
\label{BipartiteALayer3} 
\end{figure}

Consider a multi-layer bipartite network, denoted as $\mathcal{N}$, composed of $L$ distinct layers. This network possesses $n_{r}$ common row nodes and $n_{c}$ common column nodes. Specifically, $\mathcal{N}_{l}$ represents the bipartite network situated in the $l$-th layer for $l\in[L]$. For each bipartite network $\mathcal{N}_{l}$ within the multi-layer structure, an adjacency matrix $A_{l}$ is defined, which belongs to the set $\{0,1\}^{n_{r}\times n_{c}}$. This matrix serves as a comprehensive record of the connectivity patterns between row and column nodes. Precisely, $A_{l}(i_{r},j_{c})$ is assigned a value of 1 whenever a directed edge exists from row node (sending node) $ i_{r}$ to column node (receiving node) $j_{c}$, and it is set to 0 in the absence of such an edge for $ i_{r}\in[n_{r}], j_{c}\in[n_{c}], l\in[L]$. Here, the subscript ``r'' represents rows, and the subscript ``c'' denotes columns. Figure \ref{BipartiteALayer3} shows the adjacency matrices corresponding to the three layers of the multi-layer bipartite network displayed in Figure \ref{BipartiteNLayer3}. It is noteworthy that when the set of row nodes coincides with the set of column nodes, the bipartite network $\mathcal{N}$ simplifies to a multi-layer directed network. Furthermore, in the special case where the set of row nodes coincides with the set of column nodes and all edges are undirected, $\mathcal{N}$ reduces to a multi-layer undirected network.

Suppose that all layers share a common row community structure and a common column community structure. Specifically, all row nodes belong to $K_{r}$ disjoint row communities, denoted as $\{\mathcal{C}_{r,1},\mathcal{C}_{r,2},\ldots, \mathcal{C}_{r,K_{r}}\}$, while all column nodes belong to $K_{c}$ disjoint column communities, represented as  $\{\mathcal{C}_{c,1},\mathcal{C}_{c,2},\ldots, \mathcal{C}_{c,K_{c}}\}$. Here, $\mathcal{C}_{r,k_{r}}$ is a collection of row nodes belonging to the $k_{r}$-th row community for $k_{r}\in[K_{r}]$ and $\mathcal{C}_{c,k_{c}}$ is a collection of column nodes belonging to the $k_{c}$-th column community for $k_{c}\in[K_{c}]$. We assume that the number of row communities $K_{r}$ and the number of column communities $K_{c}$ are known in this paper.

Let $Z_{r}$ be an $n_{r} \times K_{r}$ matrix, with entries in $\{0,1\}$, representing the common row membership matrix of row nodes. Specifically, $Z_{r}(i_{r}, k_{r}) = 1$ if the row node $i_{r}$ belongs to the $k_{r}$-th row community, $\mathcal{C}_{r,k_{r}}$, and 0 otherwise, for each $i_{r} \in [n_{r}]$. Since we assume mutual exclusivity among row communities, each row of $Z_{r}$ features exactly one entry as 1, while the remaining $K_{r}-1$ entries are zeros. Assuming that each row community encompasses at least one row node, we infer that the rank of $Z_{r}$ is $K_{r}$. Analogously, let $Z_{c}$ be an $n_{c} \times K_{c}$ matrix, also with entries in $\{0,1\}$, serving as the common column membership matrix of column nodes. Here, $Z_{c}(j_{c},k_{c}) = 1$ if the column node $j_{c}$ belongs to the $k_{c}$-th column community, and 0 otherwise, for each $j_{c} \in [n_{c}]$. Similar to $Z_{r}$, each row of $Z_{c}$ contains exactly one 1, with the remaining entries as zeros, and the rank of $Z_{c}$ is $K_{c}$.

Let $\theta_{r}$ be an $n_{r}\times1$ vector such that $\theta_{r}(i_{r})$ is the common degree heterogeneity shared by all layers of row node $i_{r}$ and $\theta_{r}(i_{r})\in(0,1]$ for $i_{r}\in[n_{r}]$. Similarly, let $\theta_{c}\in(0,1]^{n_{c}\times1}$ be the common degree heterogeneity vector shared by all layers of column nodes. Define $\Theta_{r}$ as an $n_{r} \times n_{r}$ diagonal matrix and $\Theta_{c}$ as an $n_{c} \times n_{c}$ diagonal matrix such that $\Theta_{r}(i_{r},i_{r}) = \theta_{r}(i_{r})$ for each $i_{r} \in [n_{r}]$, and $\Theta_{c}(j_{c},j_{c}) = \theta_{c}(j_{c})$ for each $j_{c} \in [n_{c}]$.

Let $B_{l}$ represent a $K_{r}\times K_{c}$ asymmetric matrix such that $B_{l}\in[0,1]^{K_{r}\times K_{c}}$ for $l\in[L]$. We refer to $B_{l}$ as the block connectivity matrix. Given the row community classification matrix $Z_{r}$, the column community classification matrix $Z_{c}$, along with the row degree heterogeneity vector $\theta_{r}$ and the column degree heterogeneity vector $\theta_{c}$, as well as the $L$ block connectivity matrices $\{B_{l}\}^{L}_{l=1}$, we assume that the adjacency matrix for each layer of the multi-layer bipartite network $\mathcal{N}$ is generated by the degree-corrected stochastic co-block model \citep{rohe2016co}. To be precise, to model a multi-layer bipartite network with true row membership matrix $Z_{r}$ and true column membership matrix $Z_{c}$, we consider the following model, which is a multi-layer version of DC-ScBM.
\begin{defin}\label{MLDC-ScBM}
(Multi-layer DC-ScBM) Let $A_{l}\in\{0,1\}^{n_{r}\times n_{c}}$ be the adjacency matrix of the $l$-th layer bipartite network $\mathcal{N}_{l}$, $Z_{r}\in\mathbb{M}_{n_{r}, K_{r}}$ be the common row membership matrix, $Z_{c}\in\mathbb{M}_{n_{c}, K_{c}}$ be the common column membership matrix, $\theta_{r}\in(0,1]^{n_{r}\times1}$ be the common row degree heterogeneity vector, $\theta_{c}\in(0,1]^{n_{c}\times1}$ be the common column degree heterogeneity vector, and $B_{l}\in[0,1]^{K_{r}\times K_{c}}$ be the $l$-th block connectivity matrix for $l\in[L]$. Our multi-layer DC-ScBM assumes that for each pair $(i_{r}, j_{c})$ (with $i_{r}\in[n_{r}]$ and $j_{c}\in[n_{c}]$) and each layer $l\in[L]$, $A_{l}(i_{r}, j_{c})$ are independently generated from a Bernoulli distribution such that
\begin{align}\label{OmegaL}
A_{l}(i_{r},j_{c})\sim\mathrm{Bernoulli}(\Omega_{l}(i_{r}, j_{c})), \mathrm{where~}\Omega_{l}:=\Theta_{r}Z_{r}B_{l}Z'_{c}\Theta_{c}.
\end{align}
\end{defin}
By Equation (\ref{OmegaL}), we see that $\mathbb{E}[A_{l}]=\Omega_{l}$ for $l\in[L]$ under the multi-layer DC-ScBM. Meanwhile, Equation (\ref{OmegaL}) gives that $\mathbb{P}(A_{l}(i_{r}, j_{c})=1)=\theta_{r}(i_{r})\theta_{c}(j_{c})Z_{r}(i_{r},:)B_{l}Z'_{c}(j_{c},:)$ for $i_{r}\in[n_{r}], j_{c}\in[n_{c}], l\in[L]$ under the multi-layer DC-ScBM, which implies that the probability of generating an edge from row node $i_{r}$ to column node $j_{c}$ does not only depend on their belonging communities but also relies on their unique characteristics through the degree heterogeneity parameters $\theta_{r}(i_{r})$ and $\theta_{c}(j_{c})$. Equation (\ref{OmegaL}) also specifies that the multi-layer DC-ScBM is parameterized by $(Z_{r}, Z_{c}, \Theta_{r}, \Theta_{c}, \{B_{l}\}^{L}_{l=1})$. Suppose that $A_{1}, A_{2}, \ldots, A_{L}$ are generated from the multi-layer DC-ScBM according to Equation (\ref{OmegaL}), the primary task of community detection in multi-layer bipartite networks is to recover the row membership matrix $Z_{r}$ and the column membership matrix $Z_{c}$.

Our multi-layer DC-ScBM includes several previous network models as special cases. For example, when all entries of $\theta_{r}$ (and $\theta_{c}$) are equal and all row nodes are the same as column nodes, our multi-layer DC-ScBM reduces to the multi-layer ScBM studied in \citep{su2023spectral}. When $Z_{r}=Z_{c}, \Theta_{r}=\Theta_{c}$, and $B_{l}$ is symmetric for $l\in[L]$ such that $\mathcal{N}$ is undirected, and all entries of $\theta_{r}$ are equal, our multi-layer DC-ScBM degenerates to the popular multi-layer SBM considered in \citep{han2015consistent,paul2016consistent,paul2020spectral,lei2020consistent,lei2023bias}. When $L=1$, the multi-layer DC-ScBM reduces to the DC-ScBM \citep{rohe2016co}. When $L=1$ and all entries of $\theta_{r}$ (and $\theta_{c}$) are the same, the multi-layer DC-ScBM reduces to the ScBM \citep{rohe2016co}. When $L=1$ and $\mathcal{N}$ is undirected, the multi-layer DC-ScBM degenerates to the well-known DCSBM \citep{karrer2011stochastic}. Furthermore, if all entries of the degree heterogeneity vector are the same, the multi-layer DC-ScBM degenerates to the well-known SBM \citep{holland1983stochastic}.
\section{Algorithm}\label{sec3}
In this section, we describe one community detection algorithm to efficiently detect both row and column communities for multi-layer bipartite networks generated from the multi-layer DC-ScBM. The following lemma is the cornerstone of our algorithm.
\begin{lem}\label{EigOsum2}
Consider the multi-layer DC-ScBM parameterized by $(Z_{r},Z_{c},\Theta_{r},\Theta_{c},\{B_{l}\}^{L}_{l=1})$.
\begin{itemize}
  \item Suppose that $\mathrm{rank}(\sum_{l\in[L]}B_{l}B'_{l})=K$, where $K\leq K_{r}$. Set $\tilde{S}_{r}=\sum_{l\in[L]}\Omega_{l}\Omega'_{l}$. Let $U_{r}\Lambda_{r}U'_{r}$ be the top $K$ eigen-decomposition of $\tilde{S}_{r}$ such that $U'_{r}U_{r}=I_{K\times K}$ and $\Lambda_{r}$ is a $K\times K$ diagonal matrix with $k$-th diagonal element being the $k$-th largest eigenvalue in magnitude of $\tilde{S}_{r}$ for $k\in[K]$. Let $U^{*}_{r}$ be the row-normalized version of $U_{r}$ such that $U^{*}_{r}(i_{r},:)=\frac{U_{r}(i_{r},:)}{\|U_{r}(i_{r},:)\|_{F}}$ for $i_{r}\in[n_{r}]$. Then $U^{*}_{r}(i_{r},:)=U^{*}_{r}(\tilde{i}_{r},:)$ if and only if $Z_{r}(i_{r},:)=Z_{r}(\tilde{i}_{r},:)$ for $i_{r}\in[n_{r}], \tilde{i}_{r}\in[n_{r}]$. Furthermore, when $K=K_{r}$, we have $\|U^{*}_{r}(i_{r},:)-U^{*}_{r}(\tilde{i}_{r},:)\|_{F}=\sqrt{2}$ for any two row nodes satisfying $Z_{r}(i_{r},:)\neq Z_{r}(\tilde{i}_{r},:)$.
    \item Suppose that $\mathrm{rank}(\sum_{l\in[L]}B'_{l}B_{l})=\tilde{K}$, where $\tilde{K}\leq K_{c}$. Set $\tilde{S}_{c}=\sum_{l\in[L]}\Omega'_{l}\Omega_{l}$. Let $U_{c}\Lambda_{c}U'_{c}$ be the top $\tilde{K}$ eigen-decomposition of $\tilde{S}_{c}$ such that $U'_{c}U_{c}=I_{\tilde{K}\times \tilde{K}}$ and $\Lambda_{c}$ is a $\tilde{K}\times \tilde{K}$ diagonal matrix with $\tilde{k}$-th diagonal element being the $\tilde{k}$-th largest eigenvalue in magnitude of $\tilde{S}_{c}$ for $\tilde{k}\in[\tilde{K}]$. Let $U^{*}_{c}$ be the row-normalized version of $U_{c}$ such that $U^{*}_{c}(j_{c},:)=\frac{U_{c}(j_{c},:)}{\|U_{c}(j_{c},:)\|_{F}}$ for $j_{c}\in[n_{c}]$. Then $U^{*}_{c}(j_{c},:)=U^{*}_{c}(\tilde{j}_{c},:)$ if and only if $Z_{c}(j_{c},:)=Z_{c}(\tilde{j}_{c},:)$ for $j_{c}\in[n_{c}], \tilde{j}_{c}\in[n_{c}]$. Furthermore, when $\tilde{K}=K_{c}$, we have $\|U^{*}_{c}(j_{c},:)-U^{*}_{c}(\tilde{j}_{c},:)\|_{F}=\sqrt{2}$ for any two column nodes satisfying $Z_{c}(j_{c},:)\neq Z_{c}(\tilde{j}_{c},:)$.
\end{itemize}
\end{lem}
By Lemma \ref{EigOsum2}, it becomes evident that $U^{*}_{r}$ possesses precisely $K_{r}$ unique rows. This observation suggests that by implementing the K-means algorithm on all rows of $U^{*}_{r}$ with $K_{r}$ clusters, we can accurately identify the row communities. Analogously, applying the K-means algorithm to all rows of $U^{*}_{c}$ with $K_{c}$ clusters will precisely reveal the column communities. To streamline our analysis, we will concentrate on the scenario where both $\sum_{l\in[L]}B_{l}B'_{l}$ and $\sum_{l\in[L]}B'_{l}B_{l}$ are of full rank, thereby implying $K=K_{r}$ and $\tilde{K}=K_{c}$. Consequently, the $l_{2}$ norm of the difference between any two distinct rows of $U^{*}_{r}$ (and similarly, $U^{*}_{c}$) equals $\sqrt{2}$.

Set $S_{r}=\sum_{l\in[L]}(A_{l}A'_{l}-D^{r}_{l})$, where $D^{r}_{l}$ is an $n_{r}\times n_{r}$ diagonal matrix with $i_{r}$-th diagonal element being $\sum_{j_{c}\in[n_{c}]}A_{l}(i_{r},j_{c})$ for $i_{r}\in[n_{r}], l\in[L]$. Set $S_{c}=\sum_{l\in[L]}(A'_{l}A_{l}-D^{c}_{l})$, where $D^{c}_{l}$ is an $n_{c}\times n_{c}$ diagonal matrix with $j_{c}$-th diagonal element being $\sum_{i_{r}\in[n_{r}]}A_{l}(i_{r},j_{c})$ for $j_{c}\in[n_{c}], l\in[L]$. According to the analysis in \citep{lei2023bias,su2023spectral}, $S_{r}$ and $S_{c}$ serve as debiased estimations of $\tilde{S}_{r}$ and $\tilde{S}_{c}$, respectively, whereas $\sum_{l\in[L]}A_{l}A'_{l}$ and $\sum_{l\in[L]}A'_{l}A_{l}$ are biased estimates of $\tilde{S}_{r}$ and $\tilde{S}_{c}$, respectively. Let $\hat{U}_{r}\hat{\Lambda}_{r}\hat{U}'_{r}$ be the top $K_{r}$ eigen-decomposition of $S_{r}$ such that $\hat{U}'_{r}\hat{U}_{r}=I_{K_{r}\times K_{r}}$ and $\hat{\Lambda}_{r}$ is a $K_{r}\times K_{r}$ diagonal matrix with $k_{r}$-th diagonal entry being the $k_{r}$-th largest eigenvalue in magnitude of $S_{r}$ for $k_{r}\in[K_{r}]$. Let $\hat{U}^{*}_{r}$ be the row-normalized version of $\hat{U}_{r}$ such that $\hat{U}^{*}_{r}(i_{r},:)=\frac{\hat{U}_{r}(i_{r},:)}{\|\hat{U}_{r}(i_{r},:)\|_{F}}$ for $i_{r}\in[n_{r}]$. Since $S_{r}$ is a debiased estimate of $\tilde{S}_{r}$, $\hat{U}^{*}_{r}$ is a good approximation of $U^{*}_{r}$ and it should have roughly $K_{r}$ distinct rows. Consequently, applying the K-means algorithm to all rows of $\hat{U}^{*}_{r}$ should return a good estimation of row communities. Similarly, letting $\hat{U}_{c}\hat{\Lambda}_{c}\hat{U}'_{c}$ be the top $K_{c}$ eigen-decomposition of $S_{c}$ and $\hat{U}^{*}_{c}$ be the row-normalized version of $\hat{U}_{c}$, applying K-means algorithm to all rows of $\hat{U}^{*}_{c}$ should provide an accurate estimation of column communities. The following four-stage algorithm which we call NcDSoS summarizes the above analysis.
\begin{algorithm}
\caption{\underline{N}ormalized spectral \underline{c}o-clustering based on \underline{d}ebiased \underline{s}um \underline{o}f
\underline{s}quared matrices (NcDSoS)}
\label{alg:NcDSoS}
\begin{algorithmic}[1]
\Require Adjacency matrices $A_{1}, A_{2}, \ldots, A_{L}$, number of row communities $K_{r}$, and number of column communities $K_{c}$, where $A_{l}\in\{0,1\}^{n_{r}\times n_{c}}$ for $l\in[L]$.
\Ensure Estimated row membership matrix $\hat{Z}_{r}\in\mathbb{M}_{n_{r},K_{r}}$ and estimated column membership matrix $\hat{Z}_{c}\in\mathbb{M}_{n_{c},K_{c}}$.
\State Compute $S_{r}=\sum_{l\in[L]}(A_{l}A'_{l}-D^{r}_{l})$ and $S_{c}=\sum_{l\in[L]}(A'_{l}A_{l}-D^{c}_{l})$.
\State Compute the $n_{r}\times K_{r}$ matrix $\hat{U}_{r}$ containing the top $K_{r}$ eigenvectors (ordered in absolute eigenvalue) of $S_{r}$ and the $n_{c}\times K_{c}$ matrix $\hat{U}_{c}$ containing the top $K_{c}$ eigenvectors (ordered in absolute eigenvalue) of $S_{c}$.
\State Compute $\hat{U}^{*}_{r}$ and $\hat{U}^{*}_{c}$ by normalizing each row of $\hat{U}_{r}$ and $\hat{U}_{c}$ to have unit length, respectively.
\State Run K-means algorithm on rows of $\hat{U}^{*}_{r}$ with $K_{r}$ clusters to obtain the estimated row membership matrix $\hat{Z}_{r}$ and run K-means algorithm on rows of $\hat{U}^{*}_{c}$ with $K_{c}$ clusters to obtain the estimated column membership matrix $\hat{Z}_{c}$.
\end{algorithmic}
\end{algorithm}
\section{Theoretical properties}\label{sec4}
We consider the \emph{Clustering error} introduced in \citep{joseph2016impact} to evaluate the performance of a community detection algorithm when the true row and column membership matrices are available. For $k_{r}\in[K_{r}]$, let $\hat{\mathcal{C}}_{r,k_{r}}$ be the estimated $k_{r}$-th row community, i.e., $\hat{\mathcal{C}}_{r,k_{r}}=\{i\in[n_{r}]: \hat{Z}_{r}(i,k_{r})=1\}$. For $k_{c}\in[K_{c}]$, let $\hat{\mathcal{C}}_{c,k_{c}}$ be the estimated $k_{c}$-th column community, i.e., $\hat{\mathcal{C}}_{c,k_{c}}=\{j\in[n_{c}]: \hat{Z}_{c}(j,k_{c})=1\}$. The clustering errors for row communities and column communities are defined as
\begin{align*}
&\hat{f}_{r}=\mathrm{min}_{\pi\in S_{K_{r}}}\mathrm{max}_{k_{r}\in[K_{r}]}(|\mathcal{C}_{r,k_{r}}\cap \mathcal{\hat{\mathcal{C}}}^{c}_{r,\pi(k_{r})}|+|\mathcal{C}^{c}_{r,k_{r}}\cap \mathcal{\hat{\mathcal{C}}}_{r,\pi(k_{r})}|)/n_{r,k_{r}},\\
&\hat{f}_{c}=\mathrm{min}_{\pi\in S_{K_{c}}}\mathrm{max}_{k_{c}\in[K_{c}]}(|\mathcal{C}_{c,k_{c}}\cap \mathcal{\hat{\mathcal{C}}}^{c}_{c,\pi(k_{c})}|+|\mathcal{C}^{c}_{c,k_{c}}\cap \mathcal{\hat{\mathcal{C}}}_{c,\pi(k_{c})}|)/n_{c,k_{c}},
\end{align*}
where $n_{r,k_{r}}$ is the number of row nodes belonging to the $k_{r}$-th row community, $n_{c,k_{c}}$ is the number of column nodes belonging to the $k_{c}$-th column community, $S_{K_{r}}$ is the set of all permutations of $\{1,2,\ldots, K_{r}\}$, $S_{K_{c}}$ is the set of all permutations of $\{1,2,\ldots, K_{c}\}$, and the superscript $c$ represents a complementary set.

Our main theoretical result provides upper bounds on Clustering errors $\hat{f}_{r}$ and $\hat{f}_{c}$ for both row and column nodes. To establish our theoretical guarantees, we need the following assumption to control the overall sparsity of the multi-layer bipartite network $\mathcal{N}$.
\begin{assum}\label{Assum2}
$\mathrm{min}(\theta_{r,\mathrm{max}}\|\theta_{r}\|_{1}\|\theta_{c}\|^{2}_{F},\theta_{c,\mathrm{max}}\|\theta_{c}\|_{1}\|\theta_{r}\|^{2}_{F})\geq\frac{\mathrm{log}(n_{r}+n_{c}+L)}{L}$, where $\theta_{r,\mathrm{max}}=\mathrm{max}_{i\in[n_{r}]}\theta_{r}(i)$ and $\theta_{c,\mathrm{max}}=\mathrm{max}_{j\in[n_{c}]}\theta_{c}(j)$.
\end{assum}

At first glance, Assumption \ref{Assum2} might be complex. To gain a deeper understanding, let's consider a particular scenario where we set $n_{r}=n_{c}=n, \Theta_{r}=\Theta_{c}=\sqrt{\rho}I_{n\times n}$ for $\rho\in(0,1]$. For this case, Assumption \ref{Assum2} simplifies to the more straightforward form $\rho n\geq\frac{\mathrm{log}(2n+L)}{L}$. Essentially, this indicates that the sparsity parameter $\rho$ should decrease at a rate slower than $\frac{\mathrm{log}(2n+L)}{L}$, which is a relatively weak requirement for $\rho$. Meanwhile, this also underscores the benefit of exploring multi-layer networks for community detection as increasing the number of layers $L$ decreases the lower bound requirement on the network's sparsity. Furthermore, when our multi-layer DC-ScBM degenerates to the multi-layer SBM, the assumption $\rho n\geq\frac{\mathrm{log}(2n+L)}{L}>\frac{\mathrm{log}(n+L)}{L}$ is consistent with the sparsity requirement in Theorem 1 of \citep{lei2023bias} under the multi-layer SBM model. This guarantees the optimality of our sparsity requirement in Assumption \ref{Assum2}.

The lemma below precisely quantifies the spectral norm of $\|S_{r}-\tilde{S}_{r}\|$ and $\|S_{c}-\tilde{S}_{c}\|$. In the core of the analysis, we take advantage of the rectangular case of matrix Bernstein inequality \citep{tropp2012user}.
\begin{lem}\label{boundSsum}
For the multi-layer DC-ScBM parameterized by $(Z_{r},Z_{c},\Theta_{r},\Theta_{c},\{B_{l}\}^{L}_{l=1})$, suppose Assumption \ref{Assum2} holds, with probability at least $1-o(\frac{1}{n_{r}+n_{c}+L})$, we have
\begin{align*}
&\|S_{r}-\tilde{S}_{r}\|=O(\sqrt{\theta_{r,\mathrm{max}}\|\theta_{r}\|_{1}\|\theta_{c}\|^{2}_{F}L\mathrm{log}(n_{r}+n_{c}+L)})+O(\theta^{2}_{r,\mathrm{max}}\|\theta_{c}\|^{2}_{F}L),\\
&\|S_{c}-\tilde{S}_{c}\|=O(\sqrt{\theta_{c,\mathrm{max}}\|\theta_{c}\|_{1}\|\theta_{r}\|^{2}_{F}L\mathrm{log}(n_{r}+n_{c}+L)})+O(\theta^{2}_{c,\mathrm{max}}\|\theta_{r}\|^{2}_{F}L).
\end{align*}
\end{lem}
For our theoretical analysis, we need to pose conditions on the block connectivity matrices $\{B_{l}\}^{L}_{l=1}$. The following assumption requires a liner growth of $|\lambda_{K_{r}}(\sum_{l\in[L]}B_{l}B'_{l})|$ and $|\lambda_{K_{c}}(\sum_{l\in[L]}B'_{l}B_{l})|$ when the number of layers $L$ increases.
\begin{assum}\label{Assum22}
$|\lambda_{K_{r}}(\sum_{l\in[L]}B_{l}B'_{l})|\geq c_{1}L$ and $|\lambda_{K_{c}}(\sum_{l\in[L]}B'_{l}B_{l})|\geq c_{2}L$  for some constants $c_{1}>0$ and $c_{2}>0$.
\end{assum}
\begin{rem}
Here, we explain why conditions in Assumption \ref{Assum22} are mild. Note that $B_{l}$ is a $K_{r}\times K_{c}$ matrix, Assumption \ref{Assum22} does not require each $B_{l}$ being full rank but only requires $\sum_{l\in[L]}B_{l}B'_{l}$ and $\sum_{l\in[L]}B'_{l}B_{l}$ being full rank.
\end{rem}
The following theorem represents our primary theoretical finding for the proposed NcDSoS approach. It establishes upper bounds for the clustering error of both row and column communities, expressed in terms of various model parameters, within the context of the multi-layer DC-ScBM model.
\begin{thm}\label{mainNcDSoS}
Let $\hat{Z}_{r}$ and $\hat{Z}_{c}$ be obtained from Step 4 of Algorithm \ref{alg:NcDSoS}. For the multi-layer DC-ScBM parameterized by $(Z_{r},Z_{c},\Theta_{r},\Theta_{c},\{B_{l}\}^{L}_{l=1})$, when Assumptions \ref{Assum2} and \ref{Assum22} hold, with probability at least $1-o(\frac{1}{n_{r}+n_{c}+L})$, we have
\begin{align*}
&\hat{f}_{r}=O(\frac{\theta^{3}_{r,\mathrm{max}}K^{2}_{r}n_{r,\mathrm{max}}\|\theta_{r}\|_{1}\|\theta_{c}\|^{2}_{F}\mathrm{log}(n_{r}+n_{c}+L)}{\theta^{6}_{r,\mathrm{min}}\theta^{4}_{c,\mathrm{min}}n^{3}_{r,\mathrm{min}}n^{2}_{c,\mathrm{min}}L})+O(\frac{\theta^{6}_{r,\mathrm{max}}K^{2}_{r}n_{r,\mathrm{max}}\|\theta_{c}\|^{4}_{F}}{\theta^{6}_{r,\mathrm{min}}\theta^{4}_{c,\mathrm{min}}n^{3}_{r,\mathrm{min}}n^{2}_{c,\mathrm{min}}})\\
&\hat{f}_{c}=O(\frac{\theta^{3}_{c,\mathrm{max}}K^{2}_{c}n_{c,\mathrm{max}}\|\theta_{c}\|_{1}\|\theta_{r}\|^{2}_{F}\mathrm{log}(n_{r}+n_{c}+L)}{\theta^{4}_{r,\mathrm{min}}\theta^{6}_{c,\mathrm{min}}n^{2}_{r,\mathrm{min}}n^{3}_{c,\mathrm{min}}L})+O(\frac{\theta^{6}_{c,\mathrm{max}}K^{2}_{c}n_{c,\mathrm{max}}\|\theta_{r}\|^{4}_{F}}{\theta^{4}_{r,\mathrm{min}}\theta^{6}_{c,\mathrm{min}}n^{2}_{r,\mathrm{min}}n^{3}_{c,\mathrm{min}}}).
\end{align*}
where $\theta_{r,\mathrm{min}}=\mathrm{min}_{i\in[n_{r}]}\theta_{r}(i),\theta_{c,\mathrm{min}}=\mathrm{min}_{j\in[n_{c}]}\theta_{c}(j), n_{r,\mathrm{max}}=\mathrm{max}_{k_{r}\in[K_{r}]}n_{r,k_{r}}, n_{r,\mathrm{min}}=\mathrm{min}_{k_{r}\in[K_{r}]}n_{r,k_{r}}, n_{c,\mathrm{max}}=\mathrm{max}_{k_{c}\in[K_{c}]}n_{c,k_{c}}$, and $n_{c,\mathrm{min}}=\mathrm{min}_{k_{c}\in[K_{c}]}n_{c,k_{c}}$.
\end{thm}

Theorem \ref{mainNcDSoS} underscores the enhanced accuracy of community partitions obtained from NcDSoS as the number of layers, $L$, increases. This finding underscores the advantage of incorporating multiple layers in community detection tasks. Conversely, Theorem \ref{mainNcDSoS} also indicates that a reduction in the size of the smallest row (and column) communities leads to increased error rates. Although Theorem \ref{mainNcDSoS} is stated in a rather complex manner, involving numerous model parameters, the subsequent corollary simplifies its complexity by imposing certain conditions on these parameters.
\begin{cor}\label{CorSsum}
Under the same conditions of Theorem \ref{mainNcDSoS}, when $K_{r}=O(1), K_{c}=O(1),  \frac{n_{r,\mathrm{min}}}{n_{r,\mathrm{max}}}=O(1), \frac{n_{c,\mathrm{min}}}{n_{c,\mathrm{max}}}=O(1), \theta_{r,\mathrm{min}}=O(\sqrt{\rho}), \theta_{r,\mathrm{max}}=O(\sqrt{\rho}), \theta_{c,\mathrm{min}}=O(\sqrt{\rho})$, and $\theta_{c,\mathrm{max}}=O(\sqrt{\rho})$ for $\rho\in(0,1]$, we have
\begin{align*}
&\hat{f}_{r}=O(\frac{\mathrm{log}(n_{r}+n_{c}+L)}{\rho^{2}n_{r}n_{c}L})+O(\frac{1}{n^{2}_{r}})\mathrm{~and~}\hat{f}_{c}=O(\frac{\mathrm{log}(n_{r}+n_{c}+L)}{\rho^{2}n_{r}n_{c}L})+O(\frac{1}{n^{2}_{c}}).
\end{align*}
If we further assume that $n_{r}=O(n_{c})=O(n)$, we have
\begin{align*}
\hat{f}_{r}=\hat{f}_{c}=O(\frac{\mathrm{log}(n+L)}{\rho^{2}n^{2}L})+O(\frac{1}{n^{2}}).
\end{align*}
\end{cor}
In Corollary \ref{CorSsum}, the condition $n_{r,\mathrm{min}}/n_{r,\mathrm{max}}=O(1)$ ensures that the sizes of row communities remain balanced. According to this corollary, to achieve sufficiently low error rates, the sparsity parameter $\rho$ should decrease slower than $\sqrt{\frac{\mathrm{log}(n_{r}+n_{c}+L)}{n_{r}n_{c}L}}$, which aligns precisely with Assumption \ref{Assum2}. Meanwhile, increasing the sparsity parameter $\rho$ decreases the error rates. Furthermore, Corollary \ref{CorSsum} demonstrates that as the number of row (or column) nodes and/or the number of layers increases to infinity, the error rates of NcDSoS converge to zero, thus establishing its consistency. Additionally, when we set $Z_{r}=Z_{c}$, $n_{r}=n_{c}=n$, $\Theta_{r}=\Theta_{c}=\sqrt{\rho}I_{n\times n}$ for $\rho\in(0,1]$, and $B_{l}=B'_{l}$ for $l\in[L]$, our multi-layer DC-ScBM simplifies to the multi-layer SBM. In this case, the theoretical upper bounds of error rates derived for NcDSoS align with the theoretical results presented in Theorem 1 of \cite{lei2023bias}, indicating the optimality of our theoretical analysis.
\section{Simulations}\label{sec5}
In this section, we investigate the empirical performance of NcDSoS for detecting communities of the multi-layer DC-ScBM model by comparing it with the following algorithms:
\begin{itemize}
 \item \textbf{NcSoS}: \underline{N}ormalized spectral \underline{c}o-clustering based on \underline{s}um \underline{o}f \underline{s}quared matrices. NcSoS uses the non-debiased matrices $\sum_{l\in[L]}A_{l}A'_{l}$ and $\sum_{l\in[L]}A'_{l}A_{l}$ to replace $S_{r}$ and $S_{c}$ in Algorithm \ref{alg:NcDSoS}. Since $\|\sum_{l\in[L]}A_{l}A'_{l}-\tilde{S}_{r}\|=\|S_{r}-\tilde{S}_{r}+\sum_{l\in[L]}D^{r}_{l}\|\leq\|S_{r}-\tilde{S}_{r}\|+\|\sum_{l\in[L]}D^{r}_{l}\|$ and $\|\sum_{l\in[L]}A'_{l}A_{l}-\tilde{S}_{c}\|=\|S_{c}-\tilde{S}_{c}+\sum_{l\in[L]}D^{c}_{l}\|\leq\|S_{c}-\tilde{S}_{c}\|+\|\sum_{l\in[L]}D^{c}_{l}\|$, following a similar proof of Theorem \ref{mainNcDSoS}, we can immediately build NcSoS's theoretical guarantees and find that the theoretical upper bounds of NcSoS's clustering errors are larger than that of NcDSoS. This analysis underscores the importance of debiasing in enhancing the performance of spectral co-clustering methods for community detection in multi-layer bipartite networks.
  \item \textbf{NcSum}: \underline{N}ormalized spectral \underline{c}o-clustering based on \underline{sum} of adjacency matrices. NcSum detects row (and column) communities by employing the K-means algorithm on the row-normalized version of the top $\mathrm{min}(K_{r},K_{c})$ left (and right) singular vectors of $\sum_{l\in[L]}A_{l}$ with $K_{r}$ (and $K_{c}$) clusters.
  \item \textbf{Sum}: this algorithm is proposed in \citep{su2023spectral} and it detects row (and column) communities by running the K-means algorithm on the top $\mathrm{min}(K_{r},K_{c})$ left (and right) singular vectors of $\sum_{l\in[L]}A_{l}$ with $K_{r}$ (and $K_{c}$) clusters.
  \item \textbf{SoG}: this method introduced in \citep{su2023spectral} estimates row (and column) communities by running K-means algorithm on the eigenvectors of the non-debiased matrix $\sum_{l\in[L]}A_{l}A'_{l}$ (and $\sum_{l\in[L]}A'_{l}A_{l}$) with $K_{r}$ (and $K_{c}$) clusters.
  \item \textbf{DSoG}: this method proposed in \citep{su2023spectral} estimates row (and column) communities by running K-means algorithm on the eigenvectors of $S_{r}$ (and $S_{c}$) with $K_{r}$ (and $K_{c}$) clusters.
\end{itemize}

To assess the performance of the aforementioned algorithms, we employ four evaluation metrics: Clustering error, Hamming error \citep{SCORE}, normalized mutual information (NMI) \citep{strehl2002cluster,danon2005comparing,bagrow2008evaluating}, and adjusted rand index (ARI) \citep{hubert1985comparing,vinh2009information}. The details of these metrics are given below::
\begin{itemize}
  \item \textbf{Clustering error}: In our simulation studies, we define Clustering error as $\mathrm{max}(\hat{f}_{r},\hat{f}_{c})$ to simplify our analysis. This metric is non-negative, and a lower value indicates better performance in community detection for both row and column communities.
  \item \textbf{Hamming error}: This metric quantifies the similarity between the predicted and actual community assignments. It is defined as $\mathrm{max}(\mathrm{min}_{P_{r}\in\mathcal{P}_{K_{r}}}\|\hat{Z}_{r}P_{r}-Z_{r}\|_{0}/n_{r}, \mathrm{min}_{P_{c}\in\mathcal{P}_{K_{c}}}\|\hat{Z}_{c}P_{c}-Z_{c}\|_{0}/n_{c})$ where $\mathcal{P}_{K_{r}}$ (and $\mathcal{P}_{K_{c}}$) represents the set of all $K_{r}\times K_{r}$ (and $K_{c}\times K_{c}$) permutation matrices. The Hamming error ranges from 0 to 1, with a lower value indicating better performance.
  \item \textbf{NMI} and \textbf{ARI}: We adopt the NMI and ARI definitions from \citep{qing2023community} to assess the quality of community detection in both row and column communities. NMI ranges from 0 to 1, while ARI ranges from -1 to 1. In both cases, a higher value indicates better performance.
\end{itemize}

By leveraging these metrics, we can comprehensively evaluate the performance of the algorithms in community detection tasks.

In all simulated multi-layer bipartite networks, we consistently assign $K_{r}=2$ and $K_{c}=3$ as fixed parameters. Subsequently, we generate $Z_{r}$ and $Z_{c}$ in a manner that ensures that each row (and column) node belongs to each row (and column) community with an equivalent probability. For $i\in[n_{r}]$ and $j\in[n_{c}]$, we set $\theta_{r}(i)=\sqrt{\rho}\cdot\mathrm{rand}(1)$ and $\theta_{c}(j)=\sqrt{\rho}\cdot\mathrm{rand}(1)$, respectively, where rand(1) represents a random value drawn from the uniform distribution on the interval (0, 1). Furthermore, we assign $B_{l}(k_{r},k_{c})=\mathrm{rand}(1)$ for $k_{r}\in[K_{r}], k_{c}\in[K_{c}],$ and $l\in[L]$. The four parameters $n_{r}, n_{c}, L$, and $\rho$ are set independently for each experiment. Notably, in Experiment 2, where we specifically consider the case where $n_{r}$ equals $n_{c}$, we simplify the notation by setting $n=n_{r}=n_{c}$ for convenience. To ensure reliability and reproducibility, we report each metric for different methods, averaging the results over 100 random runs for each parameter setting.

\begin{figure}
\centering
\resizebox{\columnwidth}{!}{
\subfigure[]{\includegraphics[width=0.2\textwidth]{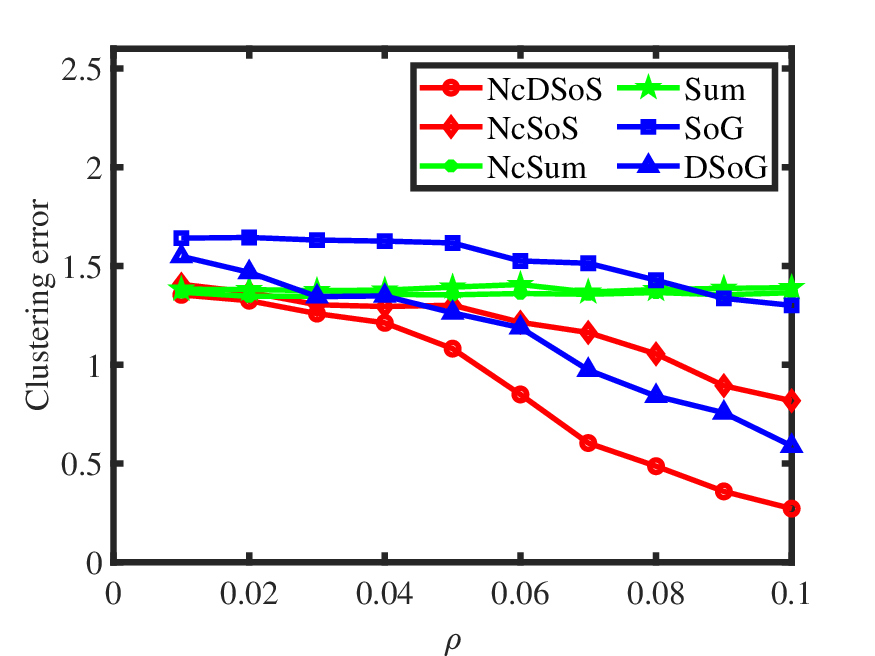}}
\subfigure[]{\includegraphics[width=0.2\textwidth]{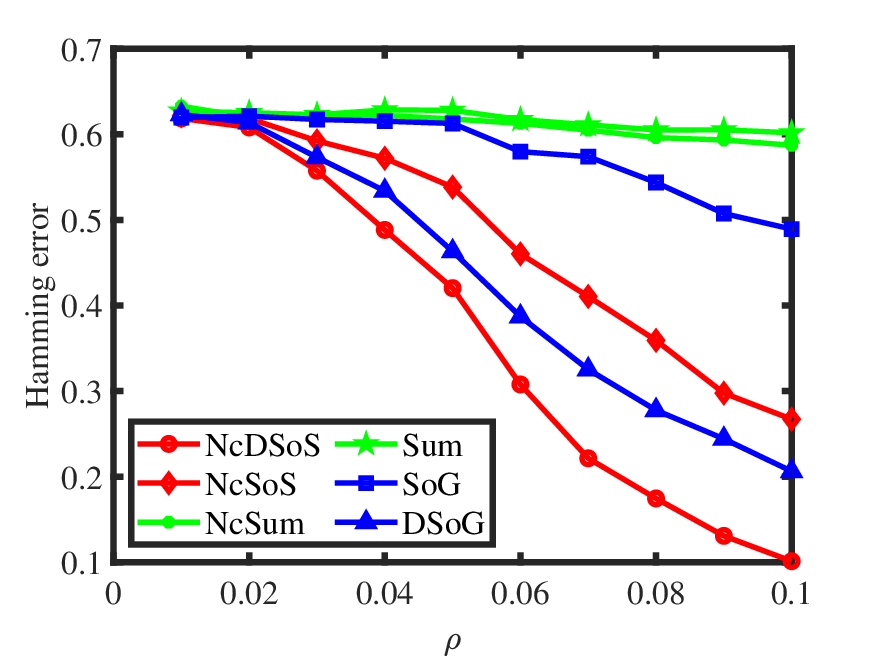}}
\subfigure[]{\includegraphics[width=0.2\textwidth]{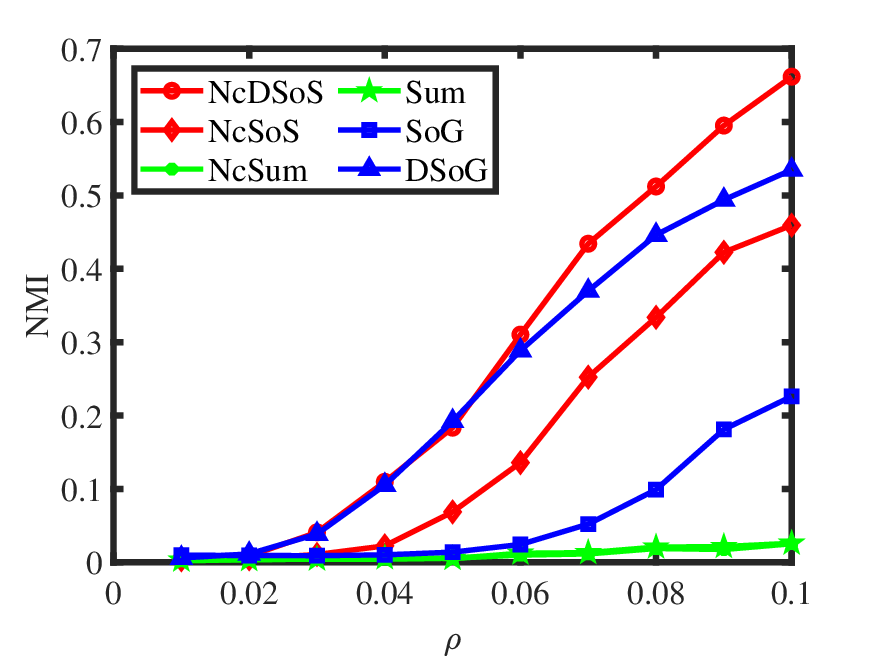}}
\subfigure[]{\includegraphics[width=0.2\textwidth]{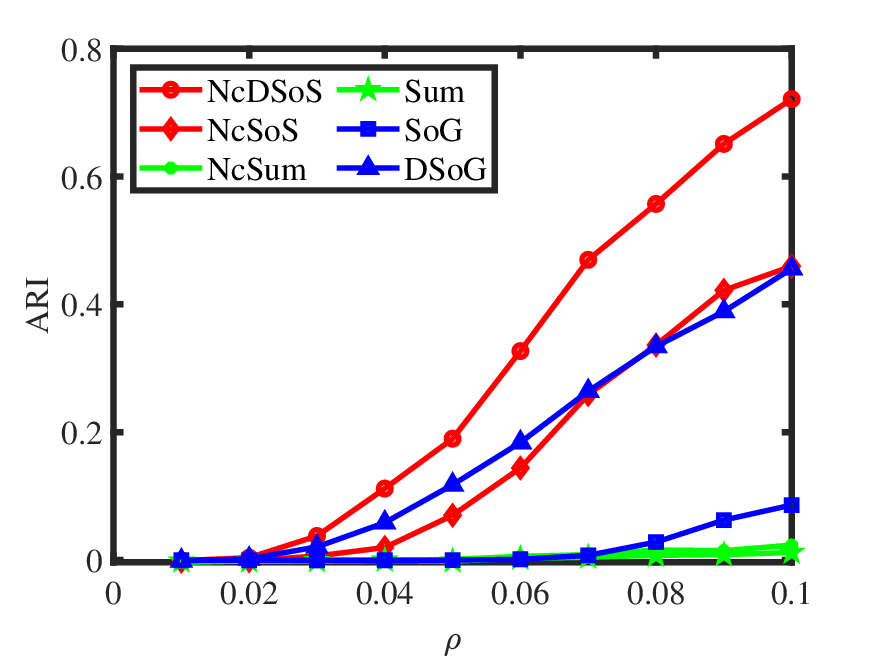}}
}
\caption{Experiment 1. Panel (a): Clustering error against increasing $\rho$. Panel (b): Hamming error against increasing $\rho$. Panel (c): NMI against increasing $\rho$. Panel (d): ARI against increasing $\rho$.}
\label{Ex1} 
\end{figure}

\textbf{Experiment 1: Changing $\rho$}. We set $n_{r}=200, n_{c}=300, L=50$, and let $\rho$ range in $\{0.01, 0.02, 0.03, \ldots, 0.1\}$. Based on the results shown in Figure \ref{Ex1}, we have the following observations: (a) Our method NcDSoS significantly outperforms its competitors. (b) The performance of NcDSoS improves as the sparsity parameter $\rho$ increases, aligning with our theoretical predictions. (c) Although both NcDSoS and DSoG incorporate debisased spectral clustering, NcDSoS surpasses DSoG because NcDSoS is designed to fit the multi-layer DC-ScBM while DSoG is for the multi-layer ScBM and our multi-layer DC-ScBM considers the degree heterogeneity of nodes while the multi-layer ScBM does not. Similar arguments hold for NcSum and SoG.

\begin{figure}
\centering
\resizebox{\columnwidth}{!}{
\subfigure[]{\includegraphics[width=0.2\textwidth]{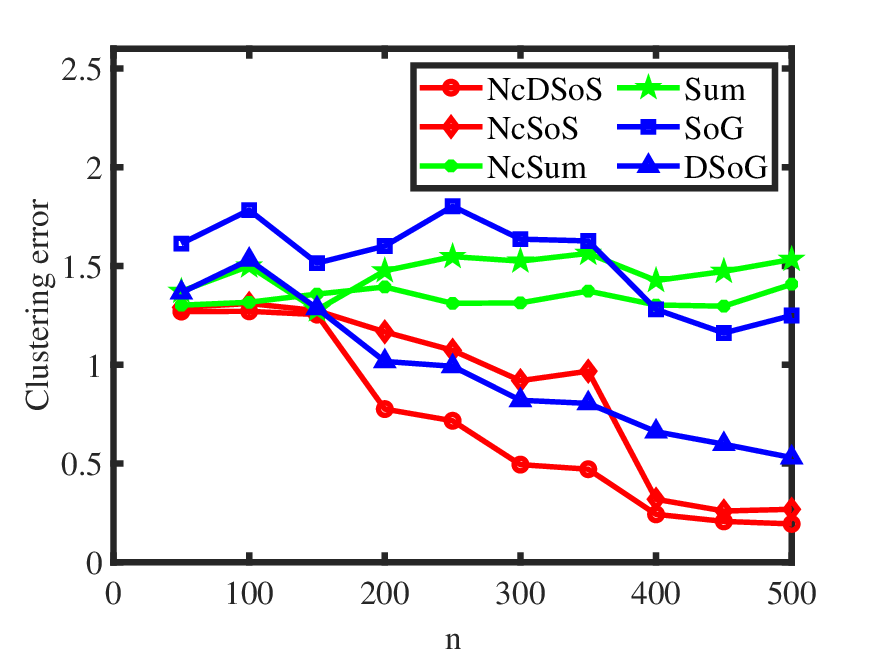}}
\subfigure[]{\includegraphics[width=0.2\textwidth]{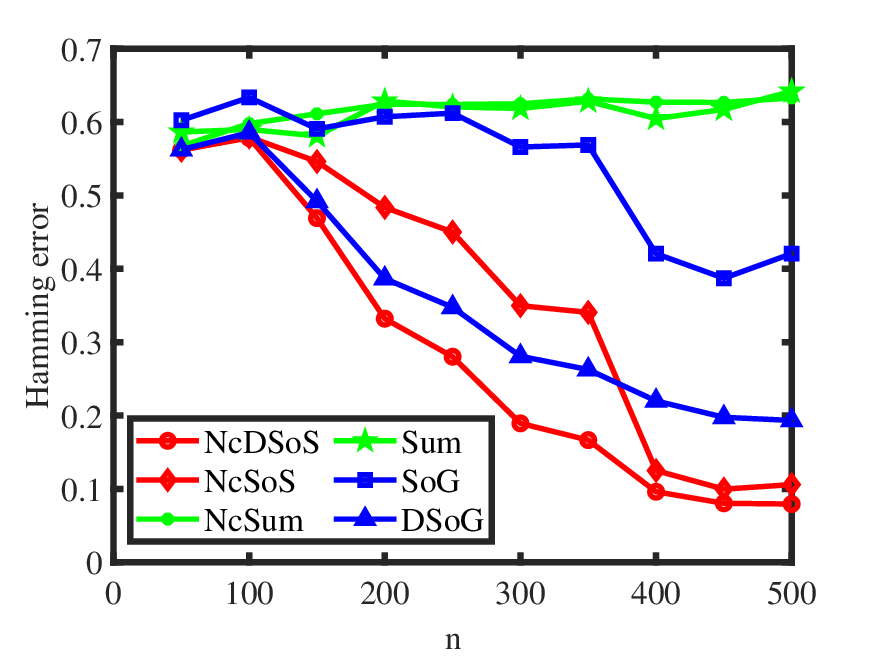}}
\subfigure[]{\includegraphics[width=0.2\textwidth]{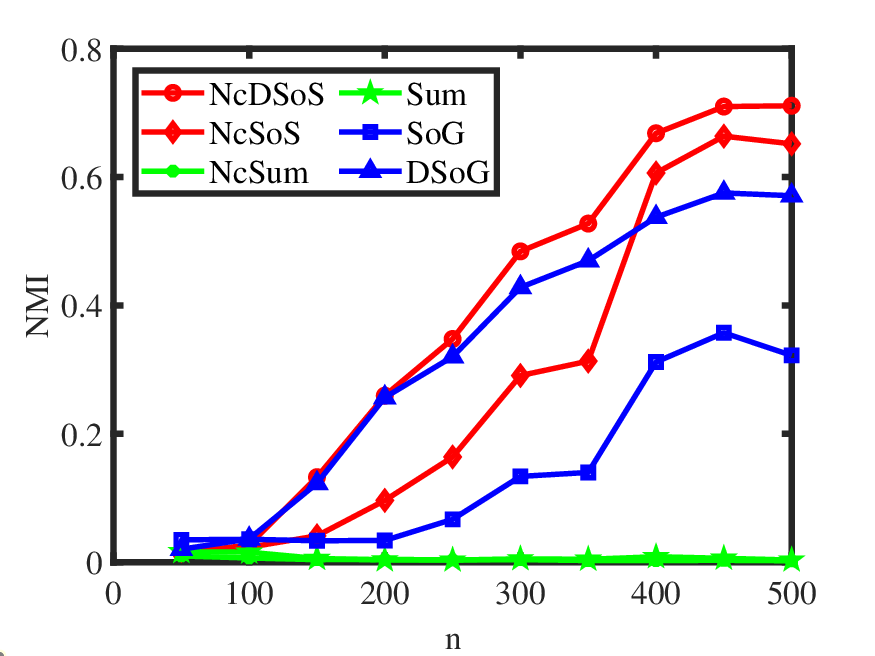}}
\subfigure[]{\includegraphics[width=0.2\textwidth]{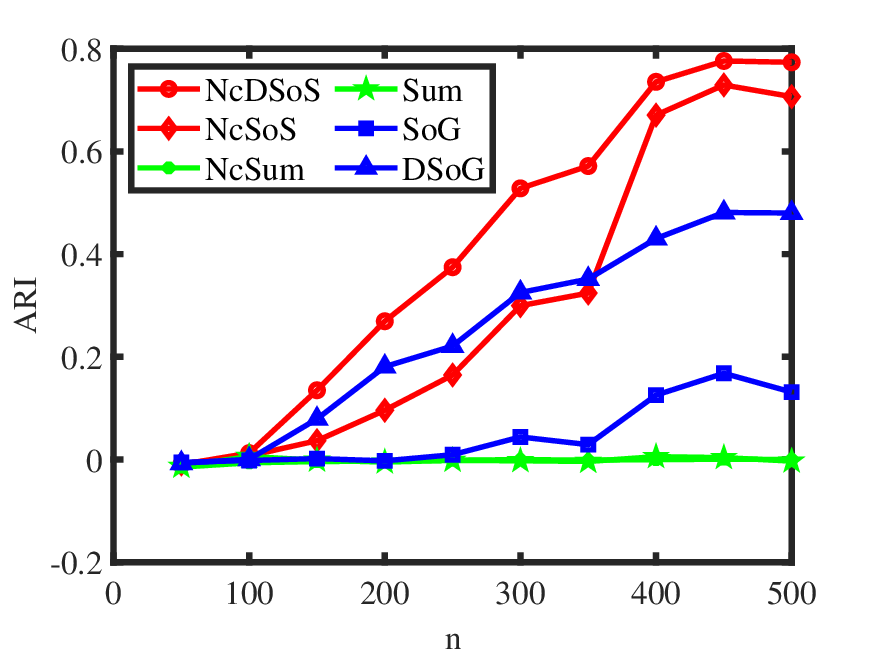}}
}
\caption{Experiment 2. Panel (a): Clustering error against increasing $n$. Panel (b): Hamming error against increasing $n$. Panel (c): NMI against increasing $n$. Panel (d): ARI against increasing $n$.}
\label{Ex2} 
\end{figure}

\textbf{Experiment 2: Changing $n$}. We set $L=20, \rho=0.1, n_{r}=n_{c}=n$, and increase $n$ from 50 to 500. Figure \ref{Ex2} displays the results. We see that (a) our NcDSoS outperforms the other five algorithms; (b) NcDSoS performs better when $n$ increases and this is consistent with our theoretical results.

\begin{figure}
\centering
\resizebox{\columnwidth}{!}{
\subfigure[]{\includegraphics[width=0.2\textwidth]{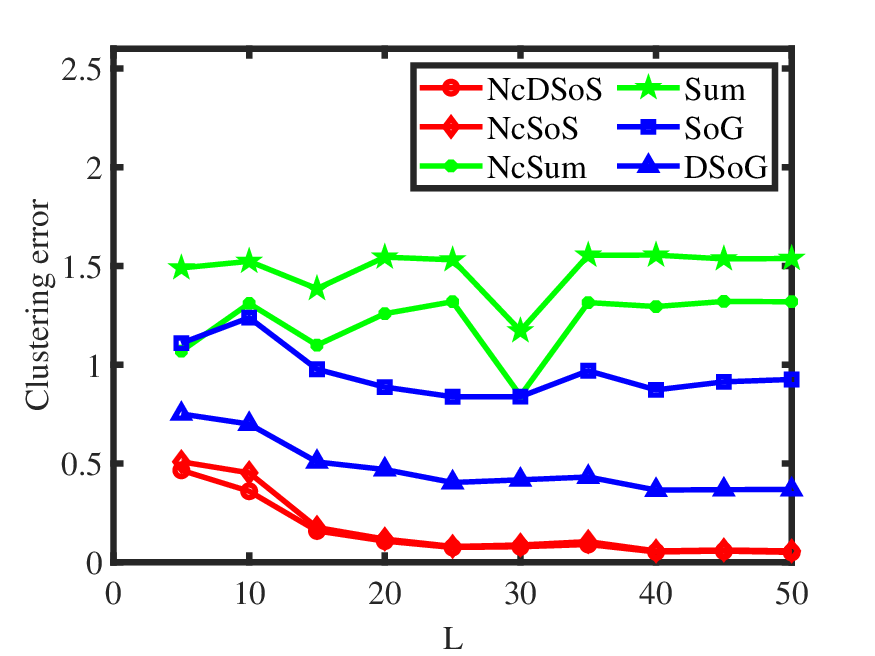}}
\subfigure[]{\includegraphics[width=0.2\textwidth]{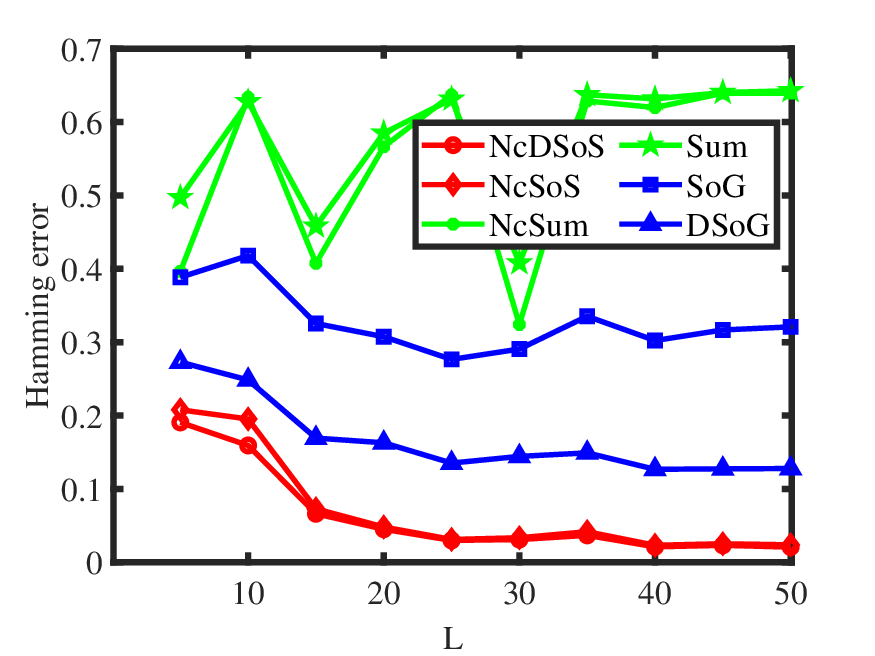}}
\subfigure[]{\includegraphics[width=0.2\textwidth]{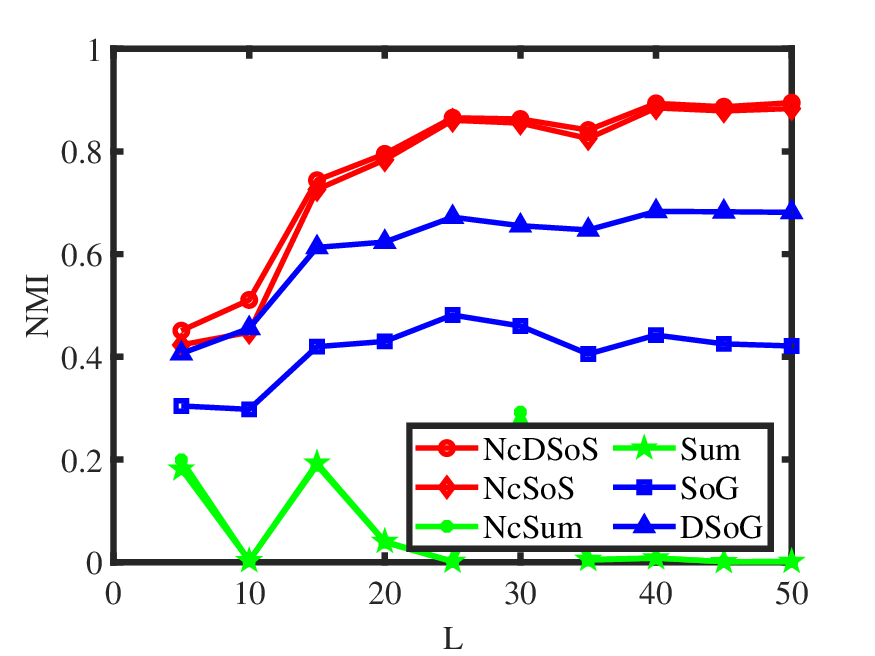}}
\subfigure[]{\includegraphics[width=0.2\textwidth]{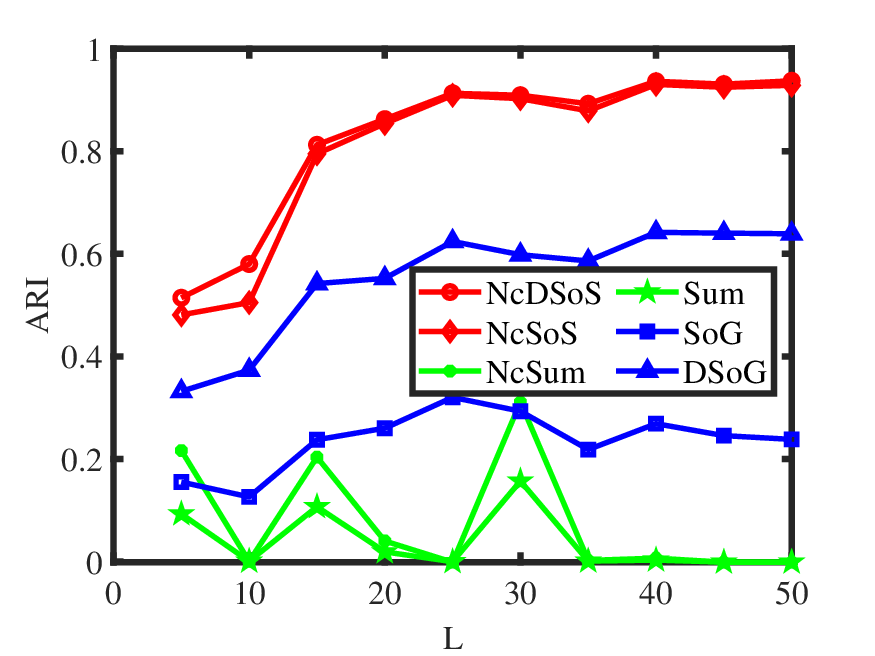}}
}
\caption{Experiment 3. Panel (a): Clustering error against increasing $L$. Panel (b): Hamming error against increasing $L$. Panel (c): NMI against increasing $L$. Panel (d): ARI against increasing $L$.}
\label{Ex3} 
\end{figure}

\textbf{Experiment 3: Changing $L$}. We set $n_{r}=500, n_{c}=600, \rho=0.1$, and let $L$ range in $\{5,10,15,\ldots,50\}$. Figure \ref{Ex3} shows the results. The data indicates that our NcDSoS outperforms its competitors across all parameter settings. Furthermore, NcDSoS exhibits better performance as the number of layers increases, aligning with our theoretical analysis.
\section{Real data applications}\label{sec6realdata}
In this section, we consider the following real-world multi-layer directed networks, which can be downloaded from\url{https://manliodedomenico.com/data.php}:
\begin{itemize}
  \item \textbf{Vickers Chan 7thGraders }: This data was collected by Vickers from 29 seventh grade students in a school in Victoria, Australia. Students were asked to nominate their classmates on three interpersonal relations (layers): Who do you get on with in the class? Who are your best friends in the class? Who would you prefer to work with?
  \item \textbf{Lazega law firm}: This data encompasses three distinct types of interactions: Co-work, Friendship, and Advice, occurring among partners and associates within a corporate law partnership.
  \item \textbf{C.elegans}: This data encompasses distinct layers that correspond to varying synaptic junctions. These layers include electric synaptic junctions ("ElectrJ"), chemical monadic synapses ("MonoSyn"), and chemical polyadic synapses ("PolySyn"), each capturing a unique aspect of neuronal connectivity within the Caenorhabditis elegans.
\end{itemize}
\begin{figure}
\centering
\resizebox{\columnwidth}{!}{
\subfigure[Vickers Chan 7thGraders]{\includegraphics[width=0.2\textwidth]{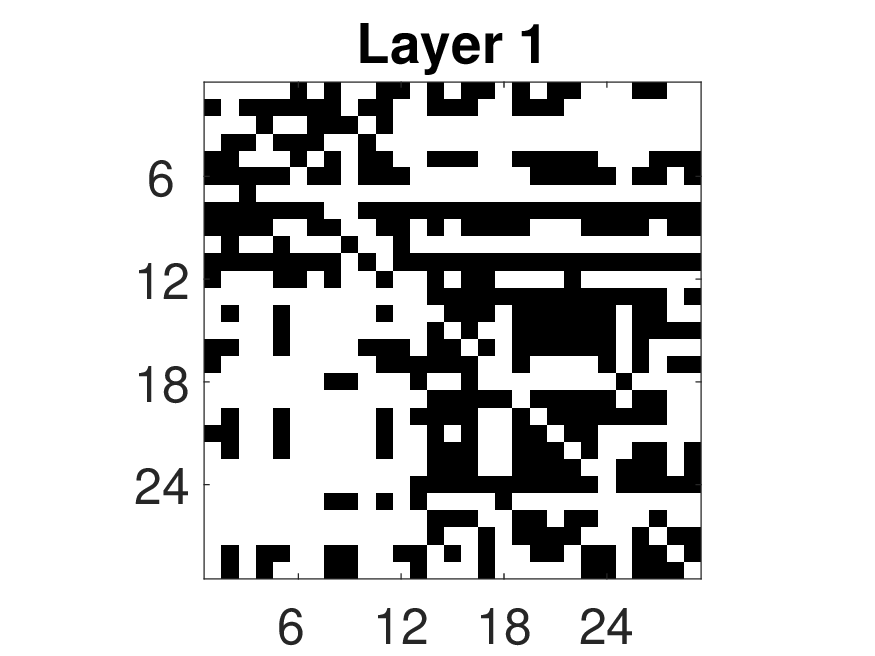}}
\subfigure[Vickers Chan 7thGraders]{\includegraphics[width=0.2\textwidth]{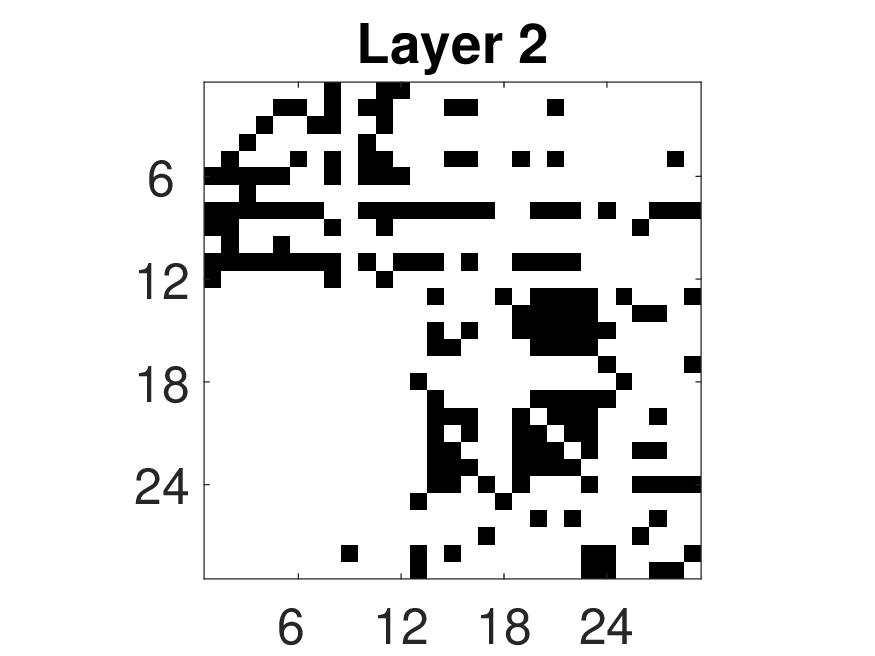}}
\subfigure[Vickers Chan 7thGraders]{\includegraphics[width=0.2\textwidth]{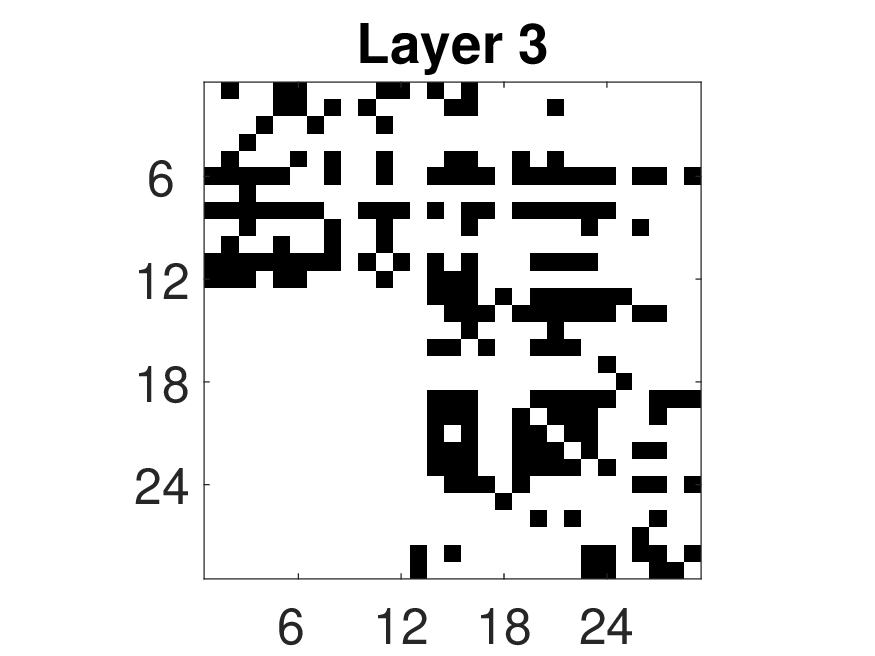}}
}
\resizebox{\columnwidth}{!}{
\subfigure[Lazega law firm]{\includegraphics[width=0.2\textwidth]{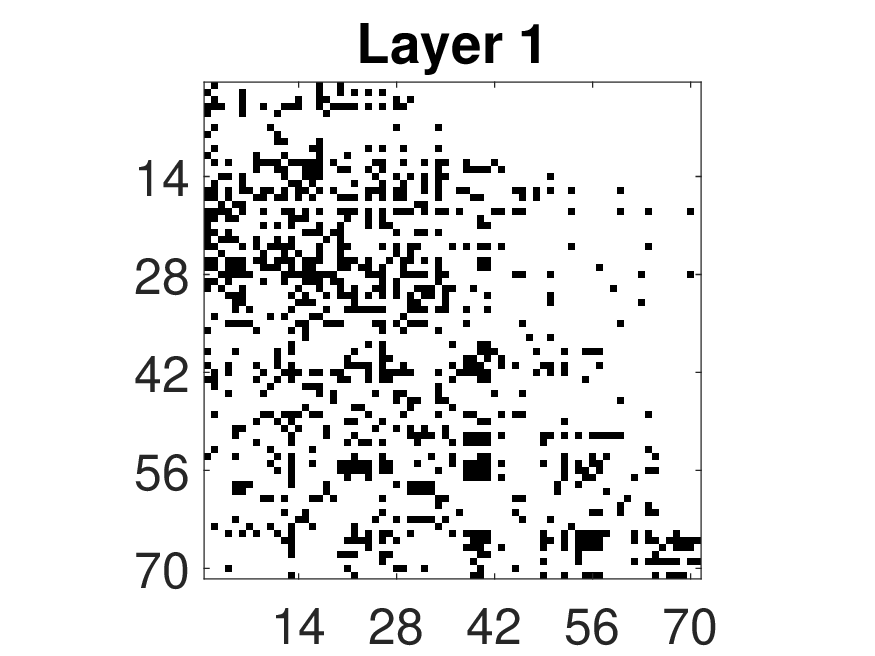}}
\subfigure[Lazega law firm]{\includegraphics[width=0.2\textwidth]{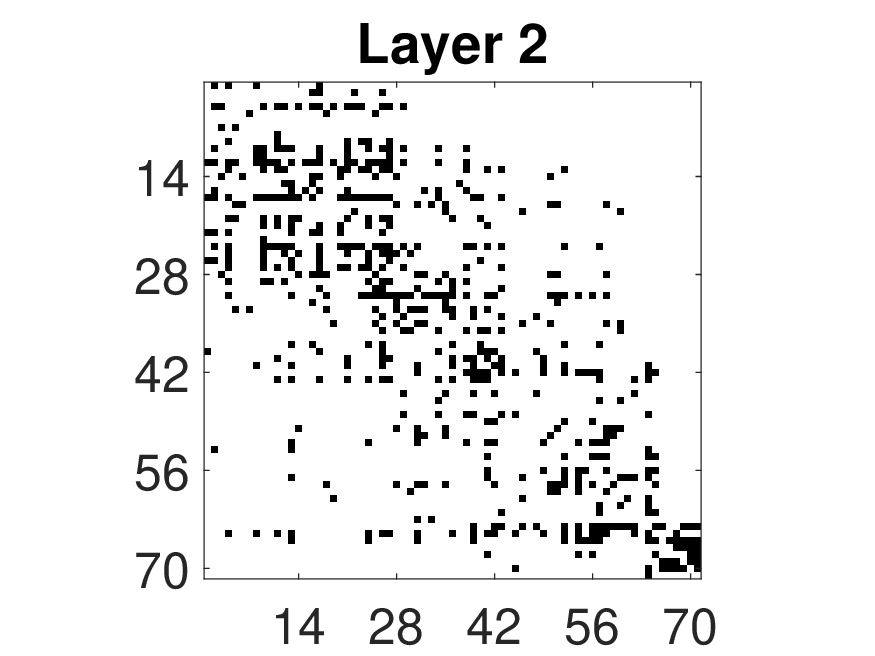}}
\subfigure[Lazega law firm]{\includegraphics[width=0.2\textwidth]{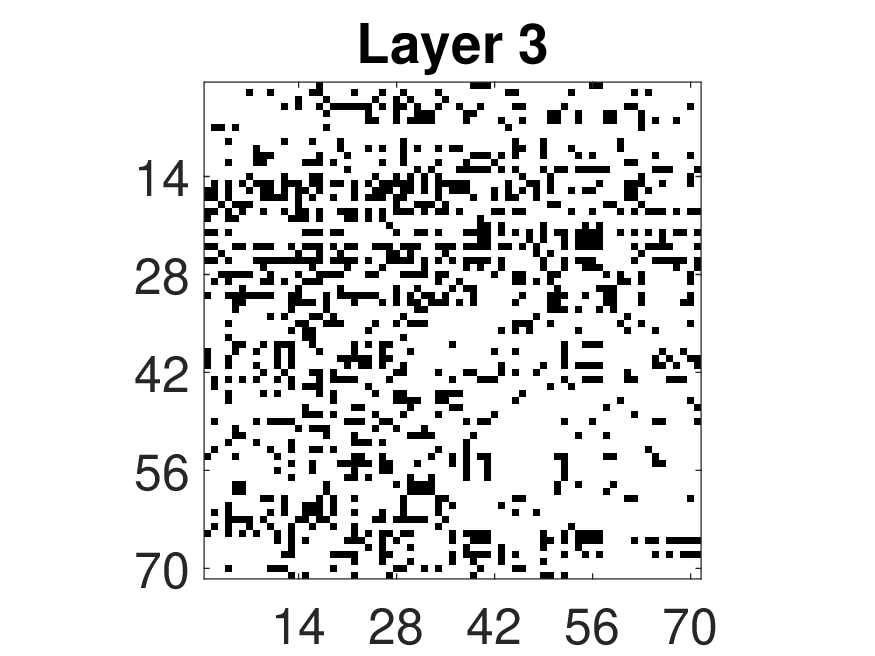}}
}
\resizebox{\columnwidth}{!}{
\subfigure[C.elegans]{\includegraphics[width=0.2\textwidth]{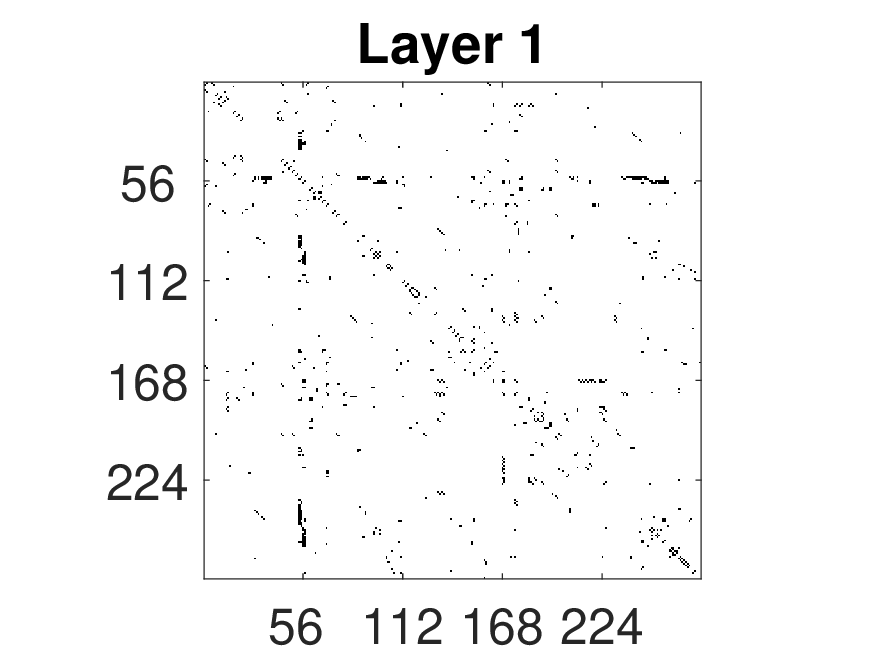}}
\subfigure[C.elegans]{\includegraphics[width=0.2\textwidth]{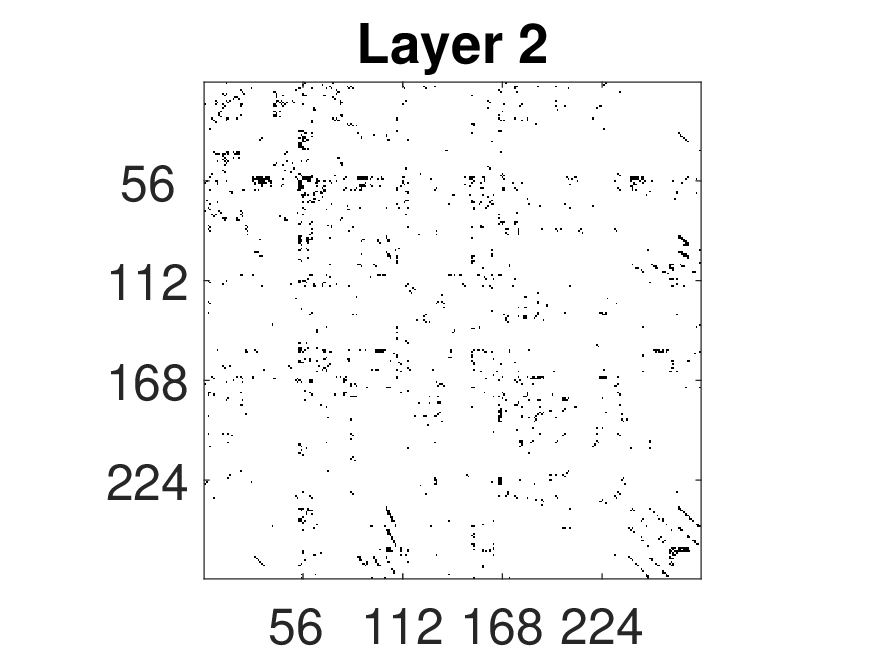}}
\subfigure[C.elegans]{\includegraphics[width=0.2\textwidth]{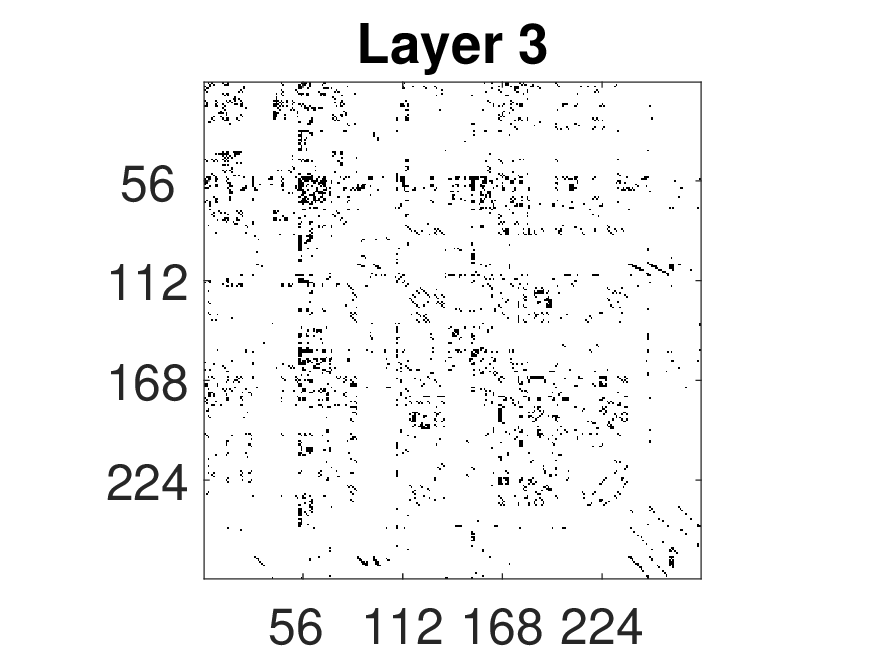}}
}
\caption{The adjacency matrices of real multi-layer directed networks used in this paper. In all matrices, black represents $1$ and white means 0.}
\label{ArealData} 
\end{figure}

Figure \ref{ArealData} and Table \ref{realdata} display the adjacency matrices and basic information of these real data sets, respectively. Notably, the adjacency matrices of Vickers Chan 7thGraders and Lazega law firm exhibit a pronounced asymmetry, indicating a significant divergence between their sending and receiving patterns. In contrast, the adjacency matrices of C.elegans display a near-symmetric pattern, suggesting a remarkable similarity in their sending and receiving behaviors.
\begin{table}[h!]
\small
	\centering
	\caption{Basic information of real-world multi-layer directed networks used in this paper.}
	\label{realdata}
	\begin{tabular}{cccccccccccc}
\hline\hline
Dataset&Source&Directed?&$\#$ Nodes&$\#$ Layers&$\#$ Edges\\
\hline
Vickers Chan 7thGraders&\citep{vickers1981representing}&Yes&29&3&740\\
Lazega law firm&\citep{lazega2001collegial,snijders2006new}&Yes&71&3&2223\\
C.elegans&\citep{chen2006wiring,de2015muxviz}&Yes&279&3&5863\\
\hline\hline
\end{tabular}
\end{table}

In any real-world multi-layer directed network comprising $n$ nodes, we define two $n\times 1$ vectors, $d^{l}_{r}$ and $d^{l}_{c}$, for each layer $l$ in the range $[L]$. Specifically, $d^{l}_{r}(i)$ represents the out-degree of node $i$ in layer $l$, calculated as the sum of all elements in the $i^{th}$ row of the adjacency matrix $A_{l}$, i.e., $d^{l}_{r}(i)=\sum_{j\in[n]}A_{l}(i,j)$. Similarly, $d^{l}_{c}(i)$ represents the in-degree of node $i$ in layer $l$, obtained by summing the elements in the $i^{th}$ column of $A_{l}$, i.e., $d^{l}_{c}(i)=\sum_{j\in[n]}A_{l}(j,i)$. For brevity, we often refer to $d^{l}_{r}$ and $d^{l}_{c}$ simply as $d_{r}$ and $d_{c}$ when discussing any particular layer.

\begin{figure}
\centering
\resizebox{\columnwidth}{!}{
\subfigure[Vickers Chan 7thGraders]{\includegraphics[width=0.2\textwidth]{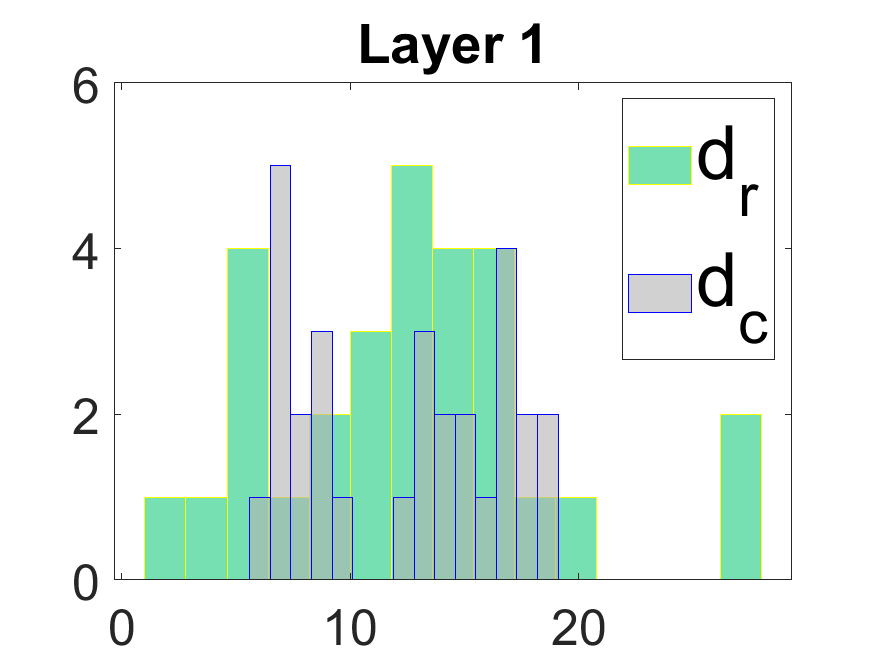}}
\subfigure[Vickers Chan 7thGraders]{\includegraphics[width=0.2\textwidth]{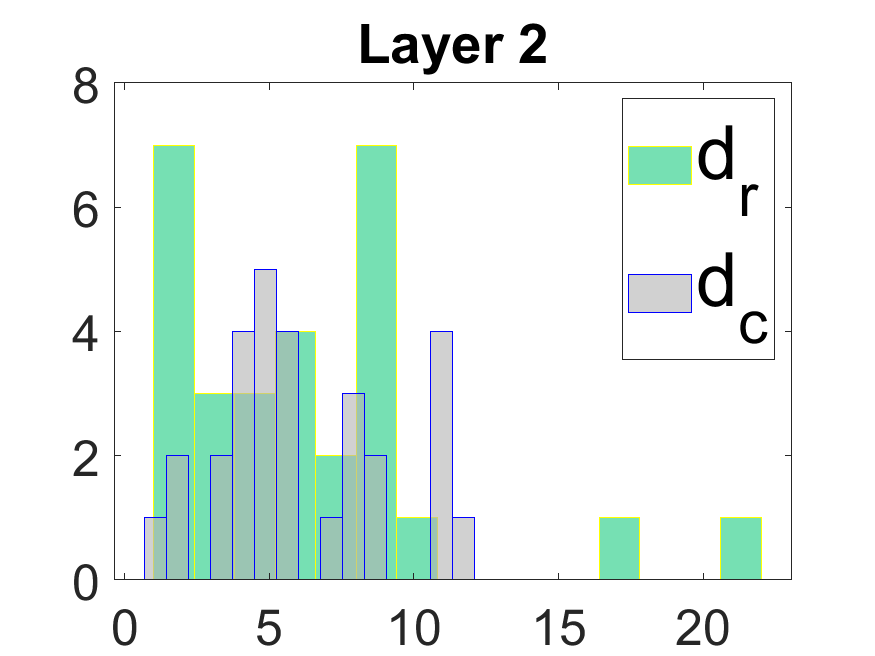}}
\subfigure[Vickers Chan 7thGraders]{\includegraphics[width=0.2\textwidth]{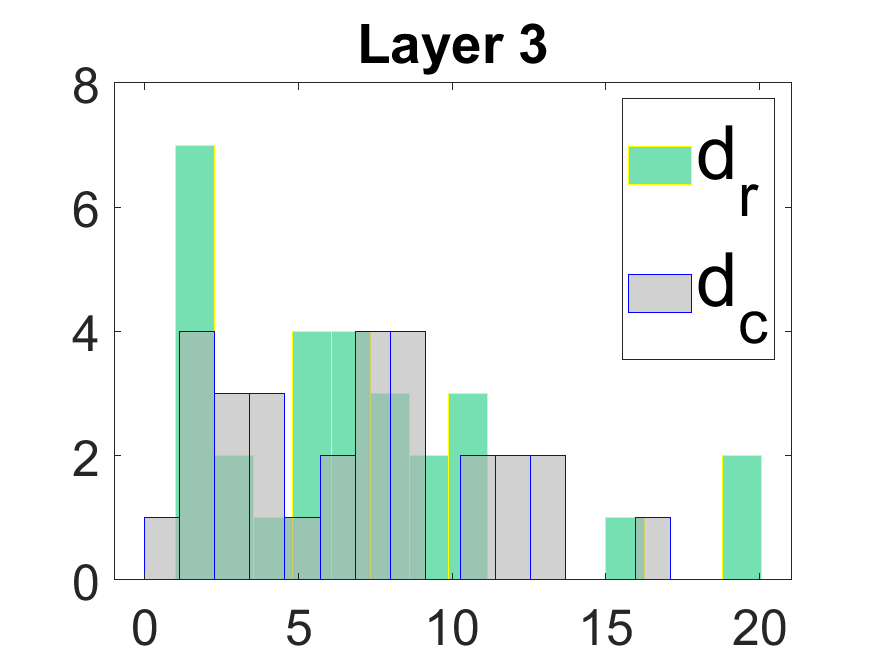}}
}
\resizebox{\columnwidth}{!}{
\subfigure[Lazega law firm]{\includegraphics[width=0.2\textwidth]{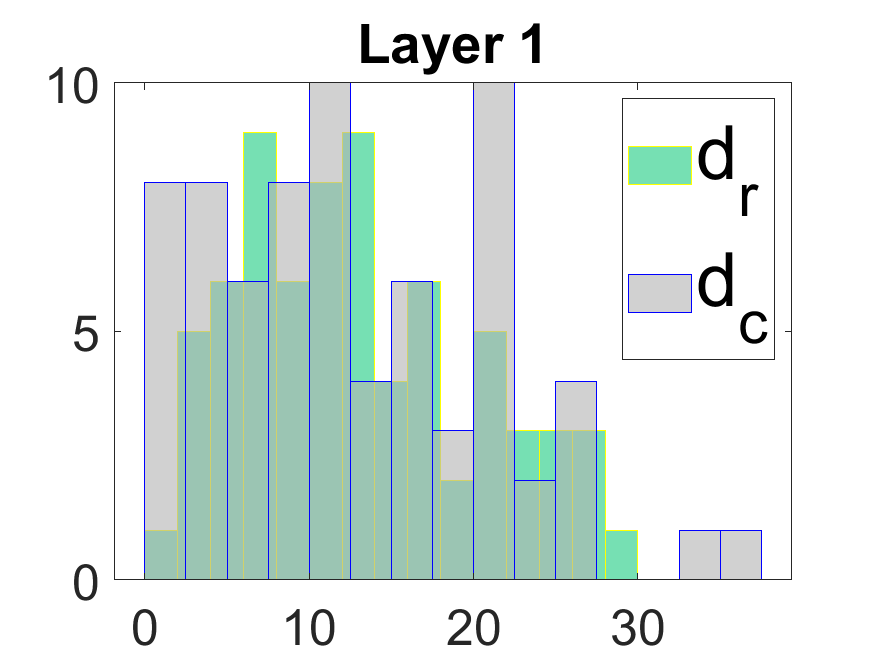}}
\subfigure[Lazega law firm]{\includegraphics[width=0.2\textwidth]{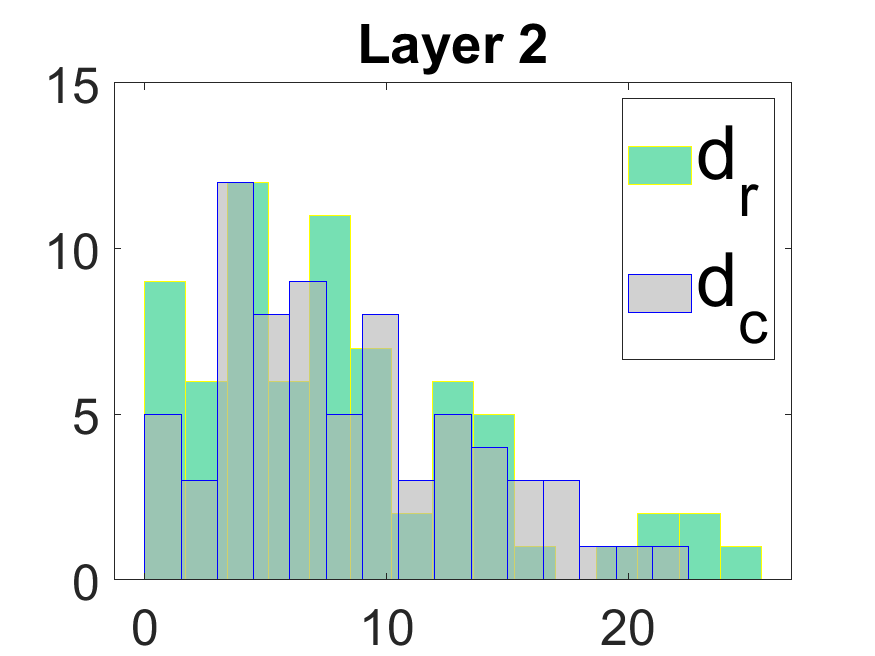}}
\subfigure[Lazega law firm]{\includegraphics[width=0.2\textwidth]{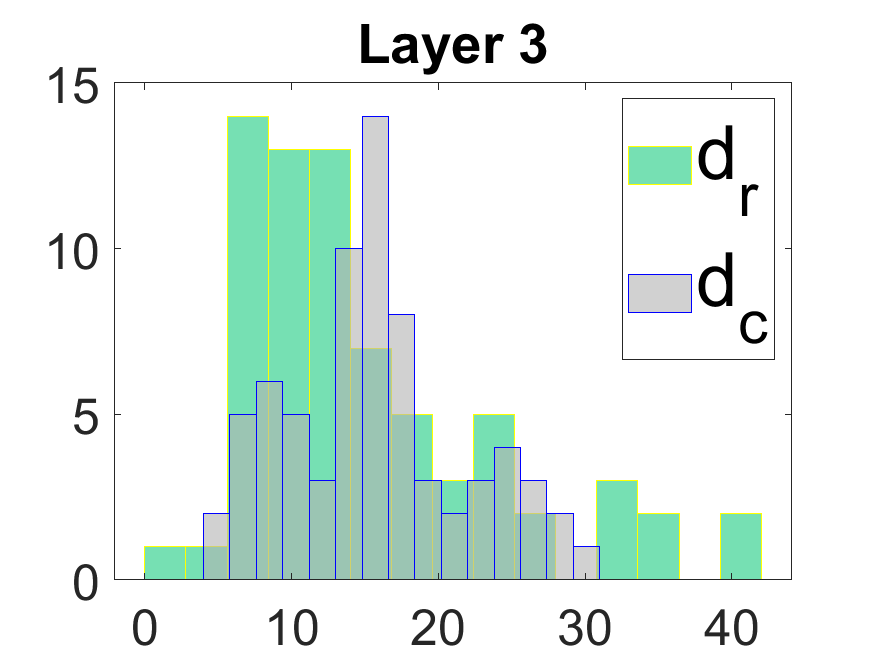}}
}
\resizebox{\columnwidth}{!}{
\subfigure[C.elegans]{\includegraphics[width=0.2\textwidth]{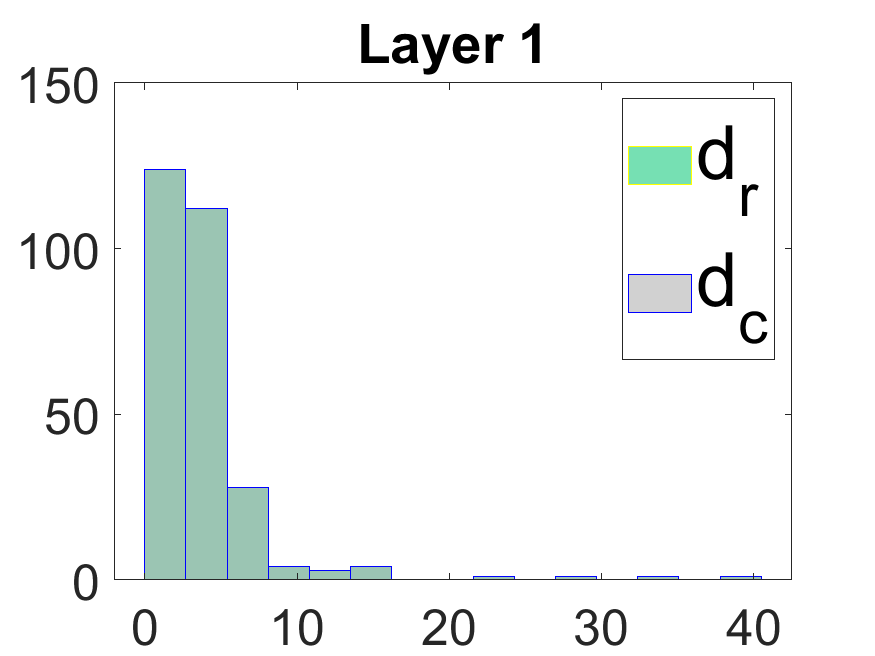}}
\subfigure[C.elegans]{\includegraphics[width=0.2\textwidth]{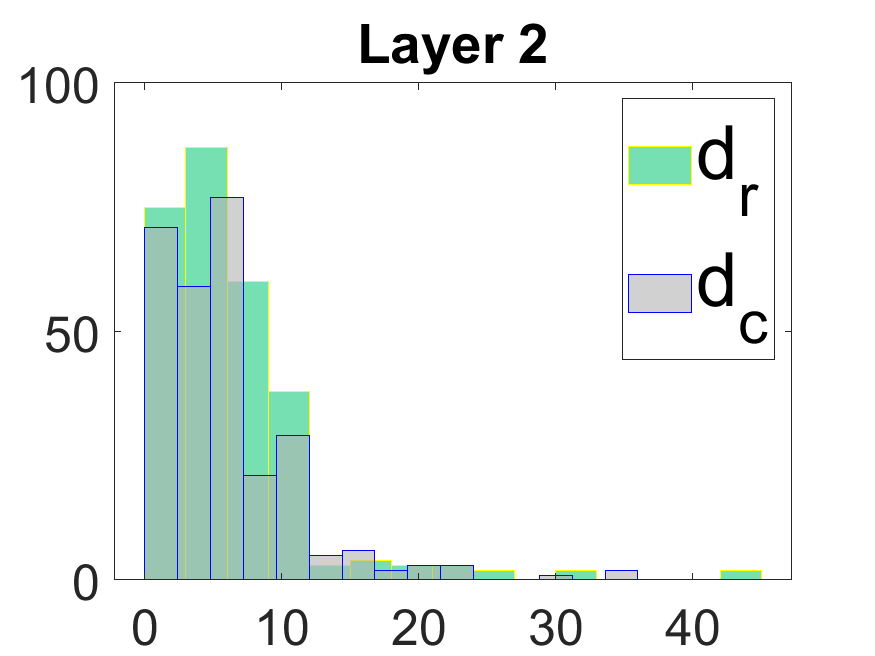}}
\subfigure[C.elegans]{\includegraphics[width=0.2\textwidth]{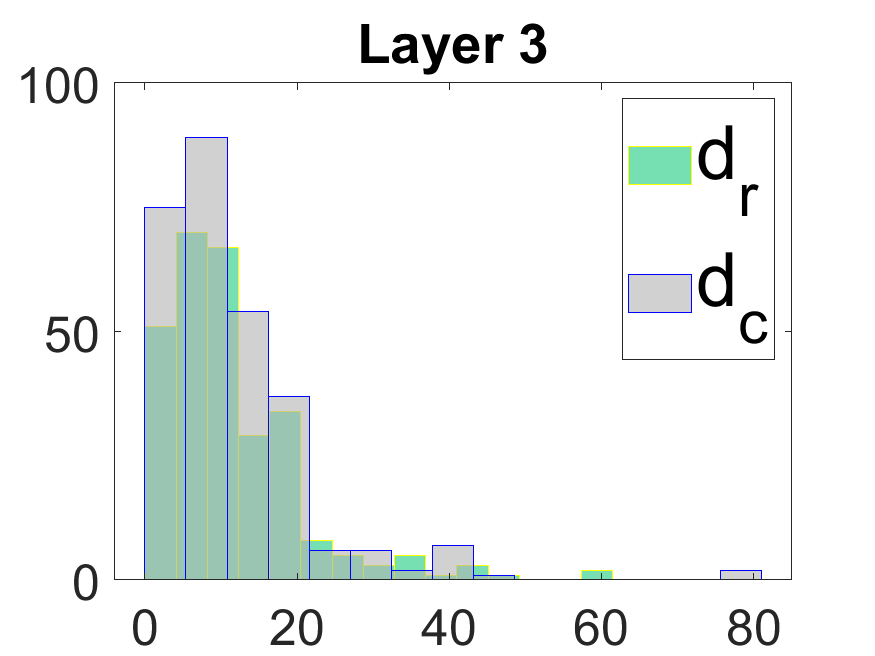}}
}
\caption{Distributions of $d_{r}$ and $d_{c}$ in each layer for real-world multi-layer directed networks considered in this paper.}
\label{HistDegree} 
\end{figure}

Figure \ref{HistDegree} depicts the distributions of out-degree and in-degree in each layer, based on the real-world datasets utilized in this study. In the case of Vickers Chan's 7th Graders network, we observe significant variations in the distributions of in-degree and out-degree across layers, indicating a substantial difference in sending and receiving patterns. A similar trend is observed in the Lazega law firm network. However, for the C.elegans network, the distributions of in-degree and out-degree in each layer exhibit a considerable overlap, suggesting a similarity in the sending and receiving behaviors. These observations align perfectly with the insights gained from Figure \ref{ArealData}.

\begin{figure}
\centering
\resizebox{\columnwidth}{!}{
\subfigure[Vickers Chan 7thGraders]{\includegraphics[width=0.2\textwidth]{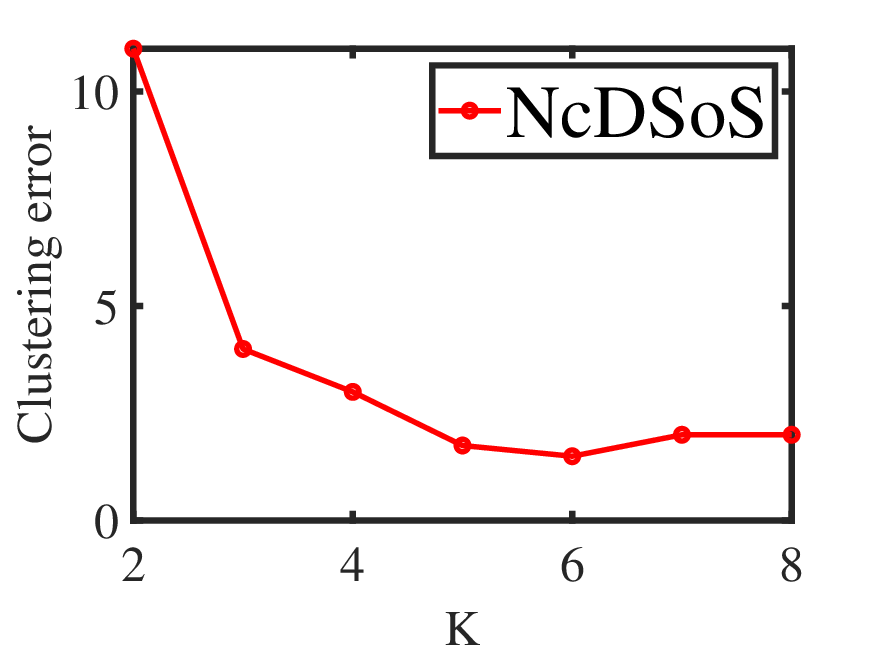}}
\subfigure[Vickers Chan 7thGraders]{\includegraphics[width=0.2\textwidth]{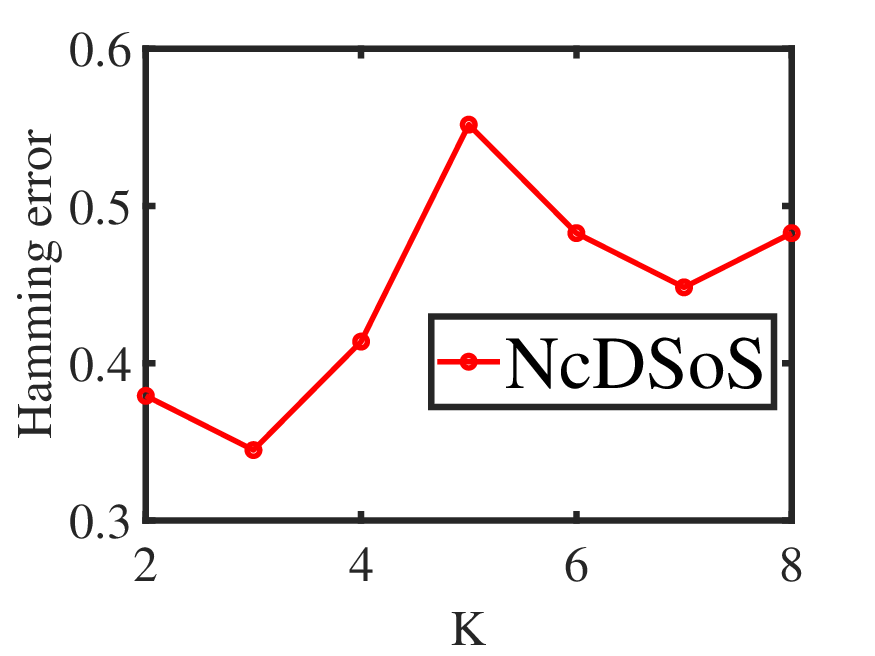}}
\subfigure[Vickers Chan 7thGraders]{\includegraphics[width=0.2\textwidth]{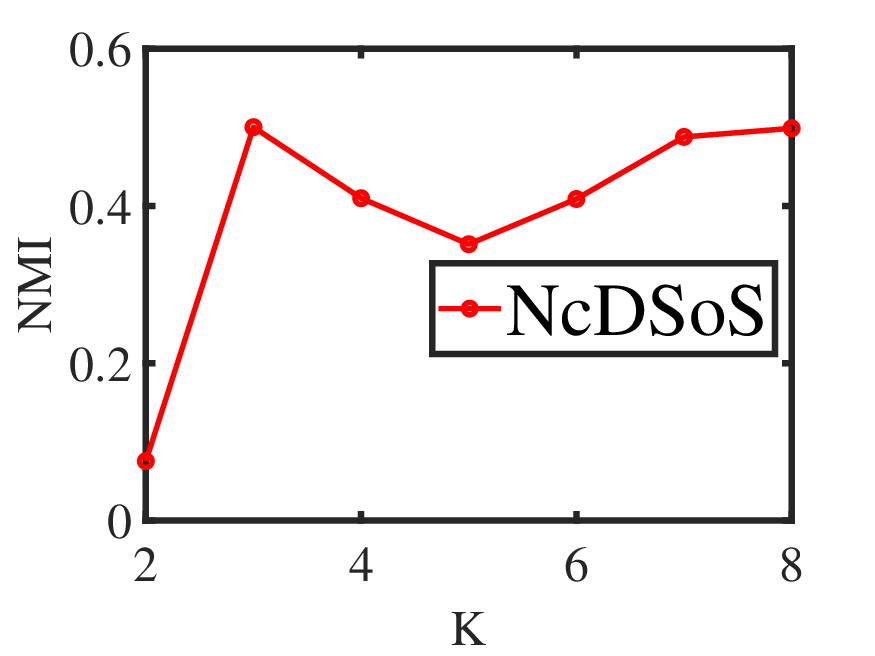}}
\subfigure[Vickers Chan 7thGraders]{\includegraphics[width=0.2\textwidth]{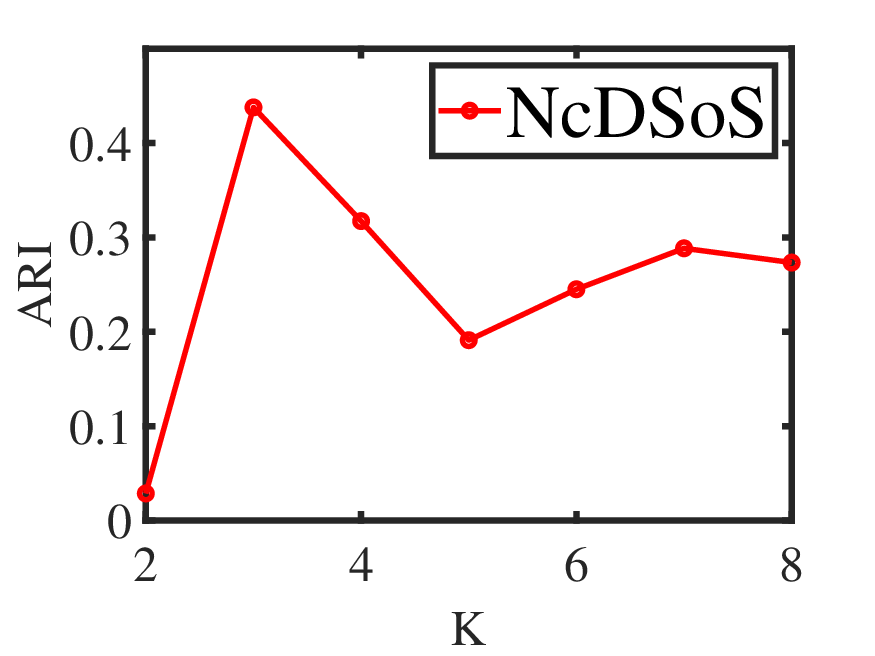}}
}
\resizebox{\columnwidth}{!}{
\subfigure[Lazega law firm]{\includegraphics[width=0.2\textwidth]{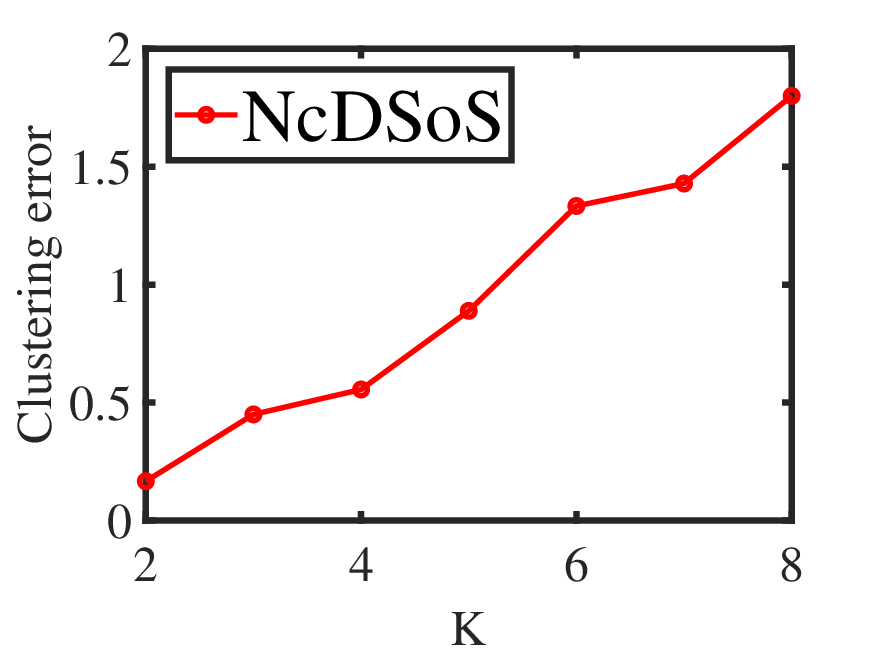}}
\subfigure[Lazega law firm]{\includegraphics[width=0.2\textwidth]{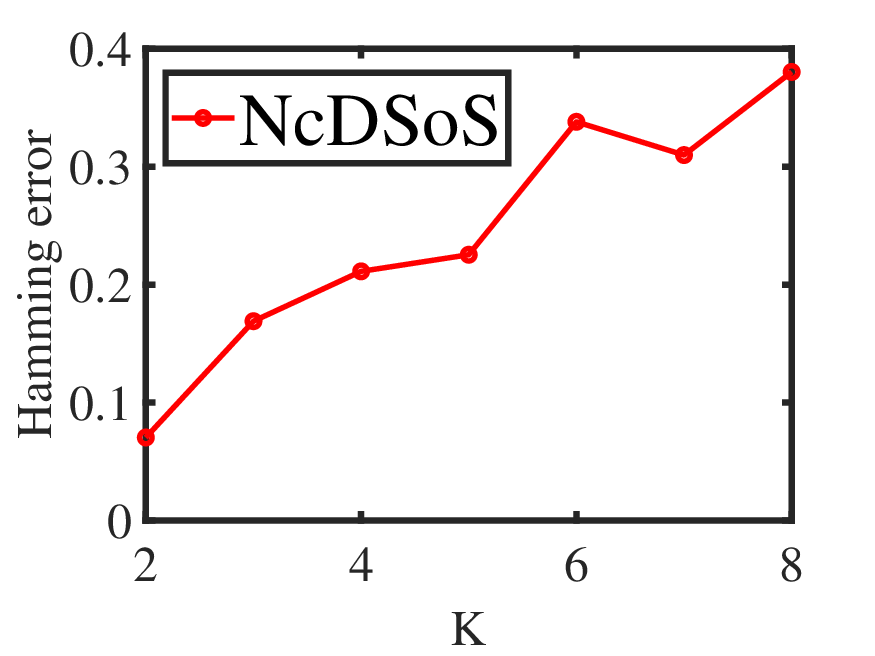}}
\subfigure[Lazega law firm]{\includegraphics[width=0.2\textwidth]{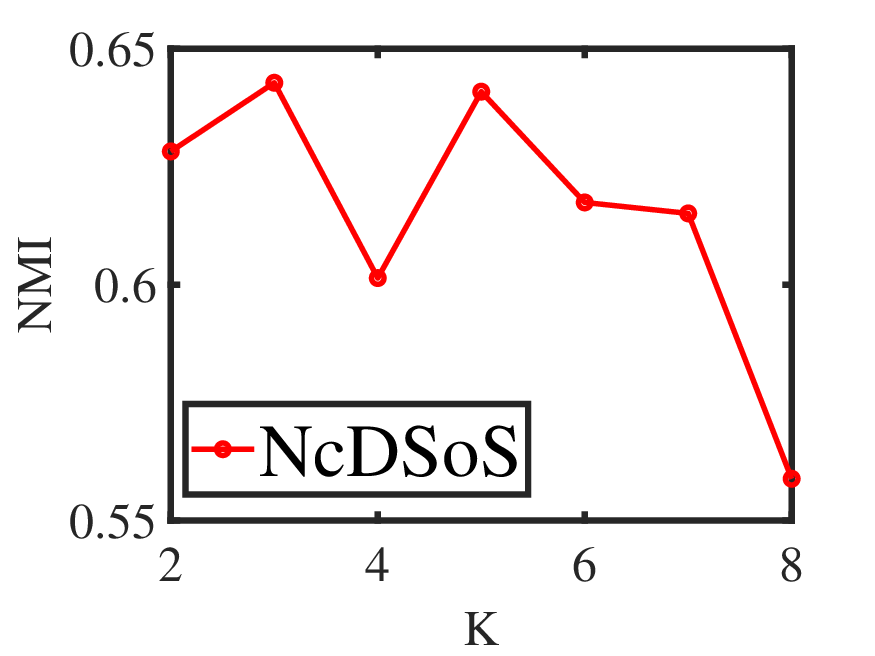}}
\subfigure[Lazega law firm]{\includegraphics[width=0.2\textwidth]{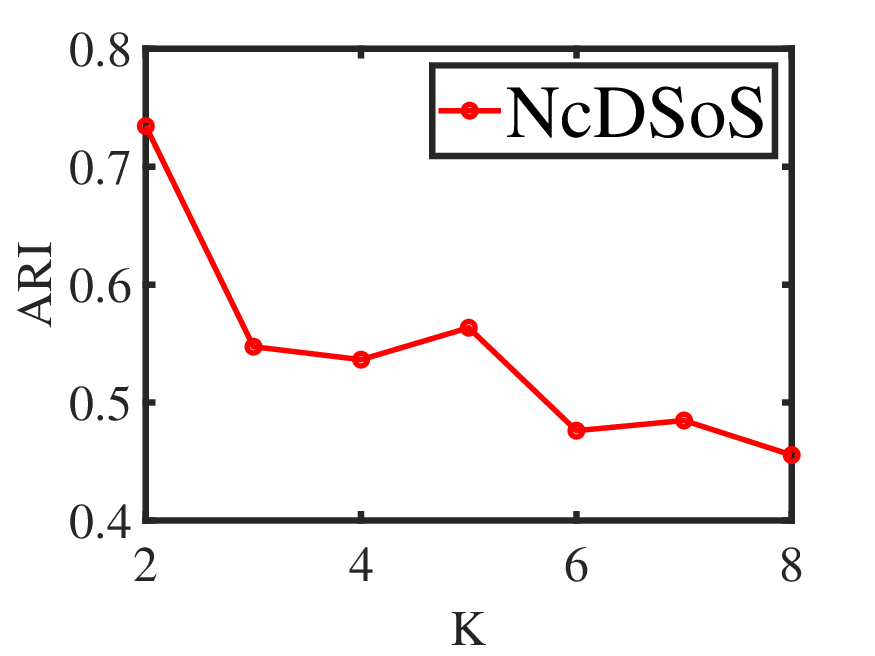}}
}
\resizebox{\columnwidth}{!}{
\subfigure[C.elegans]{\includegraphics[width=0.2\textwidth]{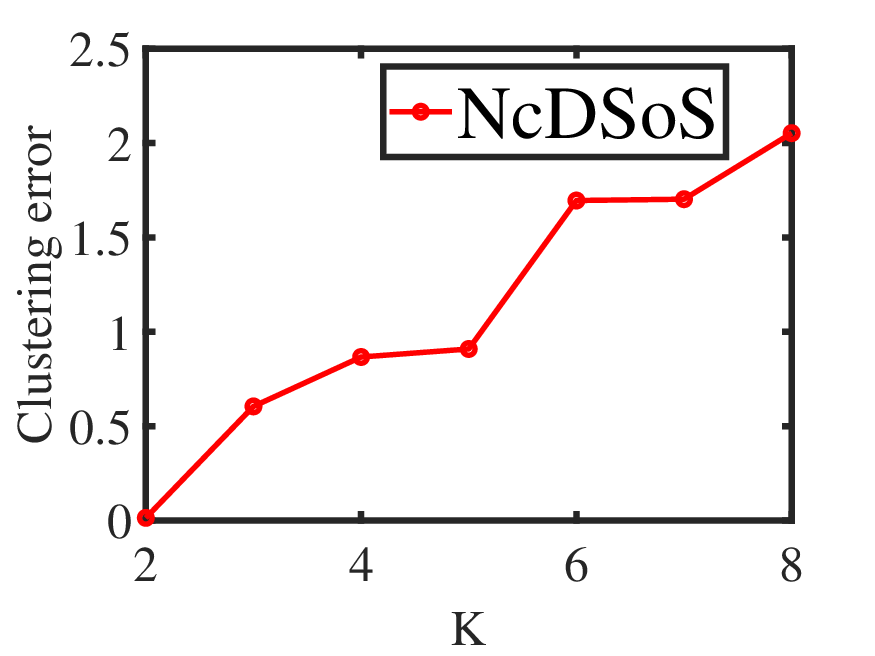}}
\subfigure[C.elegans]{\includegraphics[width=0.2\textwidth]{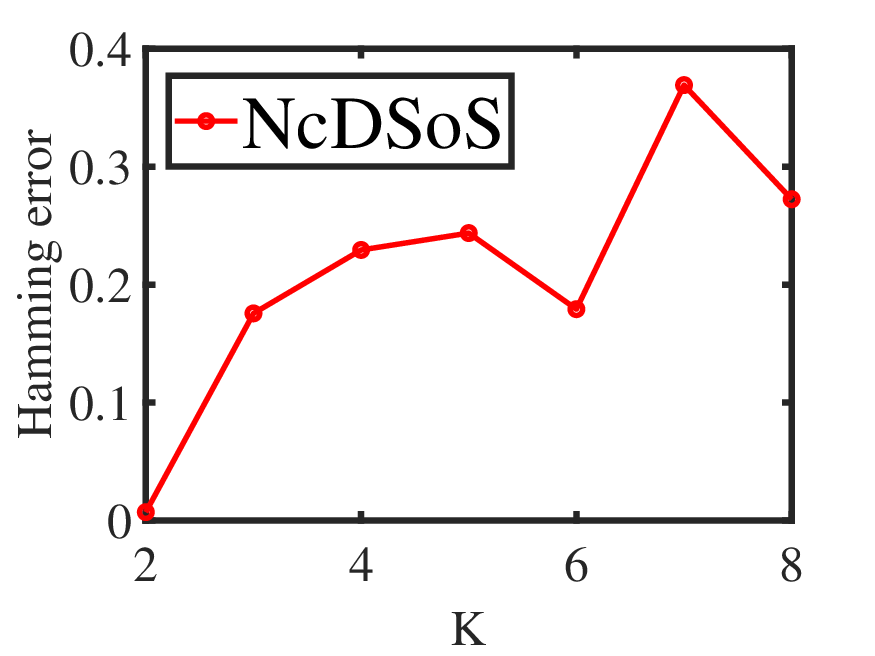}}
\subfigure[C.elegans]{\includegraphics[width=0.2\textwidth]{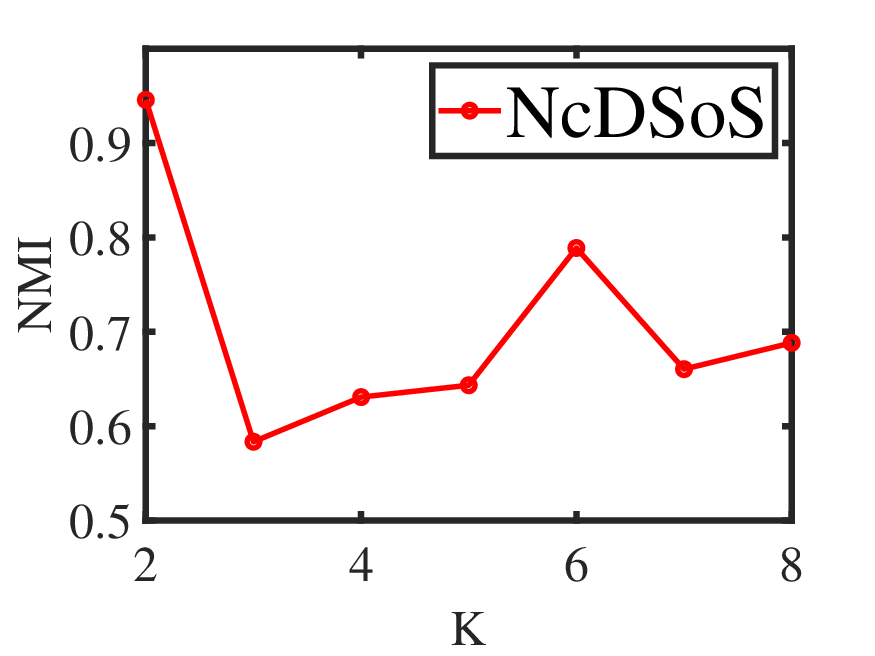}}
\subfigure[C.elegans]{\includegraphics[width=0.2\textwidth]{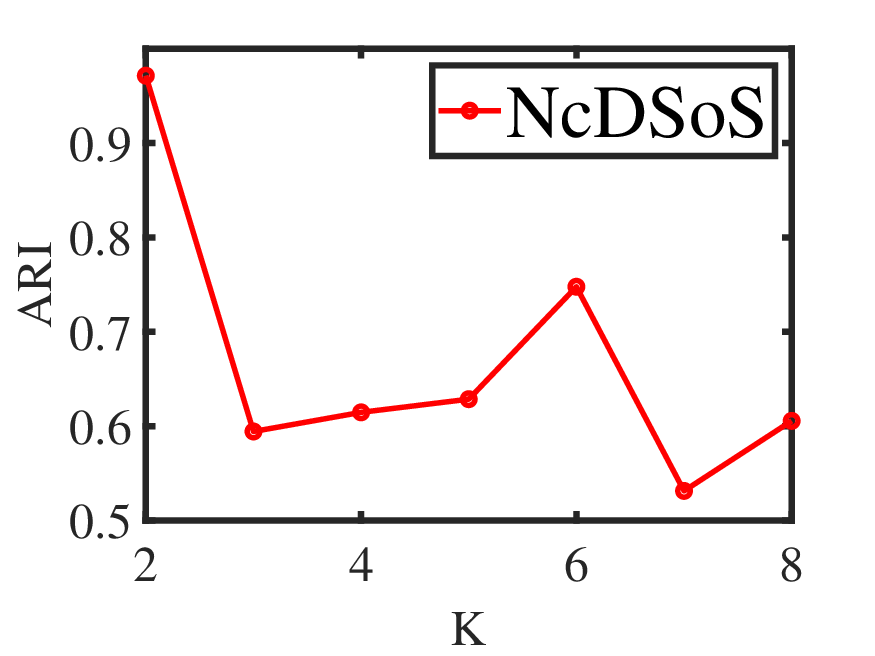}}
}
\caption{Four indices evaluating the difference between row clusters and column clusters against K for real data used in this paper, where row clusters and column clusters are obtained from our NcDSoS method.}
\label{ErrorAgainstK} 
\end{figure}

When dealing with real-world multi-layer directed networks, we face a challenge as the ground-truth community memberships and the exact number of row (column) communities remain unknown. To simplify our analysis, we assume that the number of row communities ($K_{r}$) is equal to the number of column communities ($K_{c}$), denoted as $K$. To quantify the disparity between the row and column clusters, we employ four indices: Clustering error, Hamming error, Normalized Mutual Information (NMI), and Adjusted Rand Index (ARI). Each of these metrics is calculated based on the estimated row community assignments ($\hat{Z}_{r}$) and column community assignments ($\hat{Z}_{c}$). Figure \ref{ErrorAgainstK} presents a graphical representation of how these four indices vary as $K$ increases. Based on this analysis, we can draw the following conclusions:
\begin{itemize}
  \item For Vickers Chan 7th Graders, the disparity between sending and receiving patterns is significant due to high Clustering and Hamming errors and low NMI and ARI values for $\hat{Z}_{r}$ and $\hat{Z}_{c}$. This aligns with the distributions of in-degree and out-degree shown in Figure \ref{HistDegree} panels (a)-(c). Notably, when $K=3$, the Hamming error is minimized, and NMI and ARI are maximized, indicating a relatively close similarity between sending and receiving patterns for this data.

  \item In the case of the Lazega law firm, the sending pattern significantly differs from the receiving pattern, which concurs with the findings in Figure \ref{HistDegree} panels (d)-(f). Notably, when $K=2$, both Clustering and Hamming errors are minimized, while ARI attains its maximum, suggesting a close resemblance between the sending and receiving patterns for this value of $K$.

  \item For C.elegans, as displayed in the last three panels of Figure \ref{HistDegree}, the sending pattern closely resembles the receiving pattern. Furthermore, Figure \ref{ErrorAgainstK} shows that for $K=2$, Clustering and Hamming errors are minimal, while NMI and ARI are maximal. Considering this, along with the last three panels of Figure \ref{HistDegree}, $K=2$ serves as an optimal choice for this data.
\end{itemize}

\begin{figure}
\centering
\resizebox{\columnwidth}{!}{
\subfigure[Vickers Chan 7thGraders]{\includegraphics[width=0.4\textwidth]{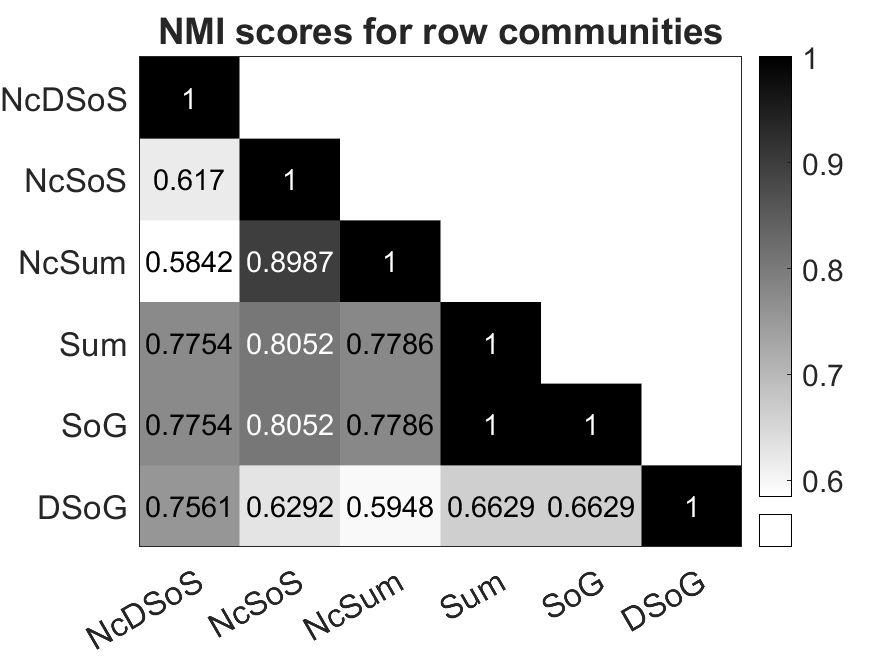}}
\subfigure[Vickers Chan 7thGraders]{\includegraphics[width=0.4\textwidth]{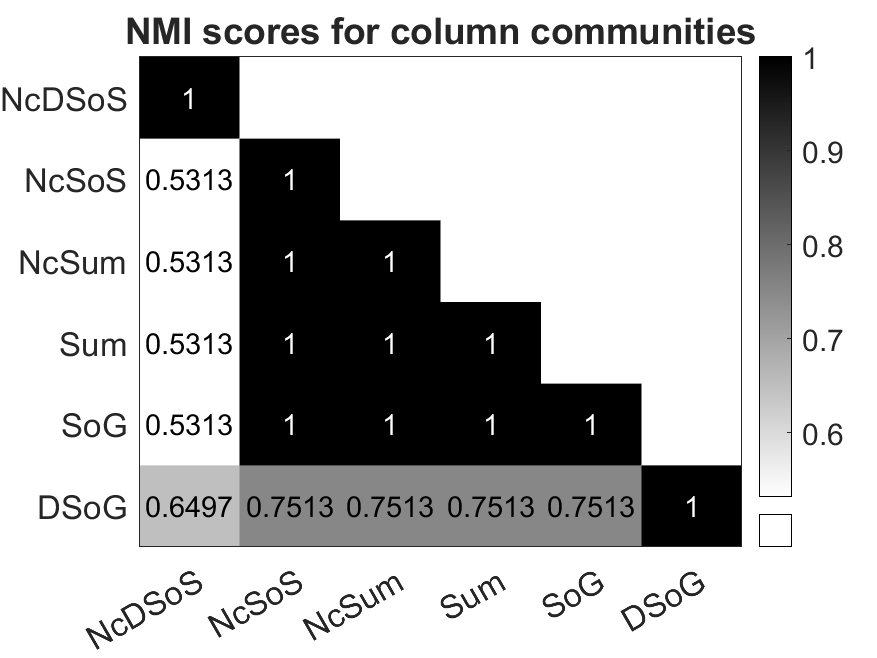}}
}
\resizebox{\columnwidth}{!}{
\subfigure[Lazega law firm]{\includegraphics[width=0.4\textwidth]{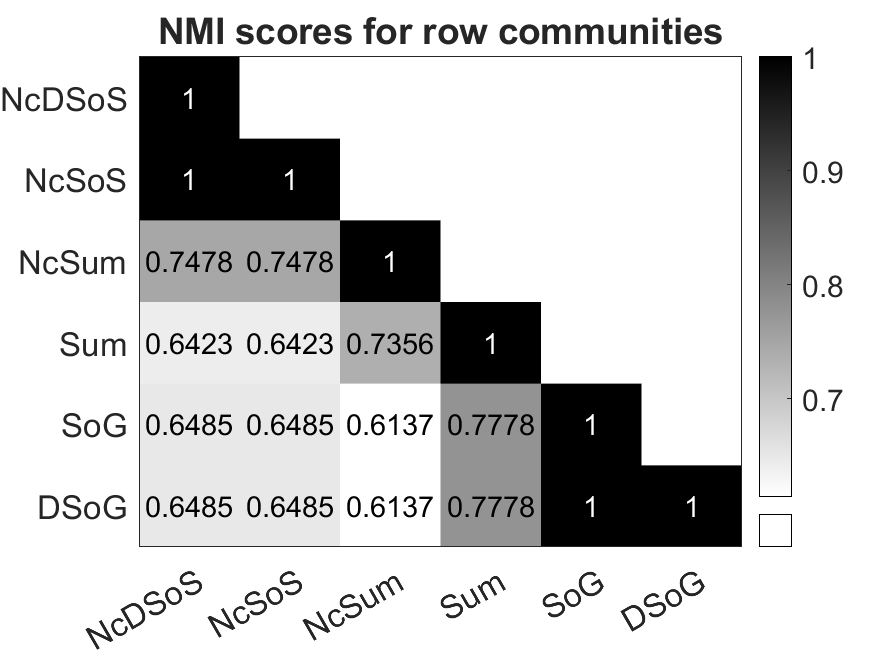}}
\subfigure[Lazega law firm]{\includegraphics[width=0.4\textwidth]{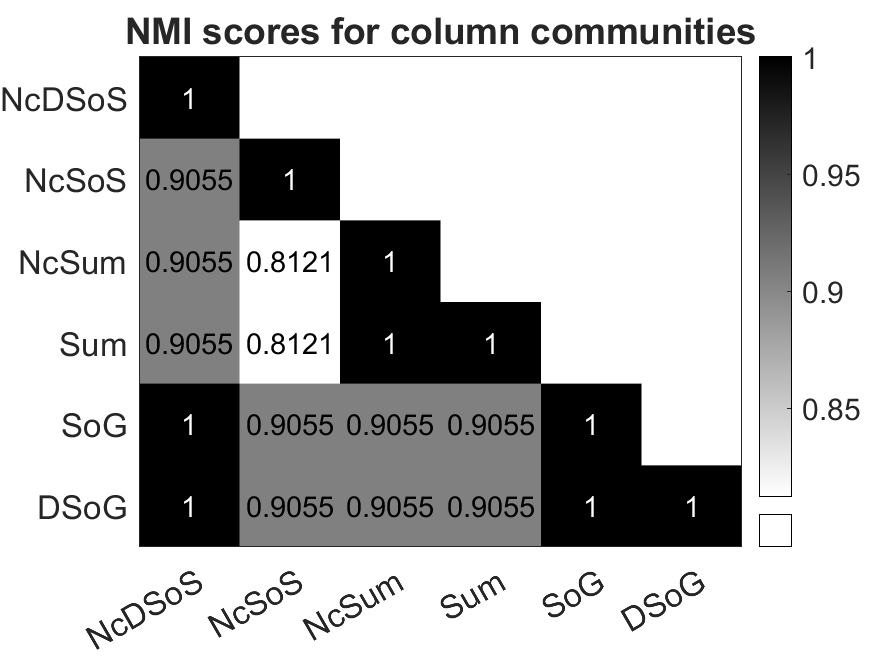}}
}
\resizebox{\columnwidth}{!}{
\subfigure[C.elegans]{\includegraphics[width=0.4\textwidth]{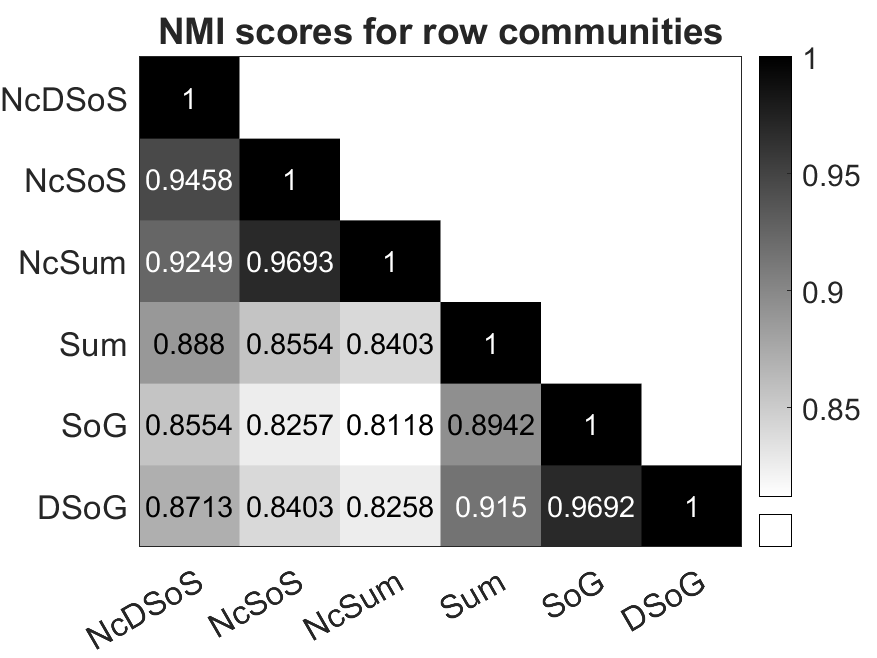}}
\subfigure[C.elegans]{\includegraphics[width=0.4\textwidth]{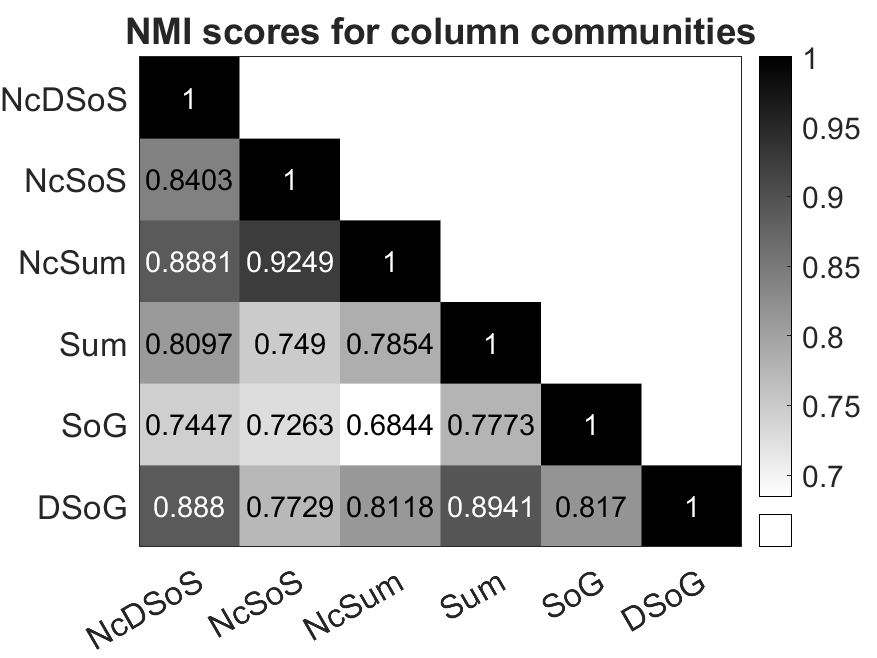}}
}
\caption{The pairwise comparison of NMI scores among 6 community detection approaches for both row and column communities of real-world multi-layer directed networks considered in this paper, where we set $K$ as 3, 2, and 2 for Vickers Chan 7thGraders, Lazega law firm, and C.elegans, respectively.}
\label{HeatNMI} 
\end{figure}

\begin{figure}
\centering
\resizebox{\columnwidth}{!}{
\subfigure[]{\includegraphics[width=0.2\textwidth]{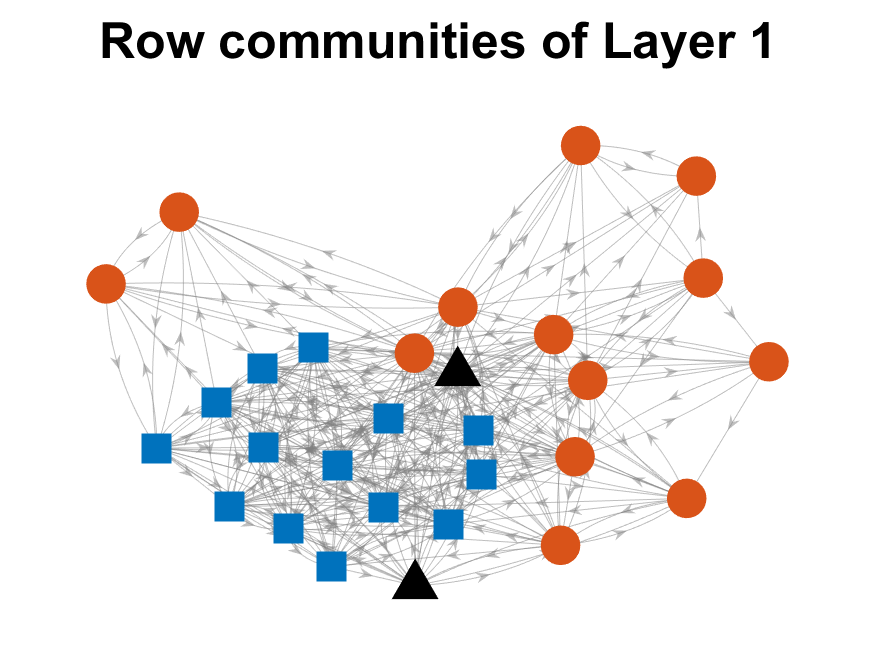}}
\subfigure[]{\includegraphics[width=0.2\textwidth]{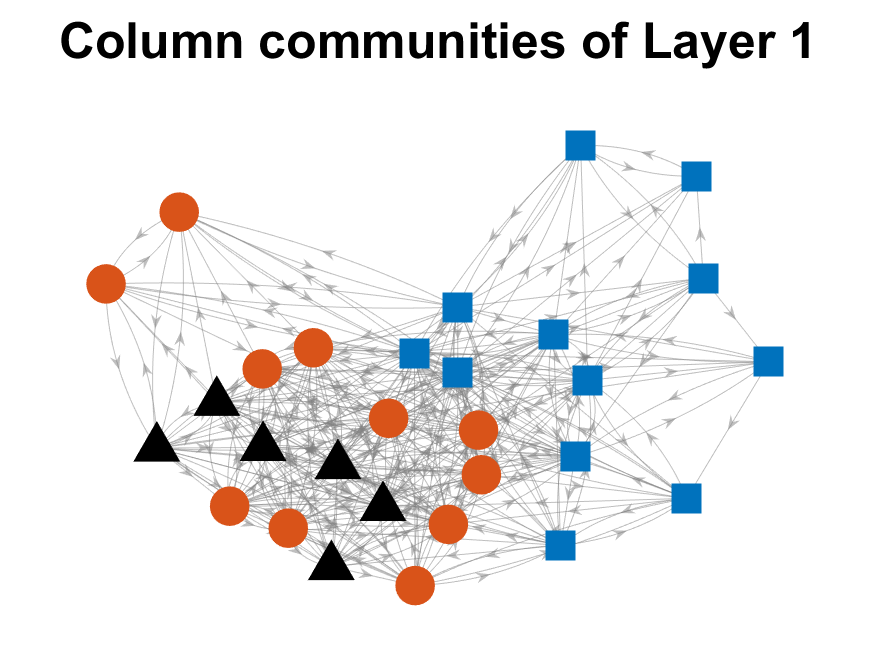}}
}
\resizebox{\columnwidth}{!}{
\subfigure[]{\includegraphics[width=0.2\textwidth]{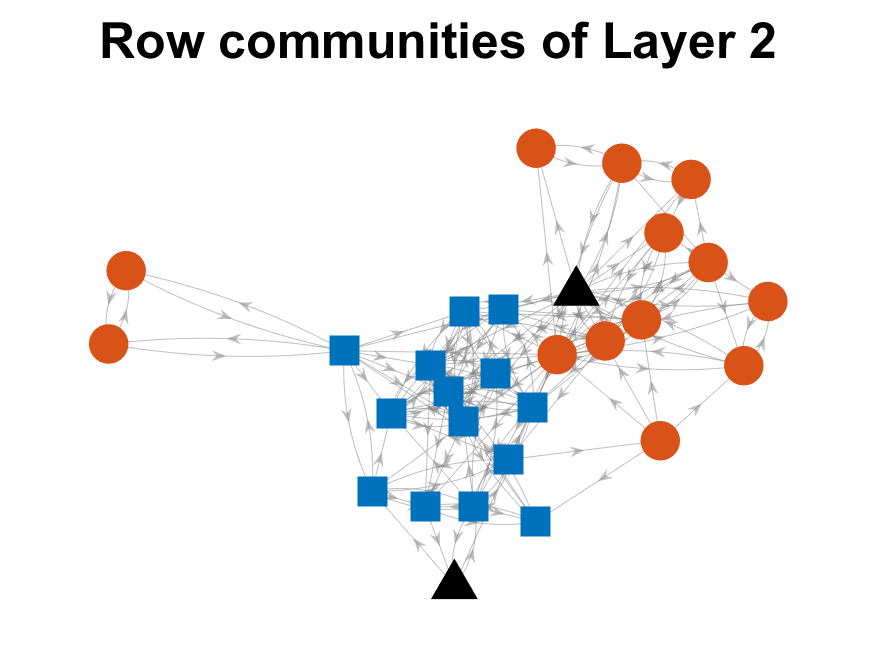}}
\subfigure[]{\includegraphics[width=0.2\textwidth]{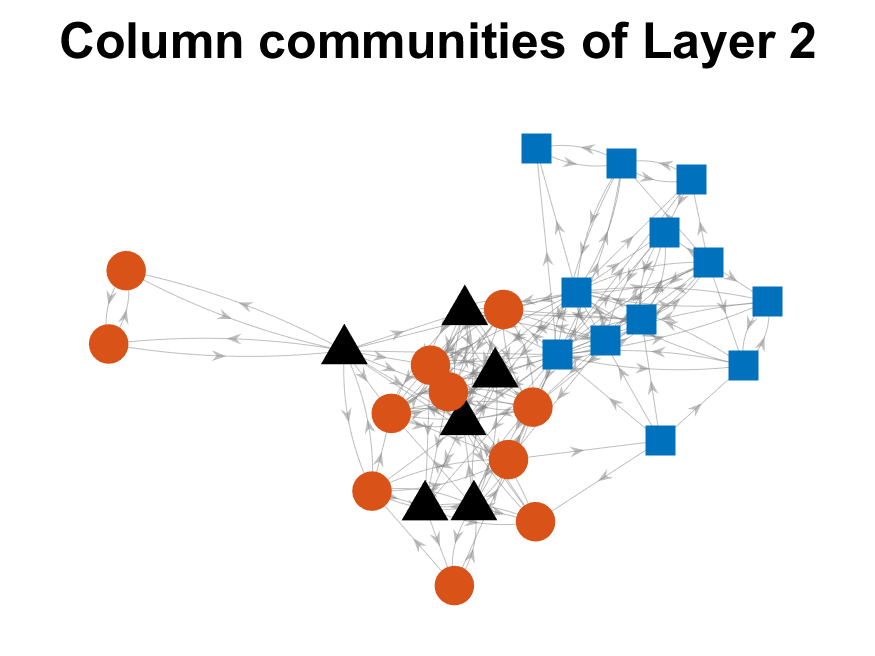}}
}
\resizebox{\columnwidth}{!}{
\subfigure[]{\includegraphics[width=0.2\textwidth]{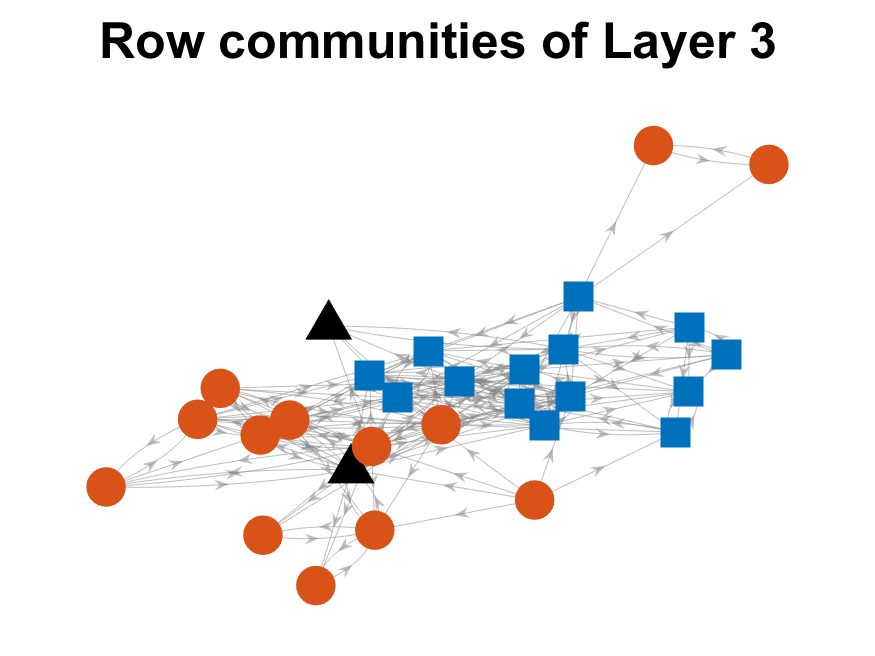}}
\subfigure[]{\includegraphics[width=0.2\textwidth]{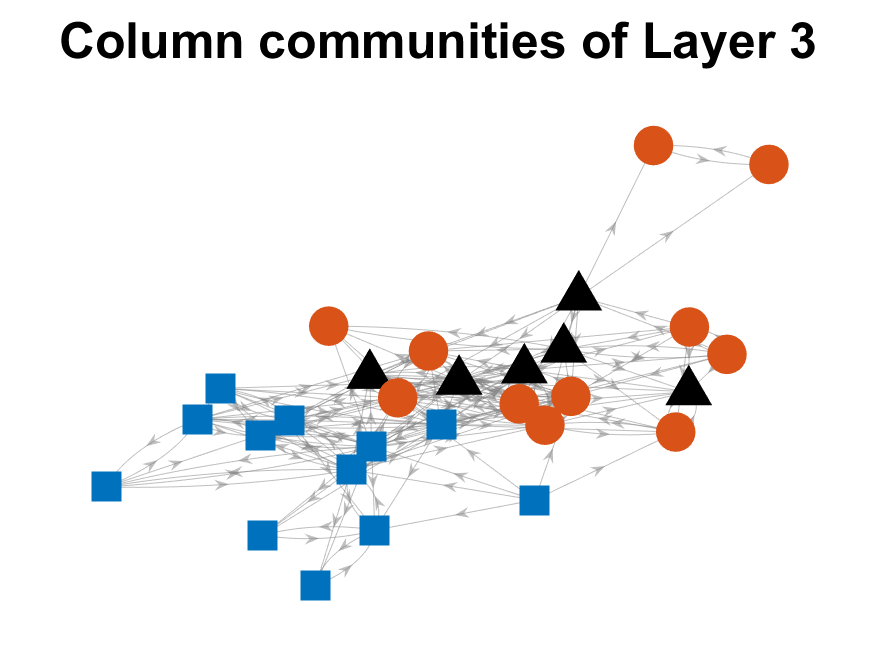}}
}
\caption{Visualization of communities detected by the NcDSoS algorithm when the number of communities $K$ is 3 for Vickers Chan 7thGraders. Nodes that possess identical color and shape belong to the same estimated community.}
\label{LabelsNetworkVC7} 
\end{figure}

\begin{figure}
\centering
\resizebox{\columnwidth}{!}{
\subfigure[]{\includegraphics[width=0.2\textwidth]{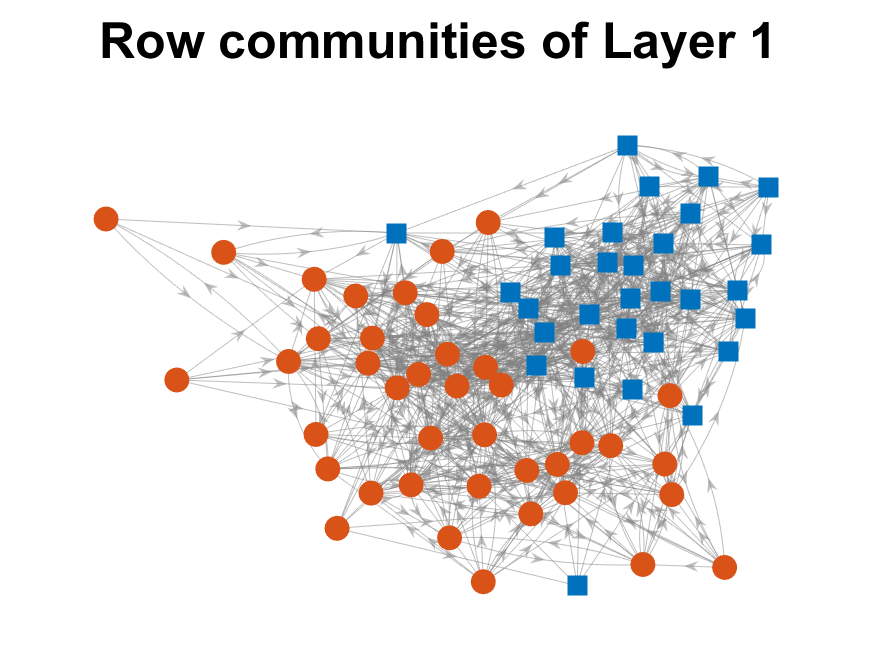}}
\subfigure[]{\includegraphics[width=0.2\textwidth]{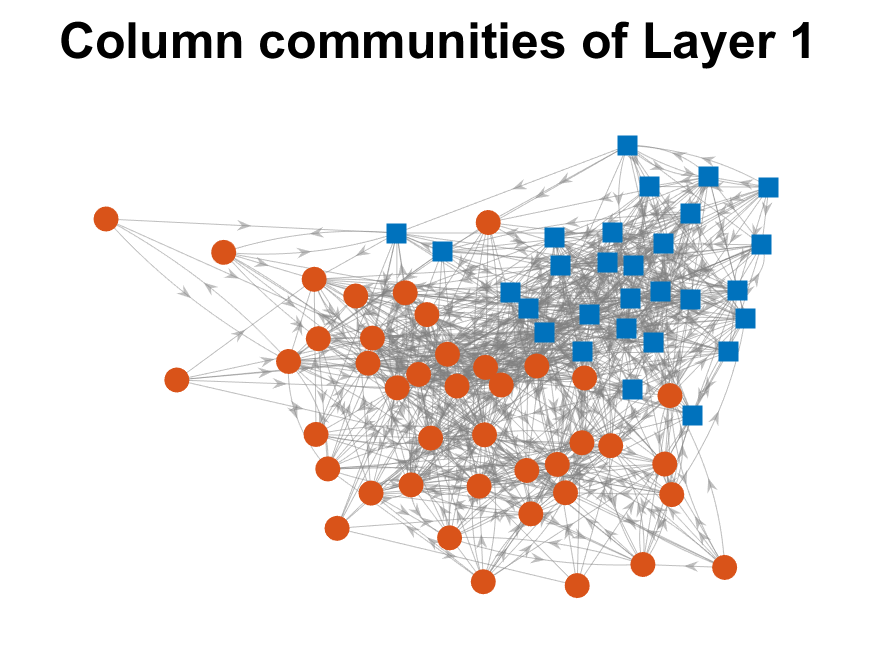}}
}
\resizebox{\columnwidth}{!}{
\subfigure[]{\includegraphics[width=0.2\textwidth]{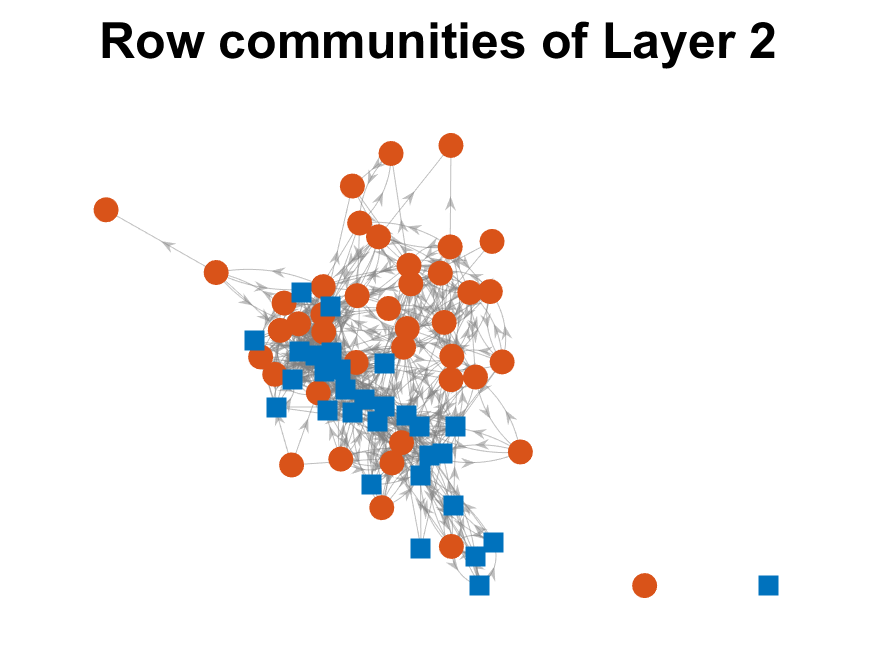}}
\subfigure[]{\includegraphics[width=0.2\textwidth]{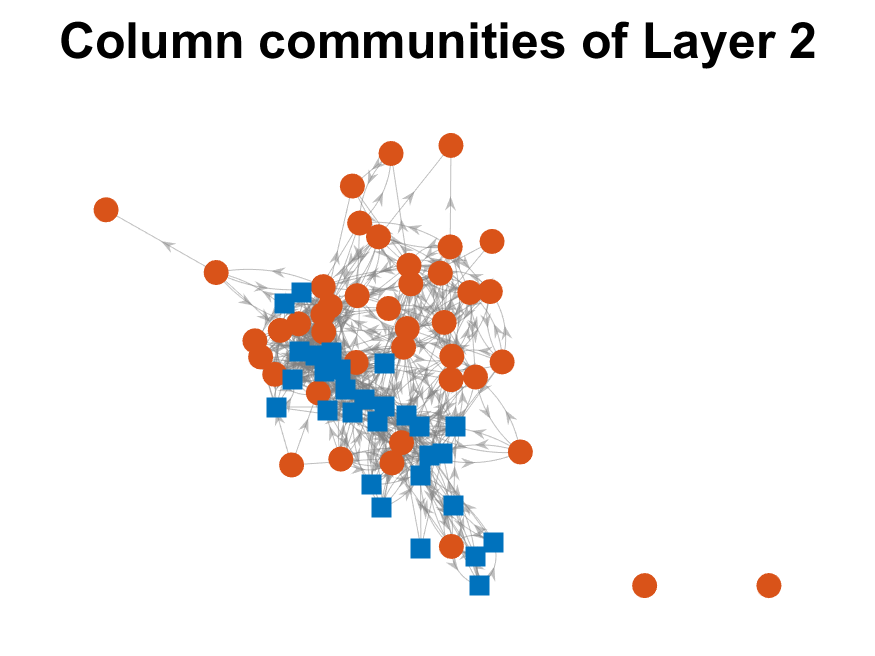}}
}
\resizebox{\columnwidth}{!}{
\subfigure[]{\includegraphics[width=0.2\textwidth]{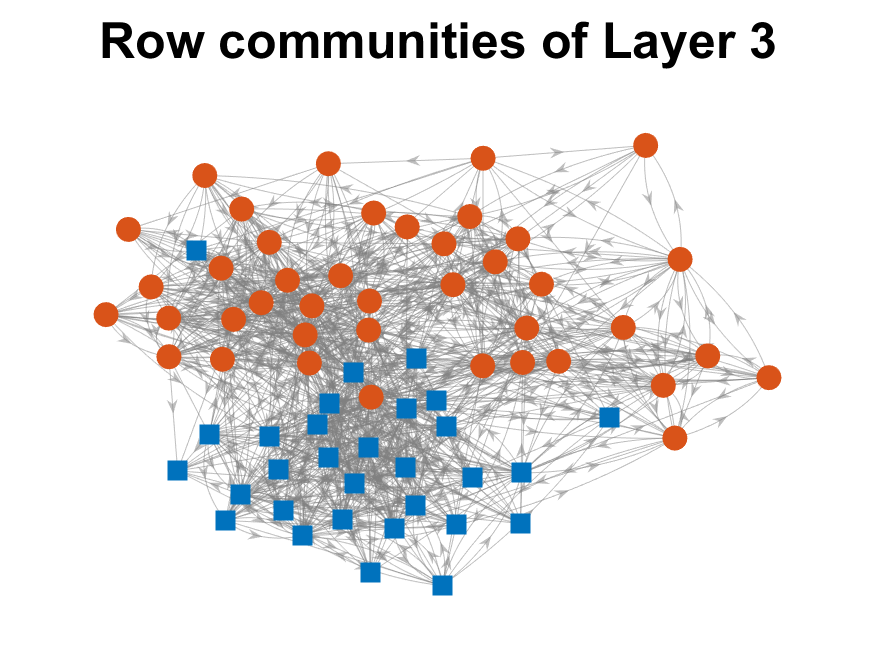}}
\subfigure[]{\includegraphics[width=0.2\textwidth]{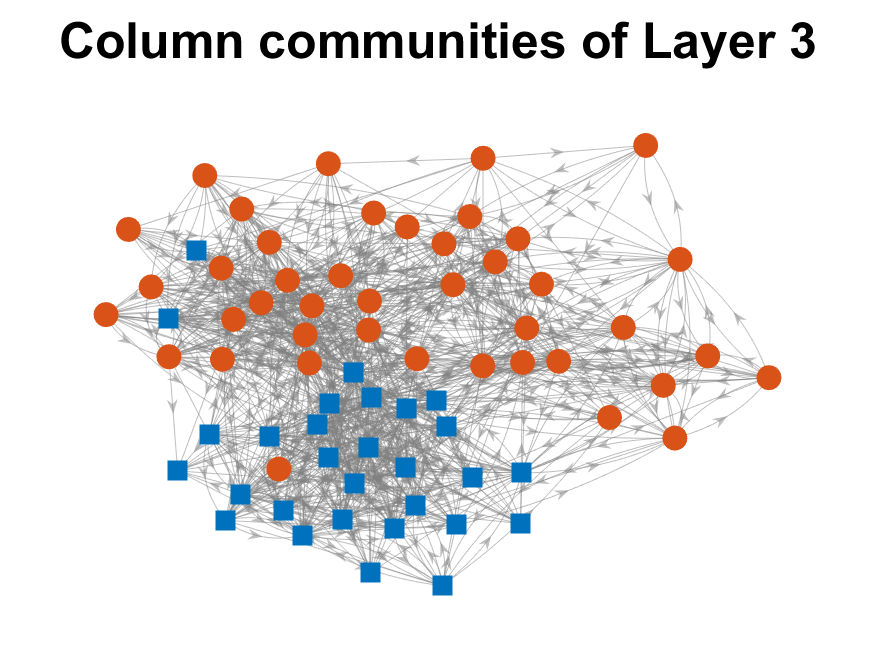}}
}
\caption{Visualization of communities detected by the NcDSoS algorithm when the number of communities $K$ is 2 for Lazega law firm. Nodes that possess identical color and shape belong to the same estimated community.}
\label{LabelsNetworkLazega} 
\end{figure}

\begin{figure}
\centering
\resizebox{\columnwidth}{!}{
\subfigure[]{\includegraphics[width=0.2\textwidth]{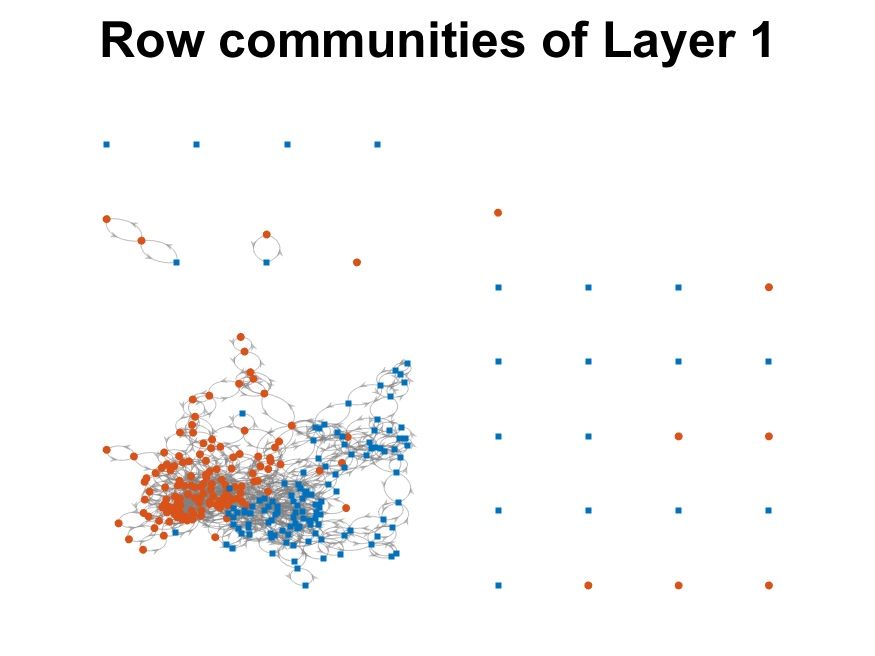}}
\subfigure[]{\includegraphics[width=0.2\textwidth]{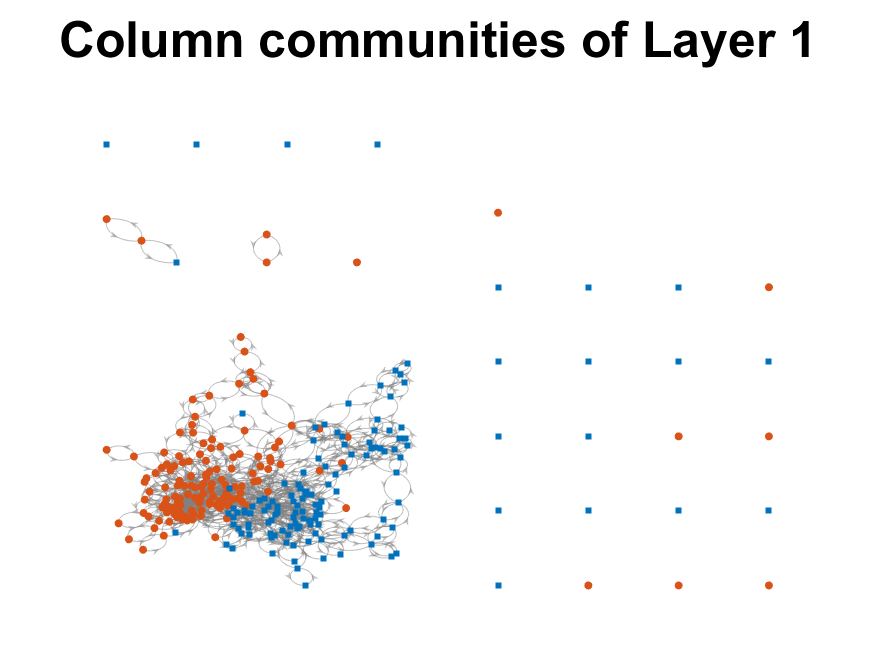}}
}
\resizebox{\columnwidth}{!}{
\subfigure[]{\includegraphics[width=0.2\textwidth]{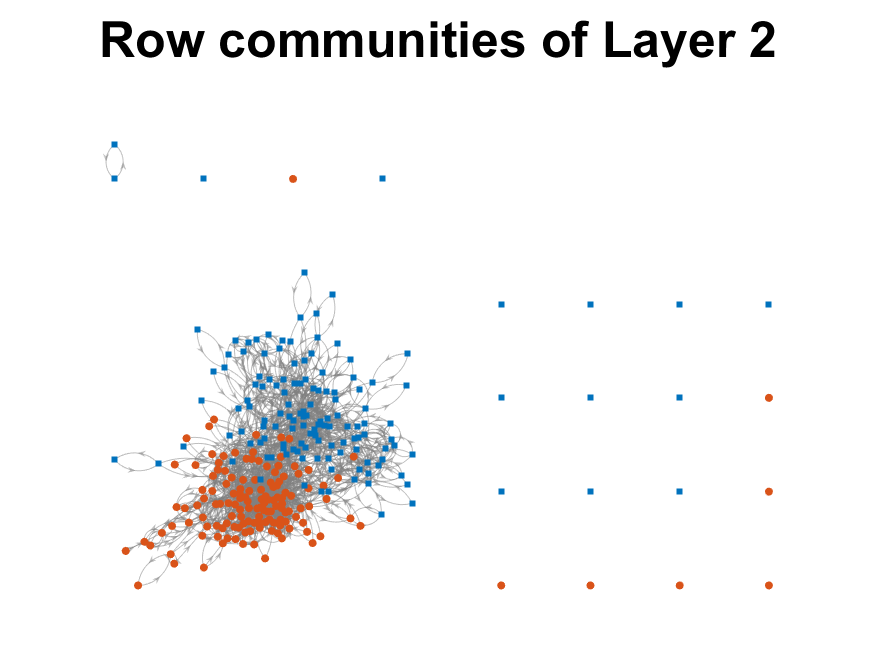}}
\subfigure[]{\includegraphics[width=0.2\textwidth]{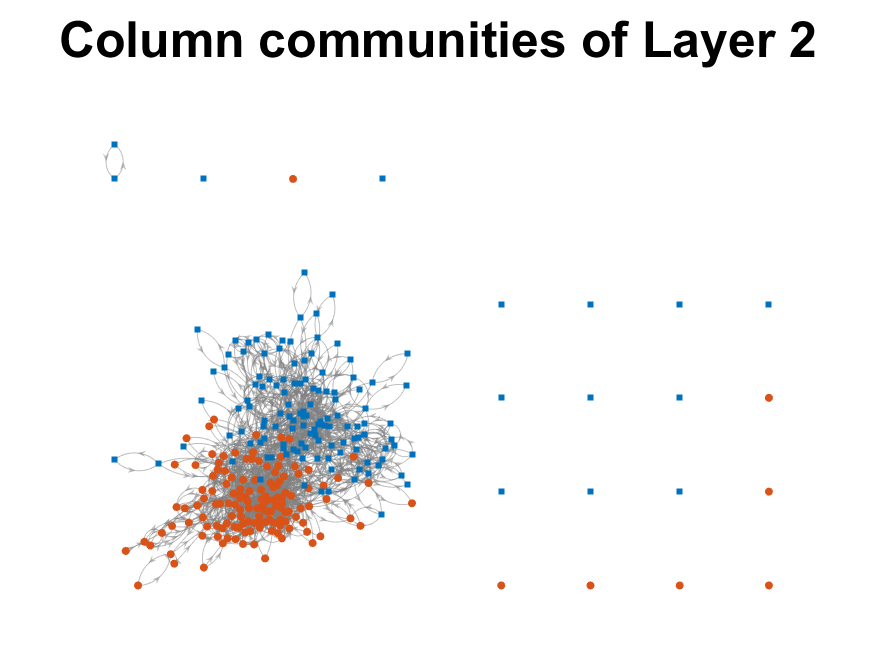}}
}
\resizebox{\columnwidth}{!}{
\subfigure[]{\includegraphics[width=0.2\textwidth]{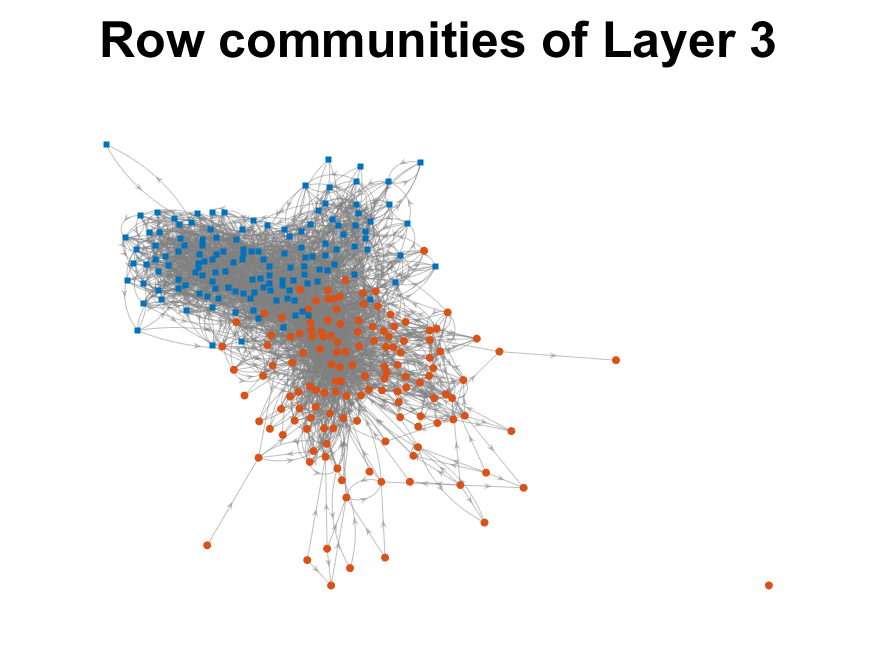}}
\subfigure[]{\includegraphics[width=0.2\textwidth]{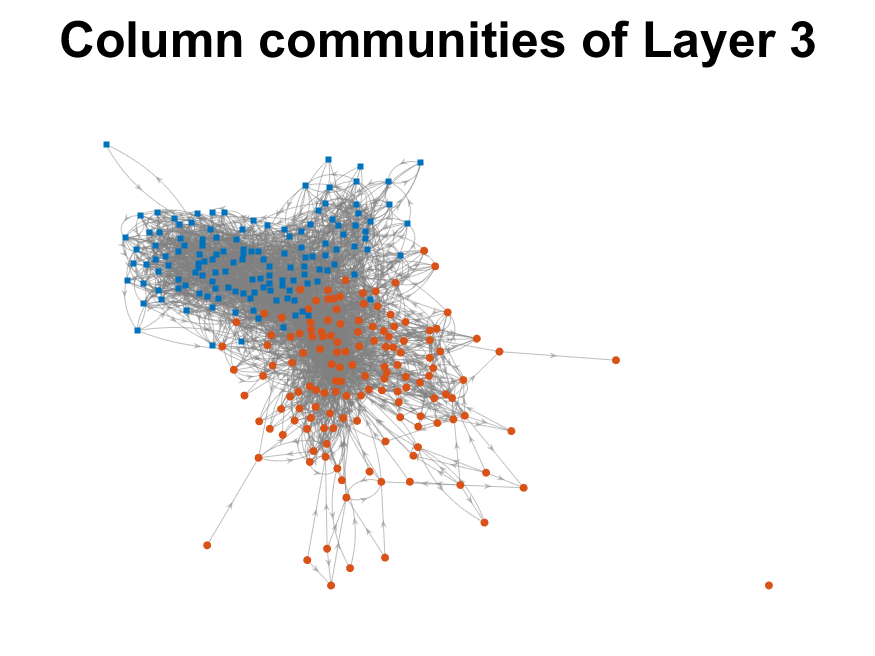}}
}
\caption{Visualization of communities detected by the NcDSoS algorithm when the number of communities $K$ is 2 for C.elegans. Nodes that possess identical color and shape belong to the same estimated community.}
\label{LabelsNetworkCelegan} 
\end{figure}

In our subsequent analysis, we specifically assign the value of $K$ as 3 for Vickers Chan 7th Graders, 2 for the Lazega law firm, and 2 for C.elegans. This choice is made to maintain a relatively similar sending and receiving pattern within these datasets. Figure \ref{HeatNMI} presents a heatmap that visualizes the pairwise NMI scores among the six community detection methods for both the row and column communities of the three real-world multi-layer directed networks. The clustering outcomes vary among different community detection methods. Given that our NcDSoS has outperformed its competitors in previous simulations, we tend to believe that our approach yields more precise community partitions. For visualization purposes, Figures \ref{LabelsNetworkVC7}, \ref{LabelsNetworkLazega}, and \ref{LabelsNetworkCelegan} illustrate the estimated row and column communities derived from our proposed method for Vickers Chan 7thGraders, Lazega law firm, and C.elegans, respectively. Notably, we observe a significant divergence between the row and column communities in the Vickers Chan 7thGraders and Lazega law firm cases, whereas, for C.elegans, the row communities closely align with the column communities. These observations align well with our previous findings.

\section{Conclusion}\label{sec7}
This paper addresses the challenge of community detection in multi-layer bipartite networks, a topic of significant importance in network science and data analysis. Unlike traditional multi-layer undirected networks, multi-layer bipartite networks consist of two distinct types of nodes, necessitating a specialized approach for community detection. To this end, we introduce a novel multi-layer degree-corrected stochastic co-block model (multi-layer DC-ScBM) that captures the latent community structure inherent in such networks. Our model extends previous network models by incorporating multi-layer bipartite networks and degree heterogeneity. We propose an efficient algorithm to detect communities within the proposed multi-layer DC-ScBM framework. The algorithm's performance is guaranteed by a rigorous theoretical analysis, which provides upper bounds on clustering errors and highlights the benefits of leveraging additional layers of bipartite networks for improved accuracy in community detection. Our numerical experiments of both synthetic and real-world datasets demonstrate the superiority of our algorithm over existing methods. In summary, this paper makes a significant contribution to the field of network science by introducing a novel model and algorithm for community detection in multi-layer bipartite networks. Our work stands at the intersection of theory and practice, offering both a solid methodological foundation and practical insights into the analysis of complex network systems.

To further enhance the accuracy and applicability of the proposed model and algorithm, future research could explore the integration of additional network features, such as node attributes. This would allow for a more comprehensive representation of the network structure, potentially leading to improved performance \citep{binkiewicz2017covariate,chunaev2020community,cai2024new}. For instance, node attributes could capture information about the type or location of nodes, providing a richer understanding of their roles and interactions within the network. Additionally, extending the framework to other types of networks, such as weighted networks, represents a promising avenue for future research. Weighted networks incorporate edge weights that represent the strength or importance of connections, providing a richer representation of network interactions \citep{newman2004analysis,opsahl2009clustering,qing2024bipartite}. Adapting the model and algorithm to handle weighted networks would enable the analysis of a wider range of network systems, such as social networks with varying connection strengths. Finally, accelerating the algorithm to handle larger-scale networks is a crucial next step. Enhancing the scalability of the algorithm would allow for the analysis of even more complex network systems, providing valuable insights into their structure. This acceleration could involve subsampling techniques and approximation algorithms \citep{deng2024subsampling,guo2023randomized}.
\section*{CRediT authorship contribution statement}
\textbf{Huan Qing:} Conceptualization; Data curation; Formal analysis; Funding acquisition; Methodology; Project administration; Resources; Software; Validation; Visualization; Writing-original draft; Writing-review $\&$ editing.
\section*{Declaration of competing interest}
The author declares no competing interests.
\section*{Data availability}
Data and code will be made available on request.
\section*{Acknowledgements}
H.Q. was sponsored by the Scientific Research Foundation of Chongqing University of Technology (Grant No: 0102240003) and the Natural Science Foundation of Chongqing, China (Grant No: CSTB2023NSCQ-LZX0048).
\appendix
\section{Proofs}\label{SecProofs}
\subsection{Proof of Lemma \ref{EigOsum2}}
\begin{proof}
Recall that $\Omega_{l}=\Theta_{r}Z_{r}B_{l}Z'_{c}\Theta_{c}$ for each $l\in[L]$, we have $\tilde{S}_{r}=\sum_{l\in[L]}\Omega_{l}\Omega'_{l}=\sum_{l\in[L]}\Theta_{r}Z_{r}B_{l}Z'_{c}\Theta^{2}_{c}Z_{c}B'_{l}Z'_{r}\Theta_{r}=\Theta_{r}Z_{r}(\sum_{l\in[L]}B_{l}Z'_{c}\Theta^{2}_{c}Z_{c}B'_{l})Z'_{r}\Theta_{r}=U_{r}\Lambda_{r}U'_{r}$, which gives that $U_{r}=\Theta_{r}Z_{r}(\sum_{l\in[L]}B_{l}Z'_{c}\Theta^{2}_{c}Z_{c}B'_{l})Z'_{r}\Theta_{r}U_{r}\Lambda^{-1}_{r}$. Set $X_{r}=(\sum_{l\in[L]}B_{l}Z'_{c}\Theta^{2}_{c}Z_{c}B'_{l})Z'_{r}\Theta_{r}U_{r}\Lambda^{-1}_{r}$, we have $U_{r}=\Theta_{r}Z_{r}X_{r}$. Thus, $U^{*}_{r}(i_{r},:)=\frac{U_{r}(i_{r},:)}{\|U_{r}(i_{r},:)\|_{F}}=\frac{\Theta_{r}(i_{r},i_{r})Z_{r}(i_{r},:)X_{r}}{\|\Theta_{r}(i_{r},i_{r})Z_{r}(i_{r},:)X_{r}\|_{F}}=\frac{Z_{r}(i_{r},:)X_{r}}{\|Z_{r}(i_{r},:)X_{r}\|_{F}}$, which suggests that $U^{*}_{r}(i_{r},:)=U^{*}_{r}(\tilde{i}_{r},:)$ if and only if $Z_{r}(i_{r},:)=Z_{r}(\tilde{i}_{r},:)$ for $i_{r}\in[n_{r}], \tilde{i}_{r}\in[n_{r}]$. Because $(\Theta_{c}Z_{c})'(\Theta_{c}Z_{c})$ is a positive definite diagonal matrix, $\mathrm{rank}(\sum_{l\in[L]}B_{l}(\Theta_{c}Z_{c})'(\Theta_{c}Z_{c})B'_{l})=\mathrm{rank}(\sum_{l\in[L]}B_{l}B'_{l})$. Therefore, when $\sum_{l\in[L]}B_{l}B'_{l}$ is full rank (i.e., $K=K_{r}$), Lemma 3 in \citep{qing2023community} gives that $\|U^{*}_{r}(i_{r},:)-U^{*}_{r}(\tilde{i}_{r},:)\|=\sqrt{2}$ if $Z_{r}(i_{r},:)\neq Z_{r}(\tilde{i}_{r},:)$. The rest proof for $U^{*}_{c}$ is similar to that of $U^{*}_{r}$, and we omit it here.
\end{proof}
\subsection{Proof of Lemma \ref{boundSsum}}
\begin{proof}
Firstly, we bound $\|S_{r}-\tilde{S}_{r}\|$. Let $E^{(ij)}_{r}$ be an $n_{r}\times n_{r}$ matrix such that its $(i,j)$-th element is 1 while all the other entries are zeros for $i\in[n_{r}], j\in[n_{r}]$. We decompose $S_{\mathrm{r}}-\tilde{S}_{r}$ into two parts:
\begin{align*}
&S_{r}-\tilde{S}_{r}=\sum_{l\in[L]}(A_{l}A'_{l}-D^{r}_{l}-\Omega_{l}\Omega'_{l})=\sum_{l\in[L]}\sum_{i\in[n_{r}]}\sum_{j\in[n_{r}]}\sum_{m\in[n_{c}]}(A_{l}(i,m)A_{l}(j,m)-\Omega_{l}(i,m)\Omega_{l}(j,m))E^{(ij)}_{r}-\sum_{l\in[L]}D^{r}_{l}\\
&=\sum_{l\in[L]}\sum_{i\in[n_{r}],j\in[n_{r}],i\neq j}\sum_{m\in[n_{c}]}(A_{l}(i,m)A_{l}(j,m)-\Omega_{l}(i,m)\Omega_{l}(j,m))E^{(ij)}_{r}+\sum_{l\in[L]}\sum_{m\in[n_{c}]}\sum_{i\in[n_{r}]}(A^{2}_{l}(i,m)-\Omega^{2}_{l}(i,m))E^{(ii)}_{r}-\sum_{l\in[L]}D^{r}_{l}\\
&=\sum_{l\in[L]}\sum_{i\in[n_{r}],j\in[n_{r}],i\neq j}\sum_{m\in[n_{c}]}(A_{l}(i,m)A_{l}(j,m)-\Omega_{l}(i,m)\Omega_{l}(j,m))E^{(ij)}_{r}+\sum_{l\in[L]}\sum_{m\in[n_{c}]}\sum_{i\in[n_{r}]}(A_{l}(i,m)-\Omega^{2}_{l}(i,m))E^{(ii)}_{r}-\sum_{l\in[L]}D^{r}_{l}\\
&=\sum_{l\in[L]}\sum_{i\in[n_{r}],j\in[n_{r}],i\neq j}\sum_{m\in[n_{c}]}(A_{l}(i,m)A_{l}(j,m)-\Omega_{l}(i,m)\Omega_{l}(j,m))E^{(ij)}_{r}-\sum_{l\in[L]}\sum_{m\in[n_{c}]}\sum_{i\in[n_{r}]}\Omega^{2}_{l}(i,m)E^{(ii)}_{r}+\sum_{l\in[L]}\sum_{i\in[n_{r}]}\sum_{m\in[n_{c}]}A_{l}(i,m)E^{(ii)}_{r}-\sum_{l\in[L]}D^{r}_{l}\\
&=\sum_{l\in[L]}\sum_{i\in[n_{r}],j\in[n_{r}],i\neq j}\sum_{m\in[n_{c}]}(A_{l}(i,m)A_{l}(j,m)-\Omega_{l}(i,m)\Omega_{l}(j,m))E^{(ij)}_{r}-\sum_{l\in[L]}\sum_{m\in[n_{c}]}\sum_{i\in[n_{r}]}\Omega^{2}_{l}(i,m)E^{(ii)}_{r}+\sum_{l\in[L]}D^{r}_{l}-\sum_{l\in[L]}D^{r}_{l}\\
&=\underbrace{\sum_{l\in[L]}\sum_{i\in[n_{r}],j\in[n_{r}],i\neq j}\sum_{m\in[n_{c}]}(A_{l}(i,m)A_{l}(j,m)-\Omega_{l}(i,m)\Omega_{l}(j,m))E^{(ij)}_{r}}_{R1}-\underbrace{\sum_{l\in[L]}\sum_{m\in[n_{c}]}\sum_{i\in[n_{r}]}\Omega^{2}_{l}(i,m)E^{(ii)}_{r}}_{R2}.
\end{align*}
$S_{r}-\tilde{S}_{r}=R1-R2$ gives
\begin{align*}
\|S_{r}-\tilde{S}_{r}\|&=\|R1-R2\|\leq\|R1\|+\|R2\|=\|R1\|+\|\sum_{l\in[L]}\sum_{m\in[n_{c}]}\sum_{i\in[n_{r}]}\Omega^{2}_{l}(i,m)E^{(ii)}_{r}\|\\
&=\|R1\|+\mathrm{max}_{i\in[n_{r}]}\sum_{l\in[L]}\sum_{m\in[n_{c}]}\Omega^{2}_{l}(i,m)\\
&\leq\|R1\|+\mathrm{max}_{i\in[n_{r}]}\sum_{l\in[L]}\sum_{m\in[n_{c}]}\theta^{2}_{r}(i)\theta^{2}_{c}(m)\leq\|R1\|+\theta^{2}_{r,\mathrm{max}}\sum_{l\in[L]}\sum_{m\in[n_{c}]}\theta^{2}_{c}(m)=\|R1\|+\theta^{2}_{r,\mathrm{max}}\|\theta_{c}\|^{2}_{F}L.
\end{align*}
Next, we bound $\|R1\|$ by using Theorem 1.6 (the rectangular case of matrix Bernstein inequality) \citep{tropp2012user}. We write this theorem below.
\begin{thm}\label{Bern}
(Theorem 1.6 of \citep{tropp2012user}) Consider a finite sequence  $\{X_{k}\}$ of independent, random matrices with dimensions $d_{1}\times d_{2}$. Suppose that each random matrix satisfies
\begin{align*}
\mathbb{E}[X_{k}]=0 \mathrm{~and~}\|X_{k}\|\leq R\mathrm{~almost~surely}.
\end{align*}
Define $\sigma^{2}:=\mathrm{max}\{\|\sum_{k}\mathbb{E}[X_{k}X'_{k}]\|,\|\sum_{k}\mathbb{E}[X'_{k}X_{k}]\|\}$.
Then, for all $t\geq0$,
\begin{align*}
\mathbb{P}(\|\sum_{k}X_{k}\|\geq t)\leq(d_{1}+d_{2})\cdot\mathrm{exp}(\frac{-t^{2}/2}{\sigma^{2}+Rt/3}).
\end{align*}
\end{thm}
Set $X^{(ijml)}_{r}=(A_{l}(i,m)A_{l}(j,m)-\Omega_{l}(i,m)\Omega_{l}(j,m))E^{(ij)}_{r}$ for $i\in[n_{r}],j\in[n_{r}], i\neq j, m\in[n_{c}],l\in[L]$. We have $R1=\sum_{l\in[L]}\sum_{i\in[n_{r}],j\in[n_{r}],i\neq j}\sum_{m\in[n_{c}]}X^{(ijml)}_{r}$. Note that although $R1$ is symmetric, $X^{(ijml)}_{r}$ is asymmetric and we need the rectangular case of matrix Bernstein to bound $\|R1\|$. The following results hold.
\begin{itemize}
  \item $\mathbb{E}[X^{(ijml)}_{r}]=0$ since $A_{l}(i,m)$ and $A_{l}(j,m)$ are independent when $i\neq j$.
  \item $\|X^{(ijml)}_{r}\|=|A_{l}(i,m)A_{l}(j,m)-\Omega_{l}(i,m)\Omega_{l}(j,m)|\leq1$. Set $R_{r}=1$.
  \item Set $\sigma^{2}_{r}=\mathrm{max}\{\|\sum_{l\in[L]}\sum_{i\in[n_{r}],j\in[n_{r}],i\neq j}\sum_{m\in[n_{c}]}\mathbb{E}[X^{(ijml)}_{r}(X^{(ijml)}_{r})']\|, \|\sum_{l\in[L]}\sum_{i\in[n_{r}],j\in[n_{r}],i\neq j}\sum_{m\in[n_{c}]}\mathbb{E}[(X^{(ijml)}_{r})'X^{(ijml)}_{r}]\|\}$. Under the multi-layer DC-ScBM, for the term $\|\sum_{l\in[L]}\sum_{i\in[n_{r}],j\in[n_{r}],i\neq j}\sum_{m\in[n_{c}]}\mathbb{E}[X^{(ijml)}_{r}(X^{(ijml)}_{r})']\|$, we have
      \begin{align*}
      &\|\sum_{l\in[L]}\sum_{i\in[n_{r}],j\in[n_{r}],i\neq j}\sum_{m\in[n_{c}]}\mathbb{E}[(A_{l}(i,m)A_{l}(j,m)-\Omega_{l}(i,m)\Omega_{l}(j,m))^{2}E^{(ii)}_{r}]\|\\
      &=\|\sum_{l\in[L]}\sum_{i\in[n_{r}],j\in[n_{r}],i\neq j}\sum_{m\in[n_{c}]}E^{(ii)}_{r}\mathbb{E}[A^{2}_{l}(i,m)A^{2}_{l}(j,m)-2A_{l}(i,m)A_{l}(j,m)\Omega_{l}(i,m)\Omega_{l}(j,m)+\Omega^{2}_{l}(i,m)\Omega^{2}_{l}(j,m)]\|\\
      &=\|\sum_{l\in[L]}\sum_{i\in[n_{r}],j\in[n_{r}],i\neq j}\sum_{m\in[n_{c}]}E^{(ii)}_{r}(\mathbb{E}[A^{2}_{l}(i,m)A^{2}_{l}(j,m)]-\Omega^{2}_{l}(i,m)\Omega^{2}_{l}(j,m))\|\\
      &=\|\sum_{l\in[L]}\sum_{i\in[n_{r}],j\in[n_{r}],i\neq j}\sum_{m\in[n_{c}]}E^{(ii)}_{r}(\mathbb{E}[A_{l}(i,m)A_{l}(j,m)]-\Omega^{2}_{l}(i,m)\Omega^{2}_{l}(j,m))\|\\
      &=\|\sum_{l\in[L]}\sum_{i\in[n_{r}],j\in[n_{r}],i\neq j}\sum_{m\in[n_{c}]}E^{(ii)}_{r}(\mathbb{E}[A_{l}(i,m)]\mathbb{E}[A_{l}(j,m)]-\Omega^{2}_{l}(i,m)\Omega^{2}_{l}(j,m))\|\\
      &=\|\sum_{l\in[L]}\sum_{i\in[n_{r}],j\in[n_{r}],i\neq j}\sum_{m\in[n_{c}]}E^{(ii)}_{r}\Omega_{l}(i,m)\Omega_{l}(j,m)(1-\Omega_{l}(i,m)\Omega_{l}(j,m))\|\\
      &=\mathrm{max}_{i\in[n_{r}]}\sum_{l\in[L]}\sum_{j\in[n_{r}],j\neq i}\sum_{m\in[n_{c}]}\Omega_{l}(i,m)\Omega_{l}(j,m)(1-\Omega_{l}(i,m)\Omega_{l}(j,m))\\
      &\leq\mathrm{max}_{i\in[n_{r}]}\sum_{l\in[L]}\sum_{j\in[n_{r}],j\neq i}\sum_{m\in[n_{c}]}\Omega_{l}(i,m)\Omega_{l}(j,m)\\
      &\leq\mathrm{max}_{i\in[n_{r}]}\sum_{l\in[L]}\sum_{j\in[n_{r}],j\neq i}\sum_{m\in[n_{c}]}\theta_{r}(i)\theta^{2}_{c}(m)\theta_{r}(j)\leq\theta_{r,\mathrm{max}}\sum_{l\in[L]}\sum_{j\in[n_{r}]}\sum_{m\in[n_{c}]}\theta^{2}_{c}(m)\theta_{r}(j)=\theta_{r,\mathrm{max}}\|\theta_{r}\|_{1}\|\theta_{c}\|^{2}_{F}L.
      \end{align*}
\end{itemize}
Similarly, we have $\|\sum_{l\in[L]}\sum_{i\in[n_{r}],j\in[n_{r}],i\neq j}\sum_{m\in[n_{c}]}\mathbb{E}[(X^{(ijml)}_{r})'X^{(ijml)}_{r}]\|\leq\theta_{r,\mathrm{max}}\|\theta_{r}\|_{1}\|\theta_{c}\|^{2}_{F}L$. Hence, $\sigma^{2}_{r}\leq \theta_{r,\mathrm{max}}\|\theta_{r}\|_{1}\|\theta_{c}\|^{2}_{F}L$. For any $t_{r}\geq0$, Theorem \ref{Bern} gives
\begin{align*}
\mathbb{P}(\|R1\|\geq t_{r})\leq 2n_{r}\cdot\mathrm{exp}(\frac{-t^{2}_{r}/2}{\sigma^{2}_{r}+R_{r}t_{r}/3})\leq 2n_{r}\cdot\mathrm{exp}(\frac{-t^{2}_{r}/2}{\theta_{r,\mathrm{max}}\|\theta_{r}\|_{1}\|\theta_{c}\|^{2}_{F}L+t_{r}/3}).
\end{align*}
Set $t_{r}=\frac{\alpha+1+\sqrt{(\alpha+1)(\alpha+19)}}{3}\sqrt{\theta_{r,\mathrm{max}}\|\theta_{r}\|_{1}\|\theta_{c}\|^{2}_{F}L\mathrm{log}(n_{r}+n_{c}+L)}$ for any $\alpha\geq0$. By Assumption \ref{Assum2}, we get
\begin{align*}
\mathbb{P}(\|R1\|\geq t_{r})\leq 2n_{r}\cdot\mathrm{exp}(-(\alpha+1)\mathrm{log}(n_{r}+n_{c}+L)\frac{1}{\frac{18}{(\sqrt{\alpha+1}+\sqrt{\alpha+19})^{2}}+\frac{2\sqrt{\alpha+1}}{\sqrt{\alpha+1}+\sqrt{\alpha+19}}\sqrt{\frac{\mathrm{log}(n_{r}+n_{c}+L)}{\theta_{r,\mathrm{max}}\|\theta_{r}\|_{1}\|\theta_{c}\|^{2}_{F}L}}})\leq\frac{2n_{r}}{(n_{r}+n_{c}+L)^{\alpha+1}}<\frac{2}{(n_{r}+n_{c}+L)^{\alpha}}.
\end{align*}
Setting $\alpha=1$, with probability at least $1-o(\frac{1}{n_{r}+n_{c}+L})$, we have
\begin{align*}
\|S_{r}-\tilde{S}_{r}\|=O(\sqrt{\theta_{r,\mathrm{max}}\|\theta_{r}\|_{1}\|\theta_{c}\|^{2}_{F}L\mathrm{log}(n_{r}+n_{c}+L)})+O(\theta^{2}_{r,\mathrm{max}}\|\theta_{c}\|^{2}_{F}L).
\end{align*}

Secondly, we bound $\|S_{c}-\tilde{S}_{c}\|$. Let $E^{(ij)}_{c}$ be an $n_{c}\times n_{c}$ matrix with its $(i,j)$-th entry being 1 and all the other entries being zeros for $i\in[n_{c}], j\in[n_{c}]$. We have
\begin{align*}
&S_{c}-\tilde{S}_{c}=\sum_{l\in[L]}(A'_{l}A_{l}-D^{c}_{l}-\Omega'_{l}\Omega_{l})=\sum_{l\in[L]}\sum_{i\in[n_{c}]}\sum_{j\in[n_{c}]}\sum_{m\in[n_{r}]}(A_{l}(m,i)A_{l}(m,j)-\Omega_{l}(m,i)\Omega_{l}(m,j))E^{(ij)}_{c}-\sum_{l\in[L]}D^{c}_{l}\\
&=\underbrace{\sum_{l\in[L]}\sum_{i\in[n_{c}],j\in[n_{c}],i\neq j}\sum_{m\in[n_{r}]}(A_{l}(m,i)A_{l}(m,j)-\Omega_{l}(m,i)\Omega_{l}(m,j))E^{(ij)}_{c}}_{C1}-\underbrace{\sum_{l\in[L]}\sum_{m\in[n_{r}]}\sum_{i\in[n_{c}]}\Omega^{2}_{l}(m,i)E^{(ii)}_{c}}_{C2},
\end{align*}
which gives
\begin{align*}
\|S_{c}-\tilde{S}_{c}\|&=\|C1-C2\|\leq\|C1\|+\|C2\|=\|C1\|+\|\sum_{l\in[L]}\sum_{m\in[n_{r}]}\sum_{i\in[n_{c}]}\Omega^{2}_{l}(m,i)E^{(ii)}_{c}\|\\
&=\|C1\|+\mathrm{max}_{i\in[n_{c}]}\sum_{l\in[L]}\sum_{m\in[n_{r}]}\Omega^{2}_{l}(m,i)\\
&\leq\|C1\|+\mathrm{max}_{i\in[n_{c}]}\sum_{l\in[L]}\sum_{m\in[n_{r}]}\theta^{2}_{c}(i)\theta^{2}_{r}(m)\leq\|C1\|+\theta^{2}_{c,\mathrm{max}}\sum_{l\in[L]}\sum_{m\in[n_{r}]}\theta^{2}_{r}(m)=\|C1\|+\theta^{2}_{c,\mathrm{max}}\|\theta_{r}\|^{2}_{F}L.
\end{align*}
As long as we obtain an upper bound of $\|C1\|$, we can bound $\|S_{c}-\tilde{S}_{c}\|$.

Setting $X^{(ijml)}_{c}=(A_{l}(m,i)A_{l}(m,j)-\Omega_{l}(m,i)\Omega_{l}(m,j))E^{(ij)}_{c}$ (for $i\in[n_{c}],j\in[n_{c}], i\neq j, m\in[n_{r}],l\in[L]$) gives $C1=\sum_{l\in[L]}\sum_{i\in[n_{c}],j\in[n_{c}],i\neq j}\sum_{m\in[n_{r}]}X^{(ijml)}_{c}$. We have the following conclusions:
\begin{itemize}
  \item $\mathbb{E}[X^{(ijml)}_{c}]=0$ because $A_{l}(m,i)$ and $A_{l}(m,j)$ are independent when $i\neq j$.
  \item $\|X^{(ijml)}_{c}\|\leq1$. Set $R_{c}=1$.
  \item Set $\sigma^{2}_{c}=\mathrm{max}\{\|\sum_{l\in[L]}\sum_{i\in[n_{c}],j\in[n_{c}],i\neq j}\sum_{m\in[n_{r}]}\mathbb{E}[X^{(ijml)}_{c}(X^{(ijml)}_{c})']\|, \|\sum_{l\in[L]}\sum_{i\in[n_{c}],j\in[n_{c}],i\neq j}\sum_{m\in[n_{r}]}\mathbb{E}[(X^{(ijml)}_{c})'X^{(ijml)}_{c}]\|\}$. For $\|\sum_{l\in[L]}\sum_{i\in[n_{c}],j\in[n_{c}],i\neq j}\sum_{m\in[n_{r}]}\mathbb{E}[X^{(ijml)}_{c}(X^{(ijml)}_{c})']\|$, we have
      \begin{align*}
      &\|\sum_{l\in[L]}\sum_{i\in[n_{c}],j\in[n_{c}],i\neq j}\sum_{m\in[n_{r}]}\mathbb{E}[(A_{l}(m,i)A_{l}(m,j)-\Omega_{l}(m,i)\Omega_{l}(m,j))^{2}E^{(ii)}_{c}]\|\\
      &=\|\sum_{l\in[L]}\sum_{i\in[n_{c}],j\in[n_{c}],i\neq j}\sum_{m\in[n_{r}]}E^{(ii)}_{c}\Omega_{l}(m,i)\Omega_{l}(m,j)(1-\Omega_{l}(m,i)\Omega_{l}(m,j))\|\\
      &=\mathrm{max}_{i\in[n_{c}]}\sum_{l\in[L]}\sum_{j\in[n_{c}],j\neq i}\sum_{m\in[n_{r}]}\Omega_{l}(m,i)\Omega_{l}(m,j)(1-\Omega_{l}(m,i)\Omega_{l}(m,j))\\
      &\leq\mathrm{max}_{i\in[n_{c}]}\sum_{l\in[L]}\sum_{j\in[n_{c}],j\neq i}\sum_{m\in[n_{r}]}\Omega_{l}(m,i)\Omega_{l}(m,j)\\
      &\leq\mathrm{max}_{i\in[n_{c}]}\sum_{l\in[L]}\sum_{j\in[n_{c}],j\neq i}\sum_{m\in[n_{r}]}\theta_{c}(i)\theta^{2}_{r}(m)\theta_{c}(j)\leq\theta_{c,\mathrm{max}}\sum_{l\in[L]}\sum_{j\in[n_{c}]}\sum_{m\in[n_{r}]}\theta^{2}_{r}(m)\theta_{c}(j)=\theta_{c,\mathrm{max}}\|\theta_{c}\|_{1}\|\theta_{r}\|^{2}_{F}L.
      \end{align*}
\end{itemize}
Following a similar analysis, we get $\|\sum_{l\in[L]}\sum_{i\in[n_{c}],j\in[n_{c}],i\neq j}\sum_{m\in[n_{r}]}\mathbb{E}[(X^{(ijml)}_{c})'X^{(ijml)}_{c}]\|\leq\theta_{c,\mathrm{max}}\|\theta_{c}\|_{1}\|\theta_{r}\|^{2}_{F}L$. Thus, $\sigma^{2}_{c}\leq \theta_{c,\mathrm{max}}\|\theta_{c}\|_{1}\|\theta_{r}\|^{2}_{F}L$. For any $t_{c}\geq0$, by Theorem \ref{Bern}, we have
\begin{align*}
\mathbb{P}(\|C1\|\geq t_{c})\leq 2n_{c}\cdot\mathrm{exp}(\frac{-t^{2}_{c}/2}{\sigma^{2}_{c}+R_{c}t_{c}/3})\leq 2n_{c}\cdot\mathrm{exp}(\frac{-t^{2}_{c}/2}{\theta_{c,\mathrm{max}}\|\theta_{c}\|_{1}\|\theta_{r}\|^{2}_{F}L+t_{c}/3}).
\end{align*}
Set $t_{c}=\frac{\alpha+1+\sqrt{(\alpha+1)(\alpha+19)}}{3}\sqrt{\theta_{c,\mathrm{max}}\|\theta_{c}\|_{1}\|\theta_{r}\|^{2}_{F}L\mathrm{log}(n_{r}+n_{c}+L)}$ for any $\alpha\geq0$. By Assumption \ref{Assum2}, we obtain
\begin{align*}
\mathbb{P}(\|C1\|\geq t_{c})\leq 2n_{c}\cdot\mathrm{exp}(-(\alpha+1)\mathrm{log}(n_{r}+n_{c}+L)\frac{1}{\frac{18}{(\sqrt{\alpha+1}+\sqrt{\alpha+19})^{2}}+\frac{2\sqrt{\alpha+1}}{\sqrt{\alpha+1}+\sqrt{\alpha+19}}\sqrt{\frac{\mathrm{log}(n_{r}+n_{c}+L)}{\theta_{c,\mathrm{max}}\|\theta_{c}\|_{1}\|\theta_{r}\|^{2}_{F}L}}})\leq\frac{2n_{c}}{(n_{r}+n_{c}+L)^{\alpha+1}}<\frac{2}{(n_{r}+n_{c}+L)^{\alpha}}.
\end{align*}
Letting $\alpha=1$, with probability at least $1-o(\frac{1}{n_{r}+n_{c}+L})$, we have
\begin{align*}
\|S_{c}-\tilde{S}_{c}\|=O(\sqrt{\theta_{c,\mathrm{max}}\|\theta_{c}\|_{1}\|\theta_{r}\|^{2}_{F}L\mathrm{log}(n_{r}+n_{c}+L)})+O(\theta^{2}_{c,\mathrm{max}}\|\theta_{r}\|^{2}_{F}L).
\end{align*}
\end{proof}
\subsection{Proof of Theorem \ref{mainNcDSoS}}
\begin{proof}
By Lemma 5.1 \citep{lei2015consistency}, there are two orthogonal matrices $Q_{r}\in\mathbb{R}^{K_{r}\times K_{r}}$ and $Q_{c}\in\mathbb{R}^{K_{c}\times K_{c}}$ such that
\begin{align*}
\|\hat{U}_{r}Q_{r}-U_{r}\|_{F}\leq\frac{2\sqrt{2K_{r}}\|S_{r}-\tilde{S}_{r}\|}{|\lambda_{K_{r}}(\tilde{S}_{r})|}\mathrm{~and~}\|\hat{U}_{c}Q_{c}-U_{c}\|_{F}\leq\frac{2\sqrt{2K_{c}}\|S_{c}-\tilde{S}_{c}\|}{|\lambda_{K_{c}}(\tilde{S}_{c})|}.
\end{align*}
Combine Assumption \ref{Assum22} with the truth that $Z'_{c}\Theta^{2}_{c}Z_{c}$ is a diagonal matrix, we get
\begin{align*}
|\lambda_{K_{r}}(\tilde{S}_{r})|&=|\lambda_{K_{r}}(\sum_{l\in[L]}\Theta_{r}Z_{r}B_{l}Z'_{c}\Theta^{2}_{c}Z_{c}B'_{l}Z'_{r}\Theta_{r})|=|\lambda_{K_{r}}(\Theta_{r}Z_{r}(\sum_{l\in[L]}B_{l}Z'_{c}\Theta^{2}_{c}Z_{c}B'_{l})Z'_{r}\Theta_{r})|\geq\lambda^{2}_{K_{r}}(\Theta_{r})\lambda_{K_{r}}(Z'_{r}Z_{r})|\lambda_{K_{r}}(\sum_{l\in[L]}B_{l}Z'_{c}\Theta^{2}_{c}Z_{c}B'_{l})|\\
&=O(\lambda^{2}_{K_{r}}(\Theta_{r})\lambda^{2}_{K_{r}}(\Theta_{c})\lambda_{K_{r}}(Z'_{r}Z_{r})\lambda_{K_{r}}(Z'_{c}Z_{c})|\lambda_{K_{r}}(\sum_{l\in[L]}B_{l}B'_{l})|)=\lambda^{2}_{K_{r}}(\Theta_{r})\lambda^{2}_{K_{r}}(\Theta_{c})\lambda_{K_{r}}(Z'_{r}Z_{r})\lambda_{K_{r}}(Z'_{c}Z_{c})O(|\lambda_{K_{r}}(\sum_{l\in[L]}B_{l}B'_{l})|)\\
&\geq \theta^{2}_{r,\mathrm{min}}\theta^{2}_{c,\mathrm{min}}n_{r,\mathrm{min}}n_{c,\mathrm{min}}O(\lambda_{K_{r}}(\sum_{l\in[L]}B_{l}B'_{l}))=O(\theta^{2}_{r,\mathrm{min}}\theta^{2}_{c,\mathrm{min}}n_{r,\mathrm{min}}n_{c,\mathrm{min}}L).
\end{align*}
Similarly, we have $|\lambda_{K_{c}}(\tilde{S}_{c})|\geq O(\theta^{2}_{r,\mathrm{min}}\theta^{2}_{c,\mathrm{min}}n_{r,\mathrm{min}}n_{c,\mathrm{min}}L)$. Thus, $\|\hat{U}_{r}Q_{r}-U_{r}\|_{F}=O(\frac{\sqrt{K_{r}}\|S_{r}-\tilde{S}_{r}\|}{\theta^{2}_{r,\mathrm{min}}\theta^{2}_{c,\mathrm{min}}n_{r,\mathrm{min}}n_{c,\mathrm{min}}L})$ and $\|\hat{U}_{c}Q_{c}-U_{c}\|_{F}=O(\frac{\sqrt{K_{c}}\|S_{c}-\tilde{S}_{c}\|}{\theta^{2}_{r,\mathrm{min}}\theta^{2}_{c,\mathrm{min}}n_{r,\mathrm{min}}n_{c,\mathrm{min}}L})$. Set $\mu_{r}=\mathrm{min}_{i\in[n_{r}]}\|U_{r}(i,:)\|_{F}$ and $\mu_{c}=\mathrm{min}_{j\in[n_{c}]}\|U_{c}(j,:)\|_{F}$. By basic algebra, we have $\|\hat{U}^{*}_{r}Q_{r}-U^{*}_{r}\|_{F}\leq\frac{2\|\hat{U}_{r}Q_{r}-U_{r}\|_{F}}{\mu_{r}}$ and $\|\hat{U}^{*}_{c}Q_{c}-U^{*}_{c}\|_{F}\leq\frac{2\|\hat{U}_{c}Q_{c}-U_{c}\|_{F}}{\mu_{c}}$. According to the proof of Lemma 7 \citep{qing2023community}, we have $\frac{1}{\mu_{r}}\leq\frac{\theta_{r,\mathrm{max}}\sqrt{n_{r,\mathrm{max}}}}{\theta_{r,\mathrm{min}}}$ and $\frac{1}{\mu_{c}}\leq\frac{\theta_{c,\mathrm{max}}\sqrt{n_{c,\mathrm{max}}}}{\theta_{c,\mathrm{min}}}$. Then, we get
\begin{align*}
\|\hat{U}^{*}_{r}Q_{r}-U^{*}_{r}\|_{F}=O(\frac{\theta_{r,\mathrm{max}}\sqrt{K_{r}n_{r,\mathrm{max}}}\|S_{r}-\tilde{S}_{r}\|}{\theta^{3}_{r,\mathrm{min}}\theta^{2}_{c,\mathrm{min}}n_{r,\mathrm{min}}n_{c,\mathrm{min}}L}) \mathrm{~and~}\|\hat{U}^{*}_{c}Q_{c}-U^{*}_{c}\|_{F}=O(\frac{\theta_{c,\mathrm{max}}\sqrt{K_{c}n_{c,\mathrm{max}}}\|S_{c}-\tilde{S}_{c}\|}{\theta^{2}_{r,\mathrm{min}}\theta^{3}_{c,\mathrm{min}}n_{r,\mathrm{min}}n_{c,\mathrm{min}}L}).
\end{align*}
By Lemma \ref{EigOsum2} and the proof of Theorem 2 \citep{qing2023community}, we have
\begin{align*}
&\hat{f}_{r}=O(\frac{K_{r}}{n_{r,\mathrm{min}}}\|\hat{U}^{*}_{r}Q_{r}-U^{*}_{r}\|^{2}_{F})=O(\frac{\theta^{2}_{r,\mathrm{max}}K^{2}_{r}n_{r,\mathrm{max}}\|S_{r}-\tilde{S}_{r}\|^{2}}{\theta^{6}_{r,\mathrm{min}}\theta^{4}_{c,\mathrm{min}}n^{3}_{r,\mathrm{min}}n^{2}_{c,\mathrm{min}}L^{2}})\\
&\hat{f}_{c}=O(\frac{K_{c}}{n_{c,\mathrm{min}}}\|\hat{U}^{*}_{c}Q_{c}-U^{*}_{c}\|^{2}_{F})=O(\frac{\theta^{2}_{c,\mathrm{max}}K^{2}_{c}n_{c,\mathrm{max}}\|S_{c}-\tilde{S}_{c}\|^{2}}{\theta^{4}_{r,\mathrm{min}}\theta^{6}_{c,\mathrm{min}}n^{2}_{r,\mathrm{min}}n^{3}_{c,\mathrm{min}}L^{2}}).
\end{align*}
Based on Lemma \ref{boundSsum}, we have
\begin{align*}
&\hat{f}_{r}=O(\frac{\theta^{3}_{r,\mathrm{max}}K^{2}_{r}n_{r,\mathrm{max}}\|\theta_{r}\|_{1}\|\theta_{c}\|^{2}_{F}\mathrm{log}(n_{r}+n_{c}+L)}{\theta^{6}_{r,\mathrm{min}}\theta^{4}_{c,\mathrm{min}}n^{3}_{r,\mathrm{min}}n^{2}_{c,\mathrm{min}}L})+O(\frac{\theta^{6}_{r,\mathrm{max}}K^{2}_{r}n_{r,\mathrm{max}}\|\theta_{c}\|^{4}_{F}}{\theta^{6}_{r,\mathrm{min}}\theta^{4}_{c,\mathrm{min}}n^{3}_{r,\mathrm{min}}n^{2}_{c,\mathrm{min}}})\\
&\hat{f}_{c}=O(\frac{\theta^{3}_{c,\mathrm{max}}K^{2}_{c}n_{c,\mathrm{max}}\|\theta_{c}\|_{1}\|\theta_{r}\|^{2}_{F}\mathrm{log}(n_{r}+n_{c}+L)}{\theta^{4}_{r,\mathrm{min}}\theta^{6}_{c,\mathrm{min}}n^{2}_{r,\mathrm{min}}n^{3}_{c,\mathrm{min}}L})+O(\frac{\theta^{6}_{c,\mathrm{max}}K^{2}_{c}n_{c,\mathrm{max}}\|\theta_{r}\|^{4}_{F}}{\theta^{4}_{r,\mathrm{min}}\theta^{6}_{c,\mathrm{min}}n^{2}_{r,\mathrm{min}}n^{3}_{c,\mathrm{min}}}).
\end{align*}
\end{proof}
\bibliographystyle{model5-names}\biboptions{authoryear}
\bibliography{refMLDCScBM}
\end{document}